\documentclass[acmsmall]{acmart}

\AtBeginDocument{%
  }

\usepackage{algorithm}
\usepackage{multirow}
\usepackage[noend]{algpseudocode}
\usepackage{xspace}
\usepackage{amsmath}
\usepackage{array}
\usepackage{graphicx}
\usepackage{subcaption}
\usepackage{listings}
\usepackage{xcolor}
\usepackage{adjustbox}
\usepackage{longfbox}

\lstdefinestyle{mystyle}{
    backgroundcolor=\color{lightgray},
    basicstyle=\ttfamily\small,
    breakatwhitespace=false,         
    breaklines=true,                 
    captionpos=b,                    
    keepspaces=true,                 
    numbers=none,                    
    showspaces=false,                
    showstringspaces=false,
    showtabs=false,                  
    tabsize=2,
    frame=single,                    
    framesep=2pt,                    
    framerule=0pt,                   
    xleftmargin=2pt,                 
    xrightmargin=2pt                 
}

\setcopyright{cc}
\setcctype{by-nc-sa}
\acmDOI{10.1145/3728916}
\acmYear{2025}
\acmJournal{PACMSE}
\acmVolume{2}
\acmNumber{ISSTA}
\acmArticle{ISSTA041}
\acmMonth{7}
\received{2024-10-31}
\received[accepted]{2025-03-31}

\begin{document}

\title{ZTaint-Havoc: From Havoc Mode to Zero-Execution Fuzzing-Driven Taint Inference}

\author{Yuchong Xie}
\orcid{0009-0008-0436-8183}
\affiliation{%
  \institution{Hong Kong University of Science and Technology}
  \city{Hong Kong}
  \country{China}
}
\email{yxiece@cse.ust.hk}

\author{Wenhui Zhang}
\orcid{0009-0004-5231-7736}
\affiliation{%
  \institution{Hunan University}
  \city{Changsha}
  \country{China}
}
\email{Zwenhui@hnu.edu.cn}

\author{Dongdong She}
\orcid{0000-0001-6655-0468}
\affiliation{%
  \institution{Hong Kong University of Science and Technology}
  \city{Hong Kong}
  \country{China}
}
\email{dongdong@cse.ust.hk}


\newcommand{\todo}[1]{\textbf{\textit{\textcolor{red}{#1}}}}
\newcommand{\sys}{ZTaint-Havoc\xspace}
\newcommand{\sysshort}{Z-H\xspace}
\newcommand{\syssolver}{ZTaint-Havoc with solver\xspace}
\newcommand{\syssolvershort}{ZH+S\xspace}
\newcommand{\aflpp}{AFL++\xspace}
\newcommand{\syscmplogdict}{ZTaint-Havoc + Cmplog + Dict\xspace}
\newcommand{\syscmplogdictshort}{ZH+CD\xspace}
\newcommand{\aflppcmplogdict}{AFL++ + Cmplog+Dict\xspace}
\newcommand{\cmplog}{CMPLOG\xspace}
\newcommand{\aflppcmplogdictshort}{AFL+CD\xspace}
\newcommand{\sysdict}{ZTaint-Havoc+D\xspace}
\newcommand{\aflppdict}{AFL++D\xspace}
\newcommand{\numstandalone}{17\xspace}
\newcommand{\numfuzzbench}{19\xspace}
\newcommand{\deleted}[1]{}                   
\newcommand{\added}[1]{{#1}}                   
\newcommand{\replaced}[2]{{#2}}
\begin{abstract}
Fuzzing is a popular software testing technique for discovering vulnerabilities. A central problem in fuzzing is identifying {\footnotesize\texttt{hot bytes}}
that can influence program behavior. 
Taint analysis can track the data flow of {\footnotesize\texttt{hot bytes}} in a white-box fashion, but it often suffers from stability issues and cannot run on large real-world programs. \added{Fuzzing-Driven Taint Inference (FTI) is a simple black-box technique to track {\footnotesize\texttt{hot bytes}} for fuzzing.} It monitors the dynamic program behaviors of program execution instances and further infers {\footnotesize\texttt{hot bytes}} in a black-box fashion. However, this method requires additional $O(N)$ program executions and incurs a large runtime overhead. 

We observe that a widely used mutation scheme in fuzzing--havoc mode can be adapted into a lightweight FTI with \emph{zero additional program execution}. In this work, we first present a computational model of the havoc mode that formally describes its mutation process. Based on this model, we show that the havoc mode can simultaneously launch FTI while generating and executing new testcases. \added{Further, we propose a novel FTI called \sys that doesn't need any additional program execution. \sys incurs minimal instrumentation overhead of 3.84\% on UniBench and 12.58\% on FuzzBench, respectively.} In the end, we give an effective mutation algorithm using the {\footnotesize\texttt{hot bytes}} identified by \sys. 


We conduct a comprehensive evaluation to investigate the computational model of havoc mode. Our evaluation result justifies that it is feasible to adapt the havoc mode to an efficient FTI without any additional program execution. We further implement our approach as a prototype \sys based on the havoc mode of AFL++. We evaluate \sys on two fuzzing datasets FuzzBench and UniBench. 
\added{Our extensive evaluation results show that \sys improves edge coverage by up to {33.71}\% on FuzzBench and {51.12}\% on UniBench over vanilla AFL++, with average improvements of {2.97}\% and {6.12}\% respectively, in 24-hour campaigns.}
\end{abstract}




\maketitle

\section{Introduction}
Fuzzing has become a powerful technique for automated software testing. \added{A core challenge in fuzzing is to identify program input bytes that can influence program behavior, commonly known as {\footnotesize\texttt{hot bytes}}. These {\footnotesize\texttt{hot bytes}} can determine variable values in the program's conditional branches. When perturbing the value of {\footnotesize\texttt{hot bytes}}, a fuzzer is more likely to flip the program's branch conditions and further lead to new code coverage or new vulnerability discovery.} 

\added{Traditional techniques to identify {\footnotesize\texttt{hot bytes}} for fuzzing tasks mainly rely upon dynamic taint analysis. Although it can effectively track {\footnotesize\texttt{hot bytes}} from program input, it suffers from large runtime overhead and fails to scale on large real-world programs\cite{rawat2017vuzzer, chen2018angora, wang2010taintscope}.}
\added{To mitigate this issue, researchers propose a lightweight approach tailored for fuzzing, called fuzzing-driven taint inference (FTI)\cite{Aschermann2019,gan2020greyone,you2019profuzzer,patafuzz}. FTI perturbs the program input, then monitors the difference in corresponding program behaviors when executing the program with the perturbed program input. Based on whether the discrepancy among the program behaviors exists after the program input perturbation, we can infer that the perturbed input bytes are {\footnotesize\texttt{hot bytes}} or not. In general, an effective FTI has two key requirements: \emph{(1) Apply a small perturbation each time to ensure the newly generated program input is still valid; (2) Conduct a large number of perturbations such that the union of perturbed bytes covers all byte locations of program input}. Despite the fact that FTI can easily scale to large real-world programs, it incurs a large runtime overhead. Because they typically employ a byte-level taint inference scheme, i.e., perturb a single byte at a time, which requires \emph{additional} $O(N)$ \emph{program executions} for an input of size $N$.} This substantial overhead raises a critical question: Is it possible to achieve effective taint inference with zero additional execution while maintaining the accuracy of existing approaches?

We observe that the havoc mode, a foundational mutation scheme that has been universally used in modern fuzzers that includes AFL\cite{ZalewskiAFL}, AFL++\cite{fioraldi2020afl++}, libFuzzer\cite{libFuzzer}, and libAFL\cite{fioraldi2022libafl}, inherently satisfies the above two key requirements of FTI through its mutation characteristics. 
The havoc mode has two key properties: First, its mutation operators can apply minimal changes from the original seed, such as single-byte perturbations. Second, these operators select mutation positions uniformly across the byte locations of program input.
These properties suggest the possibility of leveraging existing havoc mutations to build a novel FTI scheme, where every mutation in havoc mutations is considered as one inference instance in FTI.

To systematically understand and leverage these properties, we first present a computational model of the havoc mode, characterizing its mutation process through uniform random selection, stackable mutations, and splicing operations. Building upon this model, we propose a zero-execution FTI technique that integrates seamlessly with the havoc mode mutations. Our approach introduces two havoc-based techniques: \emph{(1)\added{an adaptive} threshold k to adjust the degree of perturbation and normalize the byte influence distribution, and (2) a cumulative taint inference to prevent overtaint while achieving a uniform distribution of influenced bytes.}

\begin{figure}[t]
\centering
\includegraphics[width=\linewidth]{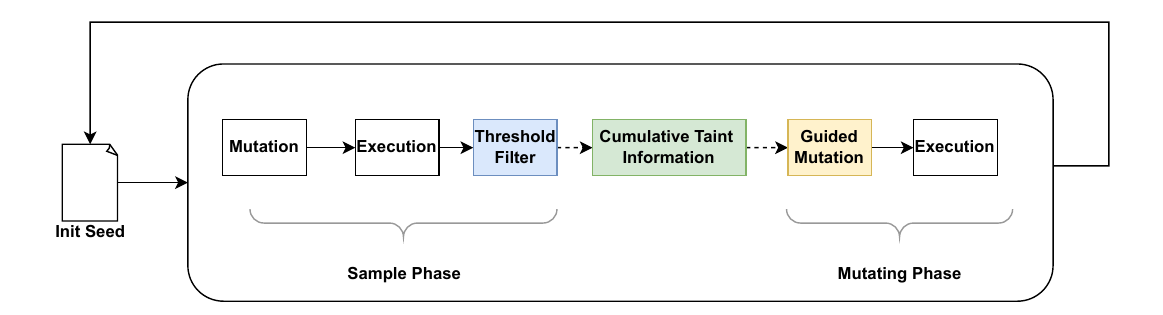}
\caption{\textbf{Workflow of \sys.}}
\label{fig:overview}
\end{figure}

To validate our approach, we implement a two-phase fuzzing algorithm \sys. As shown in Figure \ref{fig:overview}, our method consists of two main phases: a sampling phase and a mutating phase. In the sampling phase, we apply our zero-execution FTI to identify {\footnotesize\texttt{hot bytes}} during normal havoc mode mutations. The mutating phase then leverages this information through a biased havoc mutation strategy focusing on these identified {\footnotesize\texttt{hot bytes}}. We evaluate our approach using two widely-adopted benchmarks: FuzzBench\cite{metzman2021fuzzbench} and UniBench\cite{li2021UniBench}. \replaced{The results demonstrate that our method consistently outperforms AFL++, achieving up to 20.38\% higher code coverage on FuzzBench and 31.65\% on UniBench, with average improvements of 1.32\% and 4.91\% respectively, in 24-hour fuzzing campaigns across diverse real-world applications.}{The results demonstrate that our method consistently outperforms AFL++, achieving up to 33.71\% higher code coverage on FuzzBench and 51.12\% on UniBench, with average improvements of 2.97\% and 6.12\%, respectively, in 24-hour fuzzing campaigns across diverse real-world applications. \sys shows a runtime overhead of 3.84\% on UniBench and 12.58\% on FuzzBench, demonstrating that the additional computational cost remains within an acceptable range while achieving notable coverage gains.}

This research makes the following contributions:
\begin{itemize}
\item We provide the first computational model of the havoc mode, revealing how its mutation characteristics naturally align with the FTI requirements.
\item We present a novel zero-execution FTI method that leverages havoc mode's inherent properties, eliminating the need for additional program execution.
\item We demonstrate a practical implementation that enhances fuzzing effectiveness through taint-guided mutations, validated through a comprehensive evaluation. The code can be available at \url{https://github.com/Yu3H0/ZTaint-Havoc}
\end{itemize}

\vspace{0cm}

\vspace{-0.2cm}
\section{Background}

\subsection{\added{Fuzzing-Driven Taint Inference (FTI)}}
\label{sec:FTI}
\added{
Fuzzing-Driven Taint Inference (FTI)~\cite{Aschermann2019, gan2020greyone, patafuzz, you2019profuzzer} is a lightweight technique to identify {\footnotesize\texttt{hot bytes}} of program input tailored for fuzzing tasks. It first instruments the program at each branch statement to intercept the value of conditional variables that can influence the branch decision. Then it chooses a program input $X$ with $n$ bytes, represented as ${X=\{x_1, x_2, ..., x_n\}}$. FTI records the value of conditional variables $Y$ when dynamically executing the program with $X$. Iteratively, it modifies the $i$-th byte to obtain a mutant program input $X_i$, where $i={1,2,n}$. In the end, it runs the program with $X_i$ and compares the corresponding values of the conditional variables $Y_i$ against $Y$. Once a discrepancy is observed, we can infer that the single-byte difference $x_i$ between $X_i$ and $X$ can influence the program's branching behaviors. Hence, FTI considers the $i$-th byte $x_i$ hot byte. FTI can effectively enhance the fuzzer's capability to solve complex branch constraints in the tested program. However, it usually incurs a large runtime overhead as a result of the additional $O(N)$ program executions.}
An effective FTI method needs to fulfill the following essential criteria:

\textbf{R1: Minimal Difference Between Program Input and Mutant Input:} The discrepancy between parent input and mutant input should be minimal. Large differences can result in divergence in program execution paths, hindering the value comparison of program's conditional variables. Ensuring the minimal difference is essential for precise FTI.
    
\textbf{R2: Thorough Program Input Byte Examination:} Every byte within the program input should be modified and investigated in a successful FTI. Omitting any byte locations could lead to under-tainting, where certain sections of the input are left unexamined. This thorough examination guarantees that no potential taint source is missed during FTI.

\vspace{-0.2cm}
\subsection{Havoc Mode}
AFL's havoc mode, introduced by Zalewski\cite{ZalewskiAFL}, comprises a suite of operators for mutating input data and has been integrated into numerous adapted AFL fuzzers\cite{fioraldi2020afl++,lyu2019mopt,bohme2016coverage}. The havoc mode is invoked after a seed is selected for further mutation. In this process, a sequence of unit mutations is applied incrementally to the seed to generate a new testcase. Each unit mutation is performed by a randomly chosen havoc operator defined by the fuzzer, which typically incorporates many havoc operators such as flipping a bit, flipping a byte, increasing a byte value, or decreasing a byte value. 
The sequence of unit mutations accumulates into a common havoc mode mutation that generates a new testcase. 
Specifically, the havoc mode first determines the total number of unit mutations that will be used to generate a new testcase, commonly denoted as the \texttt{havoc stack}. Then it randomly selects a sequence of havoc operators and incrementally applies them to the seed to generate the new test case. Furthermore, it implements a special mutation stage named the \texttt{splicing}, where the havoc mode mutation is not applied directly to an existing seed from the corpus, but to a newly generated seed combining two random seeds from the corpus.



\section{White Box Interpretation}
\label{sec:whitebox}
In this section, we provide a comprehensive white-box interpretation of havoc mode, diving into its key features and presenting our computational model. We begin by examining the three main components of havoc mode: \texttt{havoc stack}, \texttt{splicing} and \texttt{havoc operator}. Then, we analyze the core mechanism of the havoc mode, revealing how its stochastic nature and uniform distribution of mutations contribute to its magical power in exploring the input space. Finally,  we give a formal description of the havoc mode's computation model.

\vspace{-0.1cm}
\subsection{Havoc Stack}
The havoc stack denotes the total number of unit mutations to be applied incrementally to an initial seed when generating a new testcase. 
The default value of havoc stack varies on modern fuzzers due to different implementation details of havoc mode\cite{libFuzzer, ZalewskiAFL, fioraldi2020afl++}. \replaced{For instance, AFL randomly selects a value from the set \([2, 4, 8, 16, 32, 64, 128]\), whereas the value in AFL++ varies depending on the fuzzing process's queue cycle.}{For instance, in AFL, the value is randomly selected from the set \([2, 4, 8, 16, 32, 64, 128]\), whereas in AFL++ (v4.10c), the value is randomly selected from the range \([1, 16]\) during the fuzzing process.}

\vspace{-0.1cm}
\subsection{Splicing}
Splicing is a common strategy in numerous fuzzers, appearing mainly as two forms: the splicing stage\cite{ZalewskiAFL} and the splicing operator\cite{libFuzzer}. In fuzzers like AFL and AFL++, the splicing stage merges two seeds to produce a new initial seed for further mutation. This generally involves pinpointing the differences between two seeds and merging them from a random position within these differences, usually connecting the head of one seed with the tail of another. On the other hand, the splicing operator, used by fuzzers such as libFuzzer, serves as an independent mutation method. Both strategies strive to generate more varied inputs, potentially enabling fuzzers to explore novel regions of the input space that could be challenging to access through incremental mutations alone.

\vspace{-0.1cm}
\subsection{Havoc Operator}
The havoc operator is a basic input mutation module in havoc mode. It takes a program input and outputs a mutant program input. Modern fuzzers implement many havoc operators, where each havoc operator applies a unique mutation scheme to the given input. A havoc operator typically selects a random position within the input byte sequence and then applies a specified mutation pattern on the selected byte locations. The following code snippet exemplifies a classic havoc operator, specifically the \texttt{MUT\_FLIPBIT} operation:
\begin{lstlisting}
    u8  bit = rand_below(afl, 8);
    u32 off = rand_below(afl, temp_len);
    out_buf[off] ^= 1 << bit;
\end{lstlisting}
We observe that a havoc mode mutation consisting of a sequence of havoc operators is essentially a Markov process, where each havoc operator depends only on the output of the prior unit mutation. Formally, we define the havoc operator as a function $f(s)$ that takes input $s$ and outputs $m$.
\begin{equation}
    \text{Havoc Operator}: f(s) = m
\end{equation}
\added{
We then define the perturbation distance $D(s, m)$ to measure the difference between two seeds $s$ and $m$. The difference can be measured by Hamming distance\cite{hamming1950error}, edit distance\cite{levenshtein1966binary}, or other sequence difference metrics. The perturbation distance between an initial seed $s$ and its mutant $m$ is approximately proportional to the havoc stack size $h$:}
\begin{equation}
    D(s,m) \approx h \cdot D(s,f(s))
\end{equation}

\vspace{-0.1cm}
\subsection{Havoc Mode's Computation Model}
\label{subsec:model}
After explaining the three major components of the havoc mode in detail, we now give a formal description of the havoc mode's computation model as follows:
\begin{equation}
 \text{Havoc Mode}: m = \left\{
 \begin{array}{ll}
 f^h(s)=(\underbrace{f(f(\cdots f(s) \cdots))}_{h \text{ times}}) & \small{\text{Normal Stage}}\\
 f^h(s')=(\underbrace{f(f(\cdots f(s')) \cdots))}_{h \text{ times}}), s'=splice(s) & \small{\text{Splicing Stage}} 
 \end{array}
 \right.
\end{equation}
The havoc mode consists of two stages: the normal stage that applies havoc mode mutation on a selected seed and the splicing stage that applies havoc mode mutation on a spliced seed.  
$m$ represents the mutant testcase generated by the havoc mode mutation when given seed $s$. The function $f(s)$ denotes the havoc operator and $f^h(s)$ denotes the recursive iterations $h$ times of the function $f(s)$ in seed $s$, representing the incremental perturbation of the havoc operators for $h$ times in the common stage of the havoc mode.   
Moreover, the splicing operator $splice(s)$ is used during the splicing stage. It takes seed $s$ as input and returns a spliced seed $s'$ as output.
\added{There are two types of splicing operators. The first type is used during the splicing stage and is applied before other operators. The second type is implemented as a splicing operator in some fuzzers, such as AFL++~\cite{fioraldi2020afl++}, and in such cases, we treat it as part of the havoc operator $f(s)$.}

After havoc stack $h$ is determined, we can represent the entire havoc mode as a Markov process. The state space is the set of all possible seeds. The transition matrix is the probability of the seed mutating to another seed. The initial state is the initial seed. The final state is the mutant. The goal of the Havoc mode is to find the mutant that can reach more basic block codes or trigger the target program to crash.

\vspace{-0.1cm}
\section{Connecting Havoc Mode with \added{Fuzzing-Driven} Taint Inference}
In this section, we illustrate how the havoc mode theoretically satisfies the requirements of FTI. In the havoc mode, each seed goes through a sequence of havoc operators and generates many mutant testcases. Although not every mutant test case is necessarily suitable for FTI, there exist sufficient mutant testcases for an effective FTI. Specifically, we explain how the mutation characteristics of the havoc mode ensure the two critical requirements of FTI, as mentioned in Section~\ref{sec:FTI} .

\begin{figure}[!h]
\centering
\includegraphics[width=\linewidth]{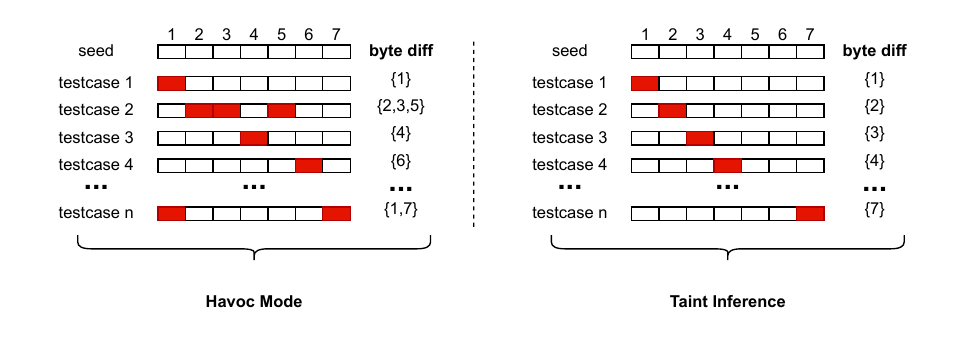}
\caption{\textbf{An example illustrating how havoc mode can be used for FTI. The left side shows the havoc mutation process where testcases are generated with different bytes modified (marked in \textcolor{red}{red}) from a 7-byte seed, with each byte represented by a block. The right side demonstrates FTI process. Through multiple iterations, havoc mode covers all byte locations similar to FTI.}}
\label{fig:havoc_taint}
\end{figure}

Our insight is that the large number of mutant testcases generated by havoc mode can be used to perform an FTI without additional program execution.    
As shown in Figure \ref{fig:havoc_taint}, the havoc mode select and perturb almost every byte position of the seed after generating a large number of mutant testcases. 
Among these mutant testcases, many only have a single-byte difference from the seed. We observe that when the havoc stack is set to a small number like 1, the generated mutant testcase would only contain one-byte differences from the original seed. Since this single-byte mutation will eventually cover all byte positions due to the uniform selection strategy, these mutant testcases naturally satisfy both \textbf{R1} (minimal perturbation) and \textbf{R2} (comprehensive coverage) requirements.

Furthermore, while single-byte mutations enable an effective FTI, they may lead to undertaint issues in some corner cases. Consider a case where program behavior is triggered by multiple bytes simultaneously - for instance, when three input bytes undergo an \texttt{AND} operation. If the input is \texttt{'000'}, 
single-byte mutations can generate mutant testcases \texttt{'100'}, \texttt{'010'} and \texttt{'001'} which might not influence the program state. 
Meanwhile, a simultaneous change to all three bytes \texttt{'111'} could significantly impact execution. In practice, the havoc stack in modern fuzzer like AFL++ is related to the mutation intensity, naturally producing larger perturbations when needed. This property ensures that our approach can capture multi-byte dependencies while still maintaining \textbf{R1}'s similarity requirement through controlled havoc stack sizes.

\vspace{-0.1cm}
\section{\added{Zero-Execution Fuzzing-Driven Taint Inference}}
 We present our methodology for leveraging the havoc mode for zero-execution FTI:
\begin{enumerate}
    \item \textbf{Mutation Threshold:} We introduce a parameter $k$ to adjust the divergence between the mutant testcase and the initial seed. Specifically, this threshold limits the number of byte locations whose value differs between the mutant testcase and the input, thus maintaining a balance between exploration and inference accuracy.
    \item \textbf{Cumulative Taint Inference:} We implement a mechanism to record the count of valid taint byte locations among the large number of mutant testcases. This approach allows for the progressive refinement of the taint data, leveraging the insights gained from each successive analysis to enhance the overall accuracy of the inference process.
\end{enumerate}
In the following subsections, we discuss the mutation threshold and cumulative taint inference mechanisms in detail.

\vspace{-0.1cm}
\subsection{Mutation Threshold}
To control the magnitude of divergence between the input seeds and their mutant test cases generated by havoc, we introduce an adaptive parameter $k$, which defines an upper bound to the number of byte differences. During normal stages, we directly use initial seeds as inputs. In contrast, in the splicing stage where two seeds are combined to create new test cases, we use the resulting spliced seeds as inputs to ensure stable starting points before havoc mutations.
Formally, we define this constraint between seed $seed$ and mutant testcases $mutant$ as follows:\added{
\begin{equation}
    D_{\text{Hamming}}(seed, mutant) \leq k
\end{equation}
where $D_{\text{Hamming}}(seed, mutant)$ denotes the Hamming distance between two $seed$ and $mutant$, representing the number of positions at which the corresponding bytes are different.}
\added{The value of $k$ is adaptively adjusted based on the seed length and mutation count in the havoc stage:}
\added{
\begin{equation}
    k = \beta \cdot \frac{L}{N_{mut}}
\end{equation}
}
\added{Here, $L$ represents the byte length of $seed$, and $N_{mut}$ is the number of mutant testcases to be generated by the havoc mode when given $seed$. The parameter $\beta$ is a scaling factor that governs how $k$ is computed based on these parameters.}

\replaced{The introduction of this threshold $k$ offers two significant advantages.}{This adaptive threshold mechanism offers several advantages}:

\begin{itemize}
    \item \added{\textbf{Length-Aware Sampling:} By scaling $k$ with the input length, we ensure havoc mode can cover almost all byte locations of the seed. A long seed typically is assigned a proportionally large $k$ value, enabling more bytes to be modified within each individual mutant testcase.}
    
    \item \added{\textbf{Mutation-Aware Adjustment:} The adjustment based on mutation count ensures that when more mutations are planned, each mutation affects fewer bytes, leading to more controlled and evenly distributed changes across the seed.}

    \item \textbf{Path Preservation:} By limiting the extent of mutations, we mitigate the risk of generating mutants that deviate excessively from the initial seed. This helps preserve the validity of execution-path comparisons, which is crucial for accurate taint inference.

    \item \textbf{Distribution Normalization:} While havoc operations are theoretically designed to modify bytes uniformly, certain fuzzers employ operators like \texttt{MUT\_OVERWRITE\_COPY} that can skew this distribution. Such operators first select a variable length and then choose a mutation position, resulting in a non-uniform distribution of influenced bytes, typically with higher frequency in the middle and lower at the extremities. By imposing a threshold $k$, we effectively normalize this distribution. 
\end{itemize}

\vspace{-0.1cm}
\subsection{Cumulative Taint Inference}



Building upon the mutation threshold mentioned above, we propose a cumulative taint inference method to further improve accuracy and mitigate overtaint issues. We denote the set of bytes in the input that truly influence program execution (true taints) as $T$, the set of input bytes that do not affect program execution (irrelevant inputs) as $I$, and the set of bytes that differ in each inference process as $D$.

Note that we cannot treat the inferred taint bytes from each iteration as equivalent, nor can we use a single aggregate set to represent tainted bytes. This is because the uniform distribution of havoc operators would eventually cover the entire input space, even if only elements of $T$ affect program execution, leading to over-tainting.

To address this challenge, we propose a data structure: an array of the same length as the input, with each element initially set to zero. During each inference iteration, we increment the value at the index corresponding to each byte in $D$.
The advantage of this approach lies in its ability to differentiate between true taints and noise. Since each inference result invariably includes elements from $T$, after multiple iterations, the values corresponding to elements in $T$ will be consistently higher than those corresponding to elements in $I$. This method allows us to obtain a more accurate approximation of $T$, thereby reducing the impact of over-tainting.

\vspace{-0.1cm}
\section{\added{Enhanced Havoc Mode with Zero-Execution FTI}}
Our approach bifurcates the havoc process for each input into two distinct phases in \autoref{alg:zero}:
\begin{enumerate}
    \item Fuzzing-driven taint inference, as described in the preceding sections.
    \item Taint-guided mutation, which leverages inferred taint information.
\end{enumerate}

\begin{algorithm}[h]
\footnotesize
\caption{\added{Enhanced Havoc Mode with Zero-Execution FTI}} 
\label{alg:zero} 
\begin{algorithmic}[1]
\State Initialize $taint\_count[1..n] \gets 0$ where $n$ is input length
\State $seed \to$ initial seed
\State \textcolor{purple}{/* Phase 1: Vanilla havoc mode with zero-execution FTI */}
\State $pb_{seed} = execute(seed)$ \Comment{\textcolor{purple}{Record program behavior for $seed$}}
\For { $i = 1, 2,...,n_1$}
    \State $t_i = havoc\_mutate(seed)$ \Comment{\textcolor{purple}{Vanilla havoc mode generates a testcase $t_i$}}
    \State $pb_i = execute(t_i)$ \Comment{\textcolor{purple}{Record program behavior for $t_i$}}
    \State $ByteDiff = diff(t_i, seed)$ 
    \If{ $|ByteDiff|\leq k$} \Comment{\textcolor{purple}{Skip testcases with too many byte diffs to avoid overtaint}}
        \If{ $pb_i \neq pb_{seed}$} 
            \For{each index $j$ in $ByteDiff$}
                \State $taint\_count[j] \gets taint\_count[j] + 1$
            \EndFor
        \EndIf
    \EndIf
\EndFor
\State \textcolor{purple}{/* Phase 2: Biased havoc mode focusing on bytes with high taint count */}
\For { $i = 1, 2,...,n_2$}
     \State $t_i = biased\_havoc\_mutate(seed, taint\_count)$ \Comment{\textcolor{purple}{Mutate bytes based on taint count}}
     \State $execute(t_i)$ 
\EndFor
\end{algorithmic}
\end{algorithm}

Upon completion of the first phase, we obtain a data structure $taint\_count$, which encapsulates the cumulative count of inferences for each byte position. This structure serves as the foundation for our taint-guided mutation strategy.

We propose a novel method that adapts the existing havoc operators to incorporate taint information. The core of this adaptation lies in the modification of the random selection process used by these operators. Specifically, we replace the uniform random function $rand()$ with a variant aware of the taint that aligns with the distribution implied by $taint\_count$.
Formally, we define our taint-guided random function as follows:
\begin{equation}
P(i) = \frac{taint\_count[i]}{\sum_{j=1}^{n} taint\_count[j]}
\end{equation}
where $P(i)$ represents the probability of selecting the $i$-th byte as the starting point for a mutation operator, and $n$ is the length of the input.

This probabilistic selection mechanism ensures that bytes with higher taint counts, those more likely to influence program behavior, have a correspondingly higher probability of being chosen as the focal points for mutation operations. Consequently, our mutation strategy becomes more targeted, concentrating on the most influential parts of the input as identified by our taint inference process.



\vspace{-0.1cm}
\section{Implementation}
In this section, we introduce \sys, our proof-of-concept implementation designed for zero-execution FTI. We use the AFL++ version 4.10c, which is the latest and most performant version of the widely recognized state-of-the-art fuzzer\cite{asprone2022comparing, fuzzcomp_report, aflpp_fuzzbench_report}, as the base fuzzer. 

\sys enhances the AFL++ LLVM-based instrumentation pass to extract the intraprocedural control flow graph (CFG) for each function and incorporates its metadata into the binary. During a fuzzing campaign, we dynamically label nodes in the CFG as visited or unvisited based on coverage data. We also maintain an updated list of frontier nodes, concentrating exclusively on frontier branches, a subset of frontier nodes that includes control instructions leading to conditional jumps. For each control instruction, we instrument an instruction or a function to detect changes in the control condition. For string comparison types, such as strcmp or memcmp, we insert a function, while for icmp types, we insert a single instruction. To reduce run-time overhead, we add an adaptive switch for each hook, only enabling the hook for frontier branches.

\replaced{In the mutation part, when a seed is selected as an initial seed for havoc, we first run the seed in the target program and record the frontier branch and the corresponding control instruction information. During the havoc stage, the first 50\% mutations are seen as sampling. We only concentrate on the mutations that differ 5\todo{change the adaptive k} bytes compared with the initial seed.}{In the mutation part, when a seed is selected as an initial seed for havoc, we first run the seed in the target program and record the frontier branch and the corresponding control instruction information. Then we calculate an adaptive threshold $k$ bounded between 1 and $\lfloor len(s)/32 \rfloor$, where $len(s)$ is the length of the initial seed. During the havoc stage, after the first run, the first 50\% mutations are seen as sampling, and we only concentrate on mutations that differ within $k$ bytes compared with the initial seed.} After these mutations are executed, we record the frontier branch and the corresponding control instruction information again. If the control instruction information is different from the initial seed, we will taint the 5 different bytes. Specifically, we will maintain a taint array; each time we taint a byte, we will increment the corresponding position in the taint array by 1. After the sampling stage, we will use the taint information to guide the mutation. We will utilize the taint array to assess the probability of mutation for each byte. A higher value indicates an increased likelihood that that byte is selected as the starting point for mutation.

\vspace{-0.1cm}
\section{Evaluation}
In this section, we aim to answer the following research questions:
\begin{itemize}

\item \textbf{RQ1:} Does havoc mode follow a uniform distribution, and how does the introduction of thresholds affect this distribution?
\item \textbf{RQ2:} How does uniform distribution compare to alternative distributions in terms of effectiveness?
\item \textbf{RQ3:} How does the havoc stack size impact the perturbation distance between the seed input and its mutants?
\item \textbf{RQ4:} What is the impact of splicing operations on perturbation distance?
\item \textbf{RQ5:} How does \sys perform in terms of code coverage compared to AFL++?
\item \textbf{RQ6:} How does the threshold parameter $k$ affect the fuzzing performance in code coverage?
\item \added{\textbf{RQ7:} What is the runtime overhead of  \sys compared to the baseline?}
\item \added{\textbf{RQ8:} How does \sys compare to \cmplog in terms of code coverage?}
\end{itemize}

\vspace{-0.1cm}
\subsection{Benchmark}
For our experiment, we chose FuzzBench\cite{metzman2021fuzzbench} and UniBench dataset\cite{li2021UniBench}. FuzzBench has established itself as a standard benchmark in fuzzing research\cite{asprone2022comparing,bohme2022reliability,liu2023sbft}, providing a controlled environment with well-designed harnesses and standardized evaluation metrics. To further enhance the comprehensiveness of our evaluation, we also incorporated standalone programs from the UniBench dataset, which features a diverse collection of real-world applications processing various data formats including images, videos, audio files, text documents, and binary files. We selected \numfuzzbench targets from the FuzzBench dataset and \numstandalone targets from the UniBench dataset, as the remaining programs have compilation issues, such as dependency incompatibility. The target list is shown in Table \ref{tab:programs}.

\begin{table}[]
\footnotesize
    \centering
    \caption{\textbf{Studied programs in our evaluation.}}
    \label{tab:programs}
    \begin{tabular}{llr|llr|llr}
    \toprule
    \textbf{Targets} & \textbf{Version} & \textbf{\# Edge} & 
    \textbf{Targets} & \textbf{Version} & \textbf{\# Edge} & 
    \textbf{Targets} & \textbf{Version} & \textbf{\# Edge} \\
    \midrule
    \multicolumn{3}{c|}{\textbf{FuzzBench Targets}} & 
    sql & c78cbf2 & 77,147 &
    tcpdump & 4.8.1 & 32,760 \\
    \cline{1-3}
    bloaty & 52948c1 & 163,038 &
    systemd & 07faa49 & 1,059 &
    jq & jq-1.5 & 6,177 \\
    zlib & zlib-1.2.13 & 4,634 &
    jsoncpp & 8190e06 & 9,820 &
    sqlite3 & SQLite-3.8.9 & 67,533 \\
    lcms & f0d9632 & 13,649 &
    libjpeg & b0971e4 & 17,783 &
    cflow & cflow-1.6 & 7,690 \\
    libpcap & c639612 & 15,450 &
    libpng & cd0ea2a & 9,032 &
    exiv2 & exiv2-0.28.0 & 122,536 \\
    libxml2 & c7260a4 & 82,498 &
    libxslt & 180cdb8 & 60,091 &
    ffmpeg & ffmpeg-4.0.1 & 586,737 \\
    openh264 & fa6d099 & 18,677 &
    openssl & b0593c0 & 85,304 &
    infotocap & ncurses-6.1 & 11,470 \\
    \cline{4-6}
    re2 & 4be2407 & 15,173 &
    \multicolumn{3}{c|}{\multirow{2}{*}{\textbf{UniBench Targets}}} &
    nm-new & binutils-2.28 & 57,709 \\
    stbi & 5736b15 & 7,655 &
    \multicolumn{3}{c|}{} & 
    objdump & binutils-2.28 & 77,318 \\
    \cline{4-6}
    vorbis & 84c0236 & 12,895 &
    mp42aac & 1.5.1-628 & 21,707 &
    pdftotext & xpdf-4.00 & 49,169 \\
    woff2 & 9476664 & 19,609 &
    lame & lame-3.99.5 & 15,776 &
    mp3gain & 1.5.2 & 3,344 \\
    curl & a20f74a & 122,327 &
    mujs & mujs-1.0.2 & 21,288 &
    tiffsplit & tiff-3.9.7 & 12,527 \\
    harfbuzz & cb47dca & 78,196 &
    flvmeta & flvmeta-1.2.1 & 6,512 &
    jhead & 3.00 & 2,092 \\
    \bottomrule
    \end{tabular}
\end{table}

\vspace{-0.1cm}
\subsection{Mutation Distance}
In this section, we assess the perturbation level using four distinct distance metrics: the L0-norm, L1-norm, L2-norm, and the edit distance (Levenshtein distance). Each of these metrics provides unique insights into the nature of mutations introduced by fuzzing techniques.


\added{Let \( x \) denote our initial seed, and \( y \) represent the post-disturbance corpus. When the lengths of \( x \) and \( y \) differ, we pad the shorter sequence with zeros to ensure equal lengths. We then transform them into two arrays, \( X \) and \( Y \), where each element ranges from 0 to 255. The length of these arrays is denoted as \( n \), which is the length of the sequences after padding. The difference array \( Diff \) is computed as:
\begin{equation}
Diff_i = X[i] - Y[i] \quad \text{for } i = 1, 2, \ldots, n
\label{eq:distance}
\end{equation}
For the L0, L1, and L2 norms, we employ the generalized Lp-norm~\cite{VectorNorm}:
\begin{equation}
\|\mathbf{Diff}\|_p = \left( \sum_{i=1}^{n} |Diff_i|^p \right)^{\frac{1}{p}}
\label{eq:lp_norm}
\end{equation}
}
where L0 counts non-zero elements, representing the number of changed bytes (note that L0 is not a true norm as it violates norm properties); L1-norm (\( p = 1 \)) sums absolute differences, measuring total change magnitude; and L2-norm (\( p = 2 \)) is the Euclidean distance, giving more weight to larger changes.

Additionally, we introduce the edit distance, defined as the minimum number of single-character edits (insertions, deletions, or substitutions) required to change one string into another. It is particularly useful for comparing strings of different lengths and capturing structural changes.
By employing these complementary metrics, we can comprehensively analyze the nature and extent of mutations introduced by various fuzzing techniques, providing a multi-faceted view of their effectiveness and characteristics.
\subsection{RQ1: Uniform Sampling}
\label{text:rq1}

\subsubsection{Experimental Setup}
To validate our hypotheses regarding mutation distributions and the effectiveness of our proposed threshold mechanism, we conducted a series of experiments that focused on two key dimensions: the distribution of mutation \textbf{starting positions} and the distribution of \textbf{influenced bytes}. 
\added{The havoc mode is essentially a string edit process using a set of fixed rules, as formally described in Section~\ref{subsec:model}. It generates many mutant inputs when given a seed input. Note that this mutant generation step is orthogonal to program logic because the variance in program logic does not influence how mutant inputs are generated. Specifically, the program logic is only used to determine the code coverage-based feedback after the mutant generation. Our evaluation collects data from the mutant generation phase. We selected 10 representative programs with varying seed lengths to maintain experimental feasibility.}
For each target, we selected an initial seed of moderate length from those provided by FuzzBench and UniBench. \autoref{tab:seed} shows the lengths of the selected seeds. Our experiments involved executing 100,000 iterations of the havoc stage using a single seed for each target. For each target program, we conducted 10 independent measurements to ensure statistical reliability.

\begin{table}[h]
\small
    \centering
    \caption{\textbf{Seed length of the selected programs. * indicates targets from UniBench.}}
    \label{tab:seed}
    \begin{tabular}{c|c||c|c}
    \hline
    \textbf{Binary} & \textbf{Length} & \textbf{Binary} & \textbf{Length} \\ \hline
    \texttt{cflow*} & 38,139 & \texttt{vorbis} & 2,603 \\ 
    \texttt{bloaty} & 22,120 & \texttt{openssl} & 1,114 \\
    \texttt{pdftotext*} & 12,567 & \texttt{tcpdump*} & 531 \\
    \texttt{tiffsplit*} & 9,834 & \texttt{libpcap} & 220 \\
    \texttt{ffmpeg*} & 6,714 & \texttt{lcms} & 128 \\ \hline
    \end{tabular}
\end{table}

Our first experiment aimed to verify the uniformity of mutation starting positions. We maintained the stack size settings of vanilla AFL++ and focused on the 28 substitution operators among AFL++'s 37 havoc operators, as these preserve seed length. By tracking the starting position of each mutation operation, we sought to demonstrate that the selection of starting positions is approximately uniform across the input space.

The second dimension of our analysis focused on the distribution of influenced bytes resulting from mutation operations, where influenced bytes are defined as bytes that differ between the original seed and the mutated input. We only consider mutants that maintain identical execution paths as their initial seeds. We conducted two sets of experiments: one without constraints and the other applying a threshold of $k = 5$. These experiments aimed to demonstrate that without a threshold, the distribution of influenced bytes is nonuniform, whereas imposing a threshold leads to a more uniform distribution across the input space.

\subsubsection{Observations}
\begin{figure}[!htb]
    \centering
    \begin{subfigure}{0.18\textwidth}
        \centering
        \includegraphics[width=\linewidth]{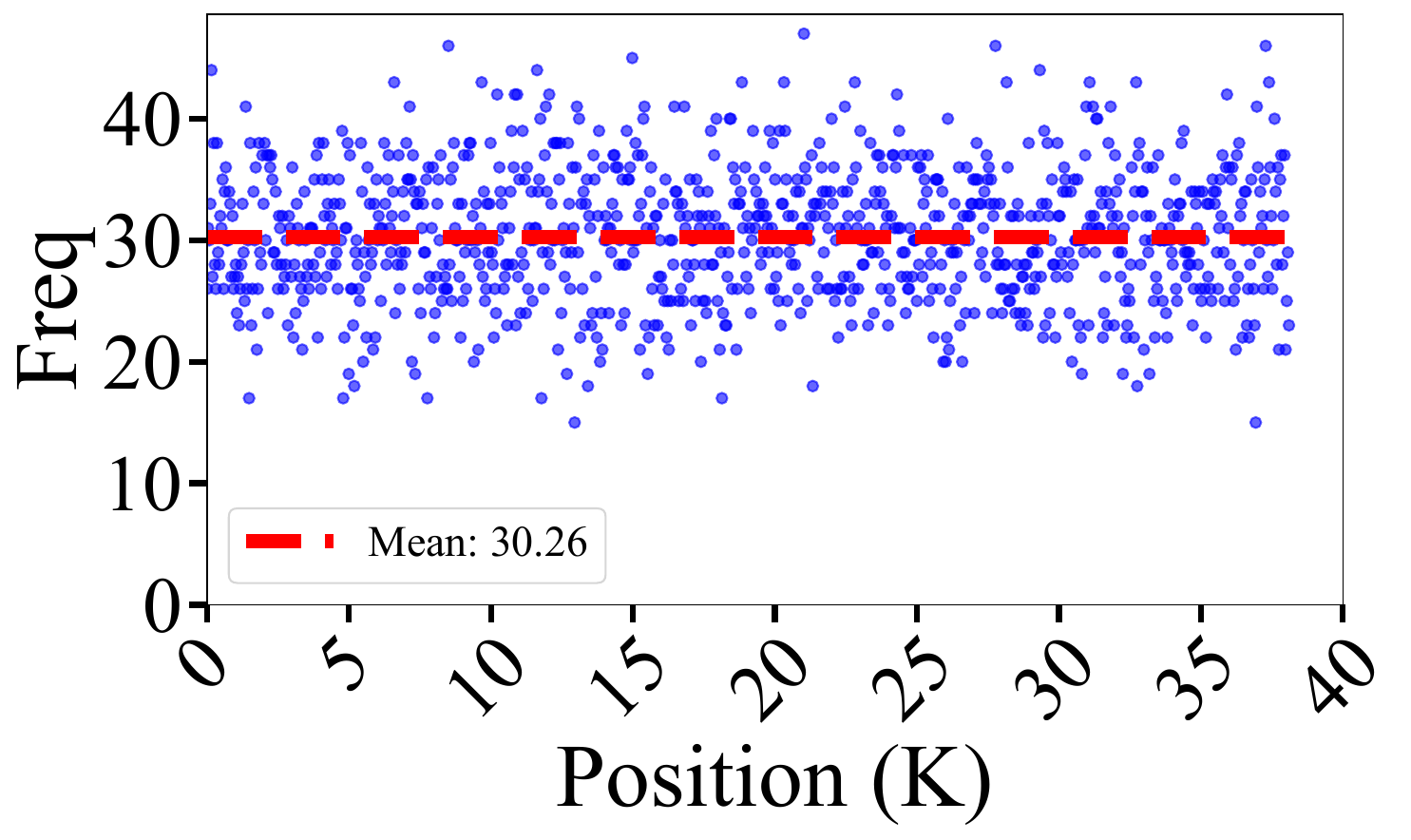}
        \caption{Cflow}
        \label{fig:cflow_1_1}
    \end{subfigure}
    \hfill
    \begin{subfigure}{0.18\textwidth}
        \centering
        \includegraphics[width=\linewidth]{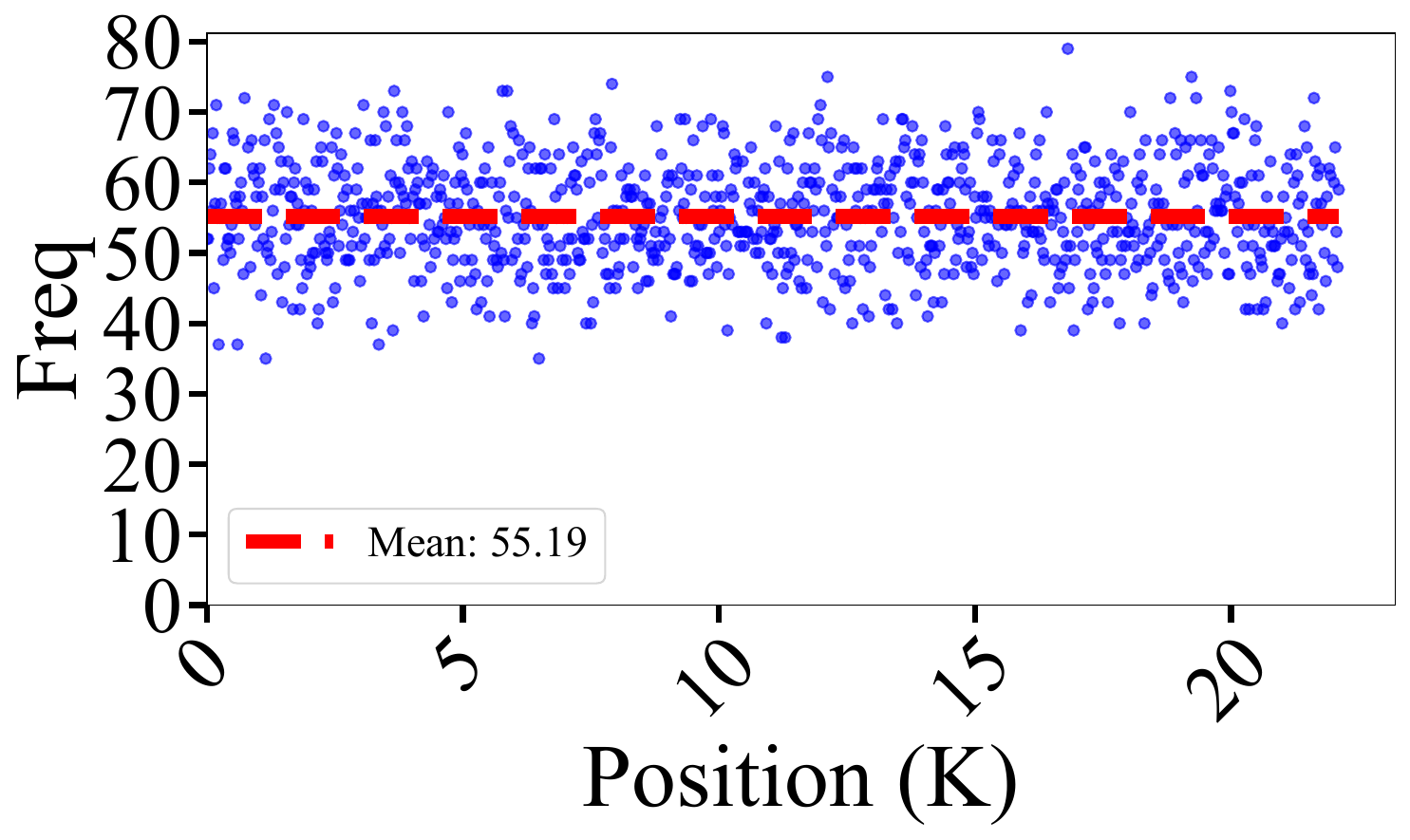}
        \caption{Bloaty}
        \label{fig:bloaty_1_1}
    \end{subfigure}
    \hfill
    \begin{subfigure}{0.18\textwidth}
        \centering
        \includegraphics[width=\linewidth]{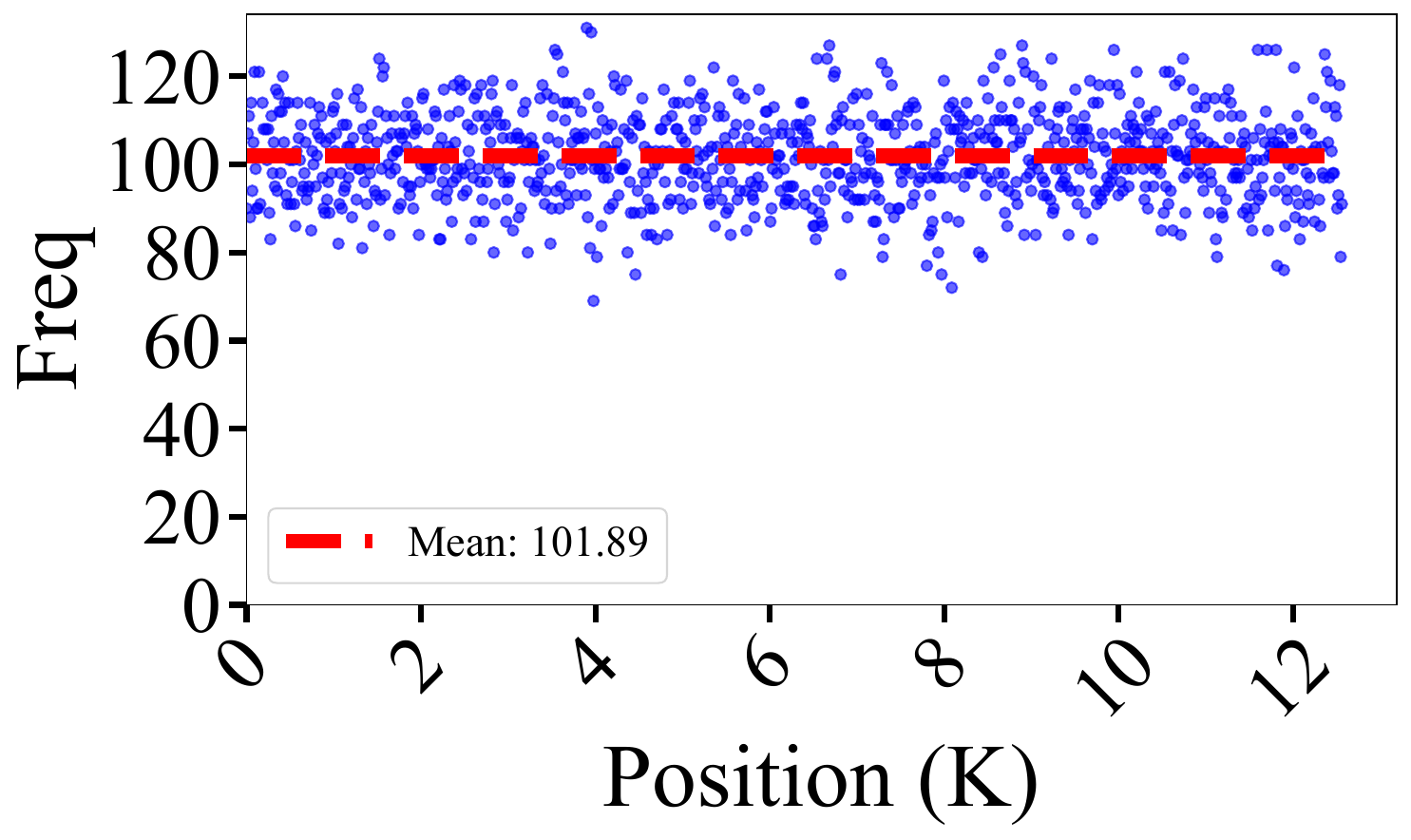}
        \caption{Pdftotext}
        \label{fig:pdftotext_1_1}
    \end{subfigure}
    \hfill
    \begin{subfigure}{0.18\textwidth}
        \centering
        \includegraphics[width=\linewidth]{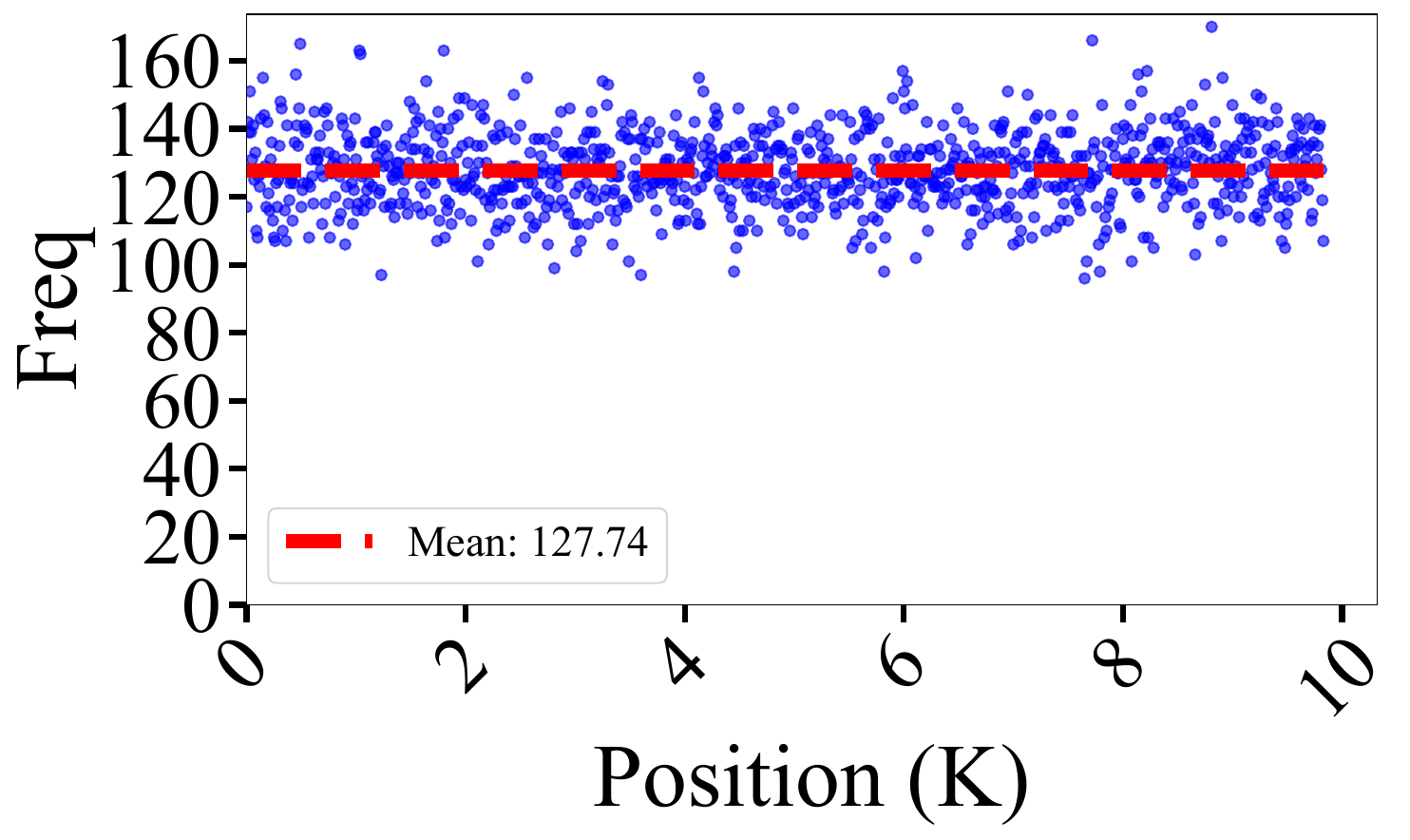}
        \caption{Tiffsplit}
        \label{fig:tiffsplit_1_1}
    \end{subfigure}
    \hfill
    \begin{subfigure}{0.18\textwidth}
        \centering
        \includegraphics[width=\linewidth]{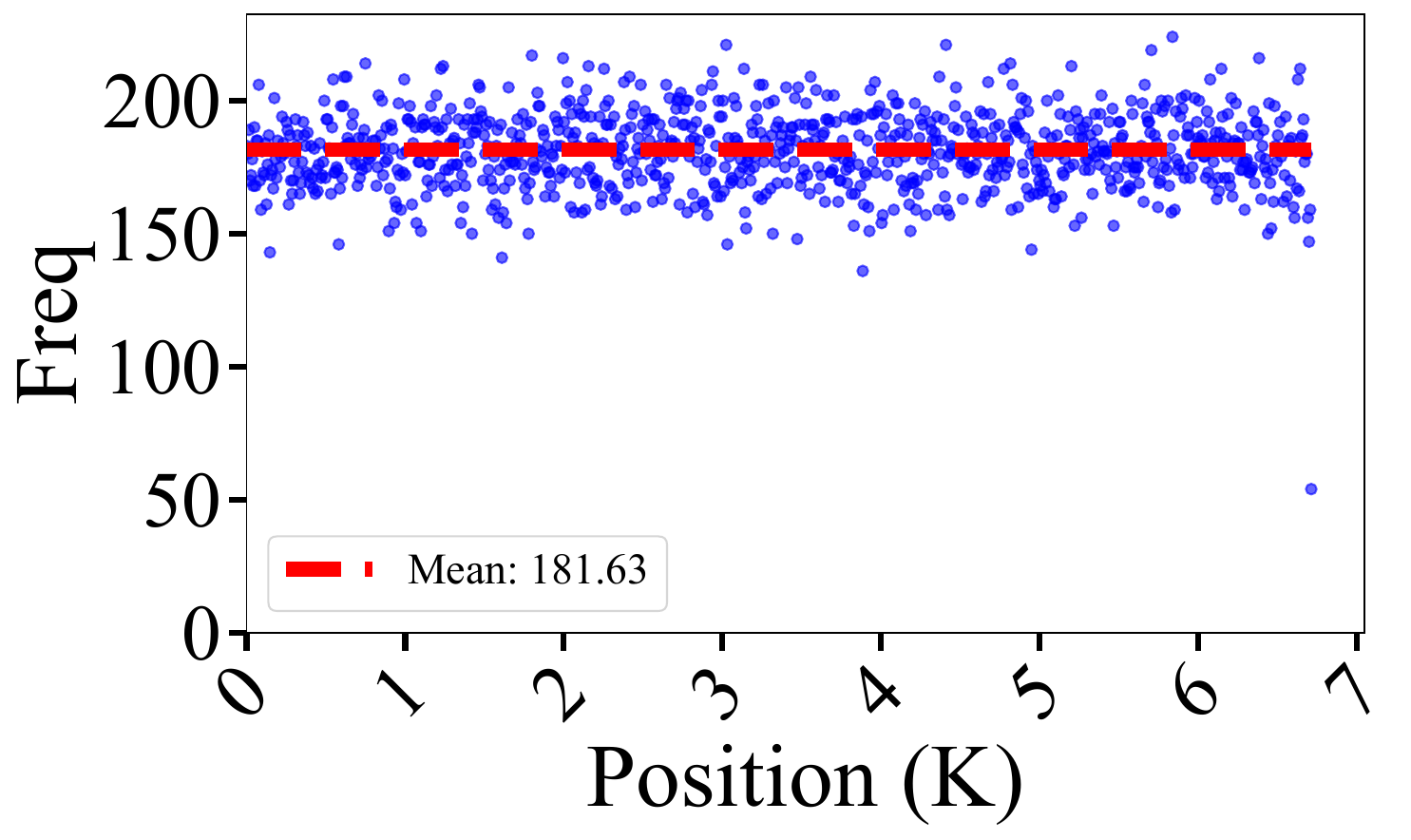}
        \caption{FFmpeg}
        \label{fig:ffmpeg_1_1}
    \end{subfigure}


    \begin{subfigure}{0.18\textwidth}
        \centering
        \includegraphics[width=\linewidth]{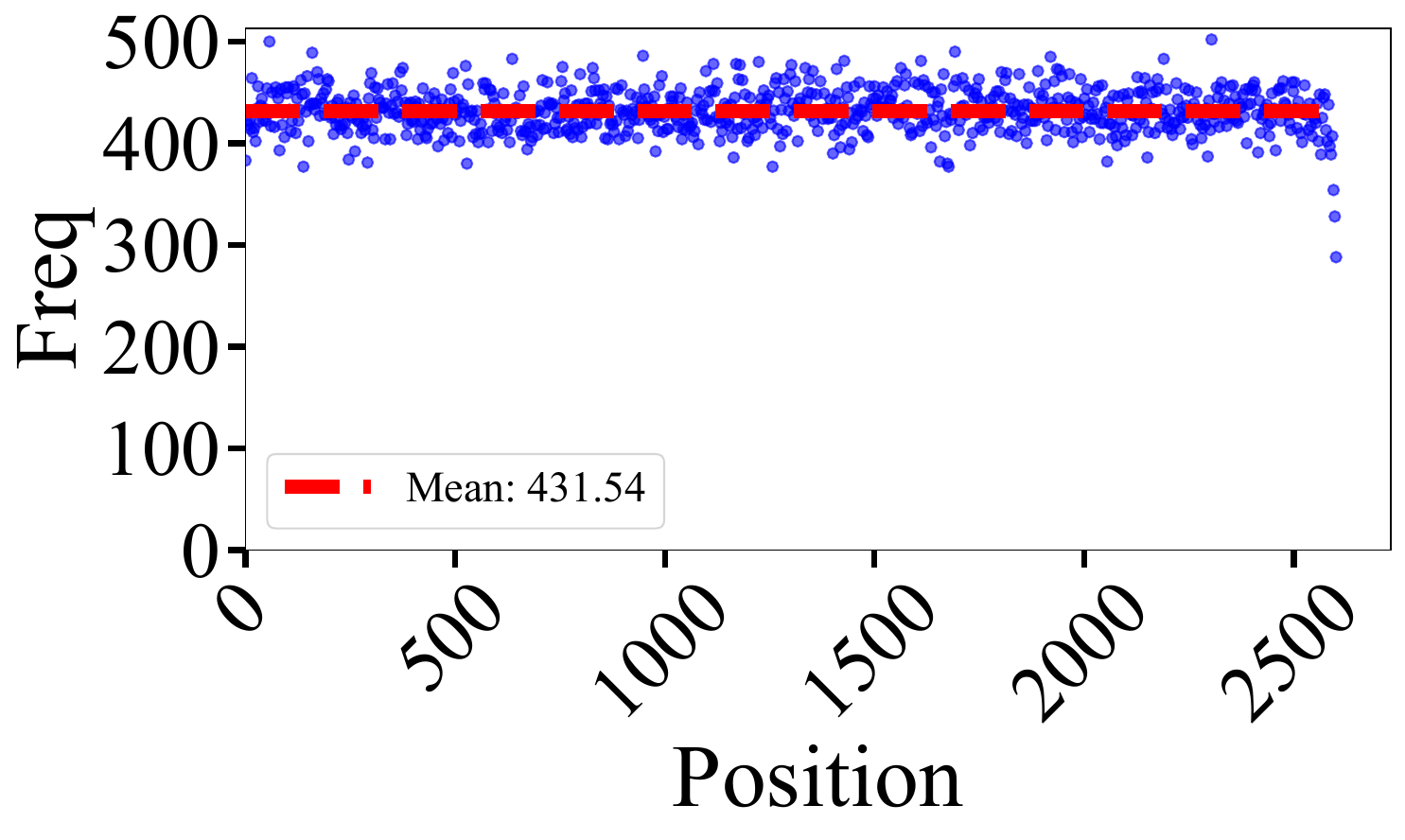}
        \caption{Vorbis}
        \label{fig:vorbis_1_1}
    \end{subfigure}
    \hfill
    \begin{subfigure}{0.18\textwidth}
        \centering
        \includegraphics[width=\linewidth]{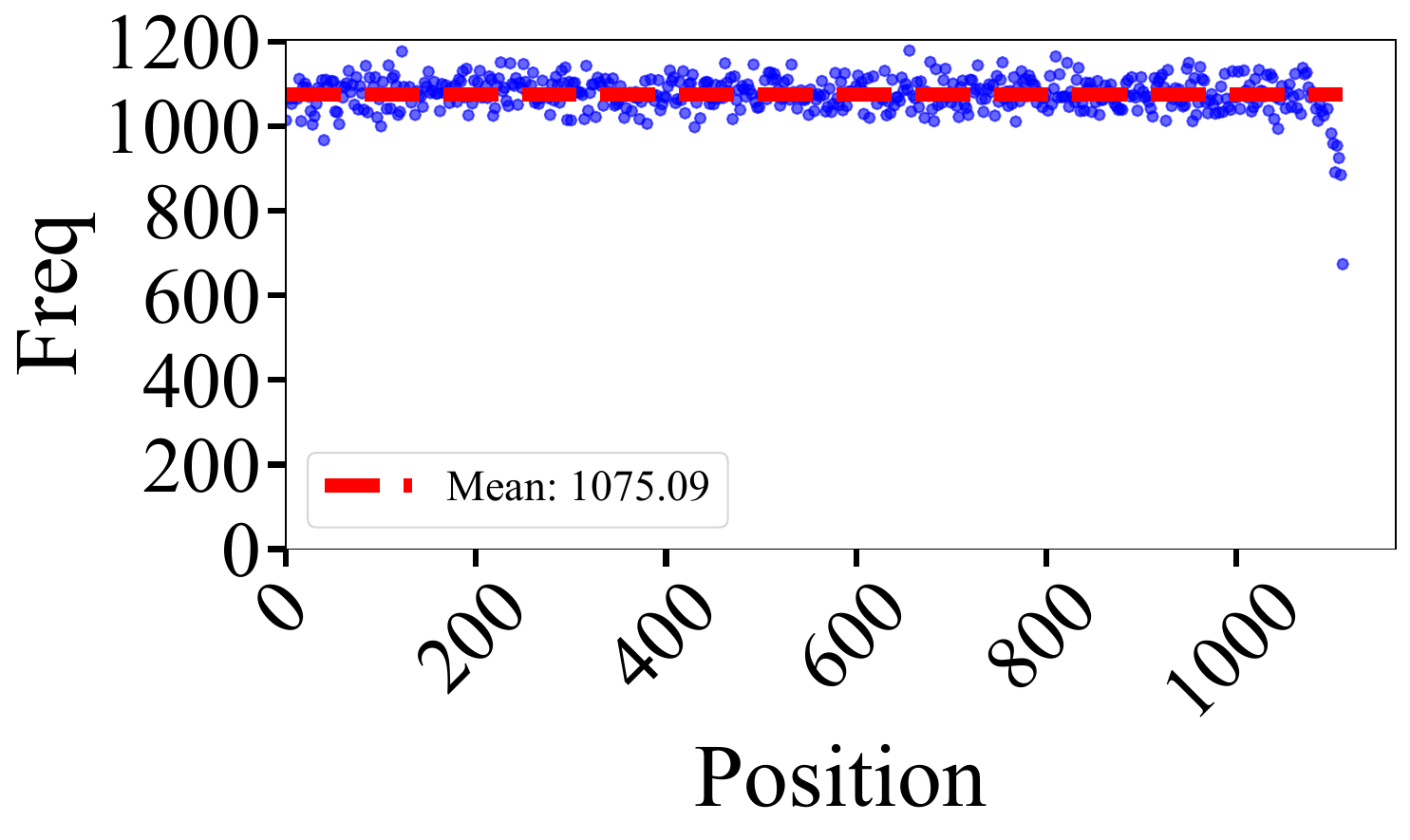}
        \caption{OpenSSL}
        \label{fig:openssl_1_1}
    \end{subfigure}
    \hfill
    \begin{subfigure}{0.18\textwidth}
        \centering
        \includegraphics[width=\linewidth]{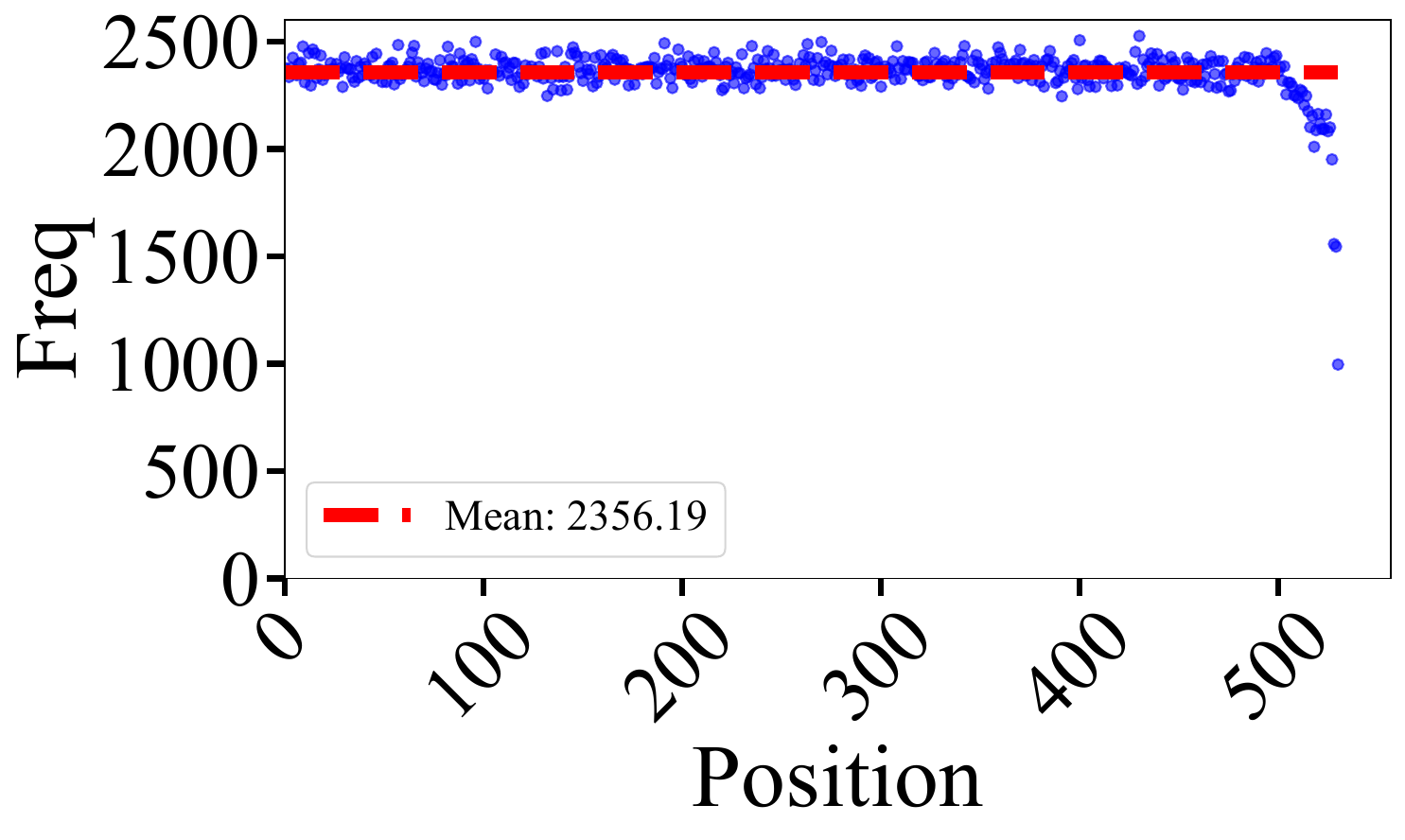}
        \caption{Tcpdump}
        \label{fig:tcpdump_1_1}
    \end{subfigure}
    \hfill
    \begin{subfigure}{0.18\textwidth}
        \centering
        \includegraphics[width=\linewidth]{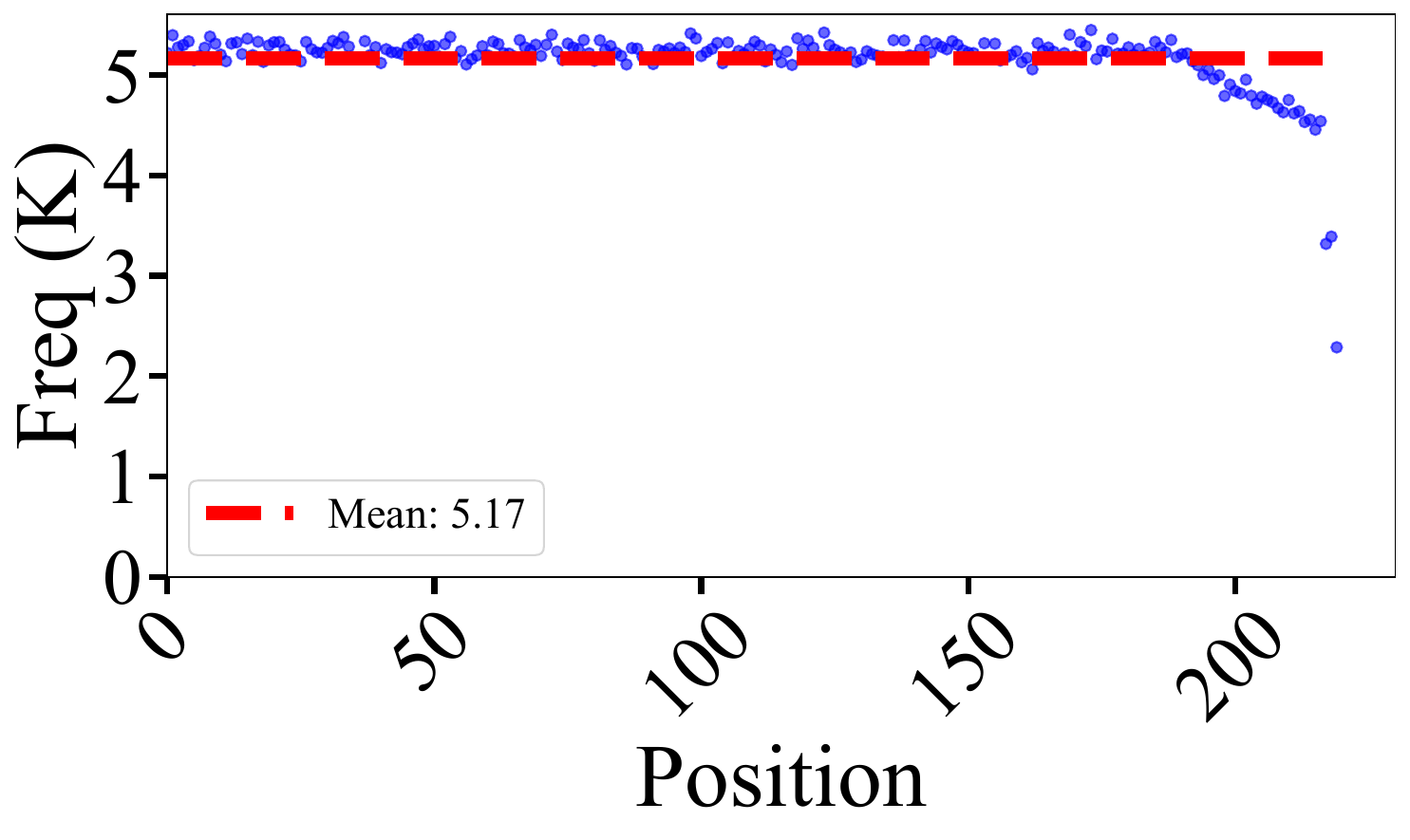}
        \caption{Libpcap}
        \label{fig:libpcap_1_1}
    \end{subfigure}
    \hfill
    \begin{subfigure}{0.18\textwidth}
        \centering
        \includegraphics[width=\linewidth]{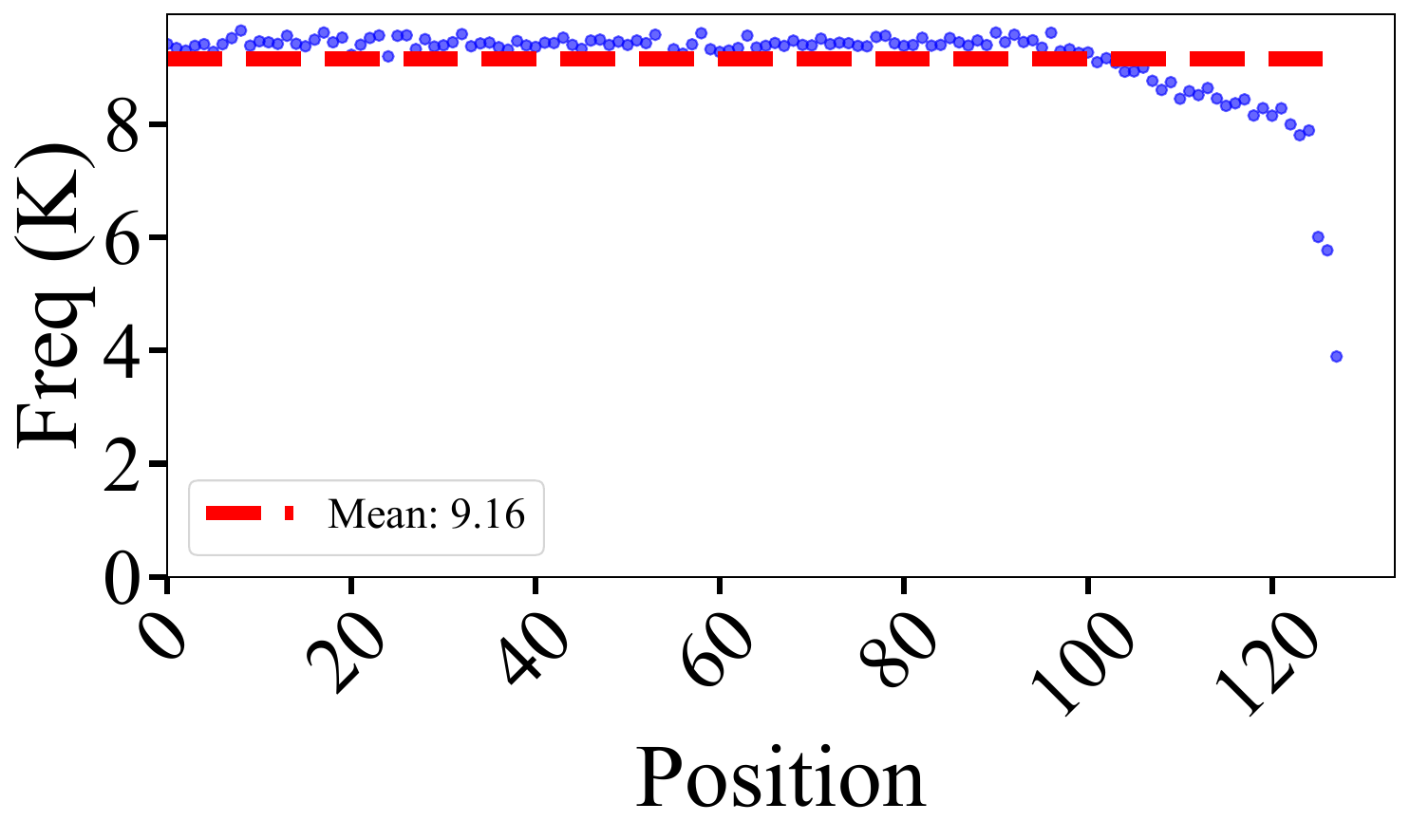}
        \caption{LCMS}
        \label{fig:lcms_1_1}
    \end{subfigure}

    \caption{\textbf{Distribution of mutation starting positions across ten target programs. The x-axis represents the position in input space where mutations begin, while the y-axis shows the accumulated selection frequency over 10 independent runs. The red line indicates the average frequency across all positions.}}
    \label{fig:even1}
\end{figure}

Figure \ref{fig:even1} presents the results of our first experiment. The data demonstrate a predominantly uniform distribution of mutation start positions across the majority of the input space. However, a notable decrease in mutation frequency is observed near the tail of the input. This tail-end decline can be attributed to constraints imposed by certain mutation operators, such as \texttt{MUT\_OVERWRITE\_FIXED}, which have limited selection options near the input's end due to length constraints.
\vspace{0cm}

\begin{figure}[!htb]
    \centering
    \begin{subfigure}{0.18\textwidth}
        \centering
        \includegraphics[width=\linewidth]{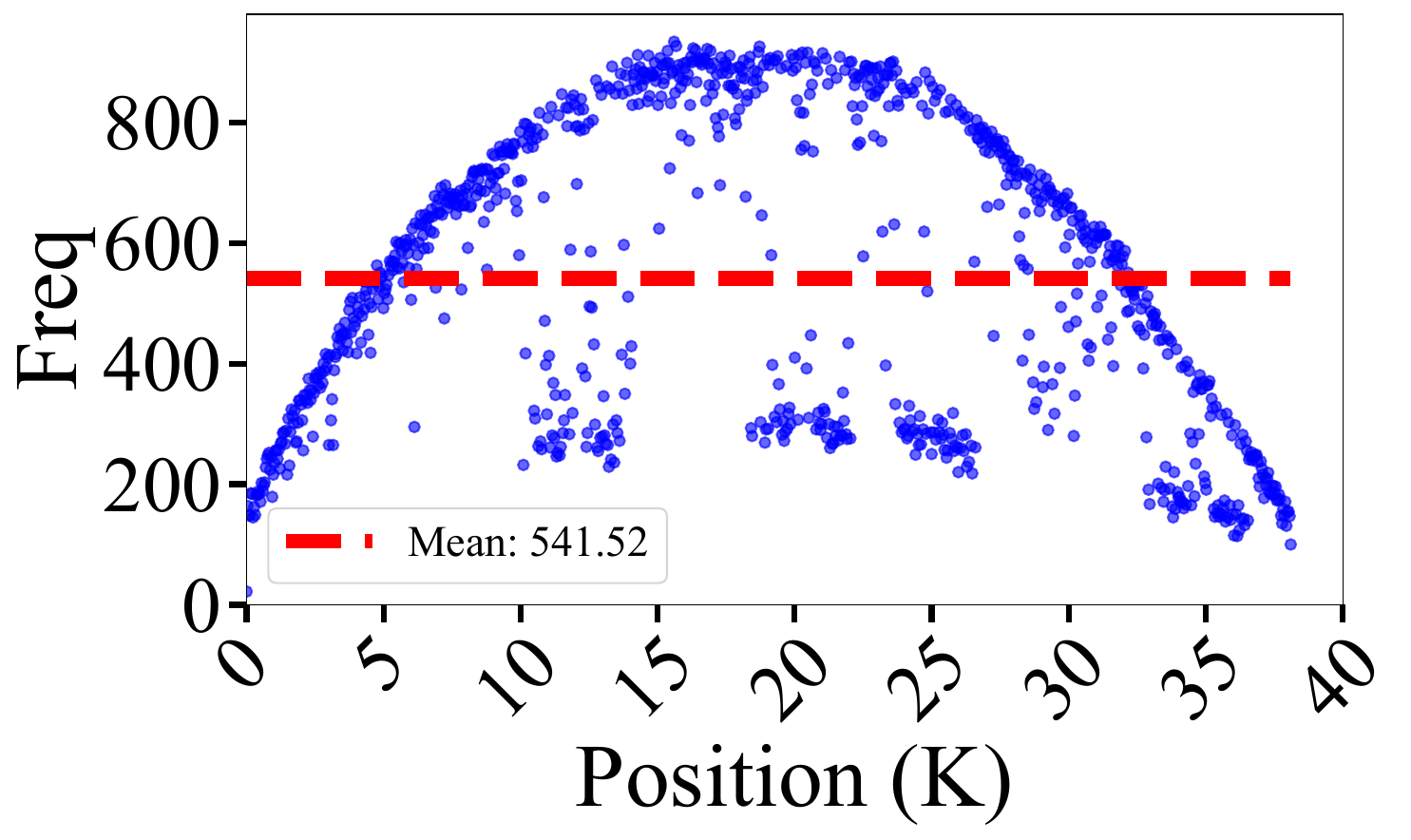}
        \caption{Cflow}
        \label{fig:cflow_1_2}
    \end{subfigure}
    \hfill
    \begin{subfigure}{0.18\textwidth}
        \centering
        \includegraphics[width=\linewidth]{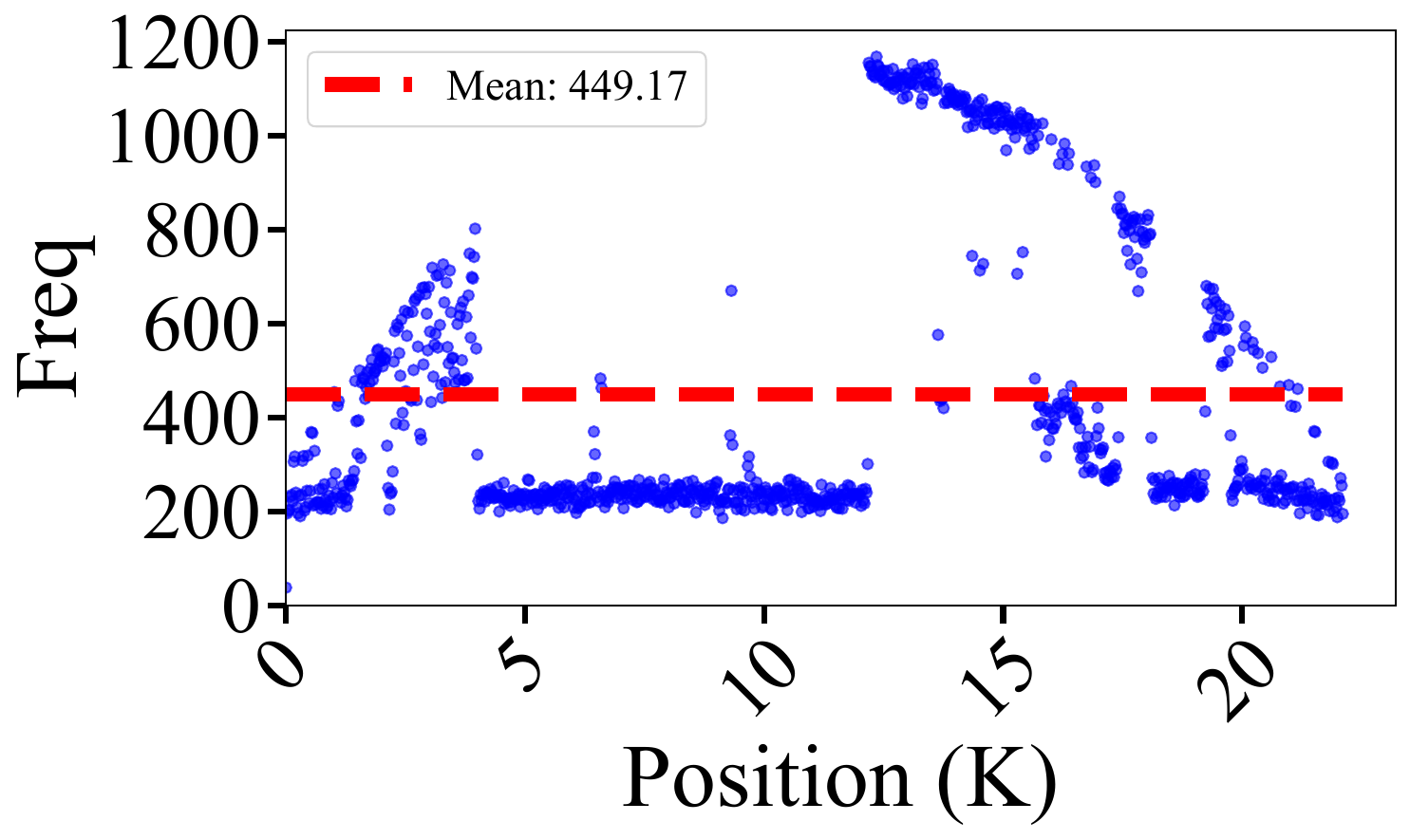}
        \caption{Bloaty}
        \label{fig:bloaty_1_2}
    \end{subfigure}
    \hfill
    \begin{subfigure}{0.18\textwidth}
        \centering
        \includegraphics[width=\linewidth]{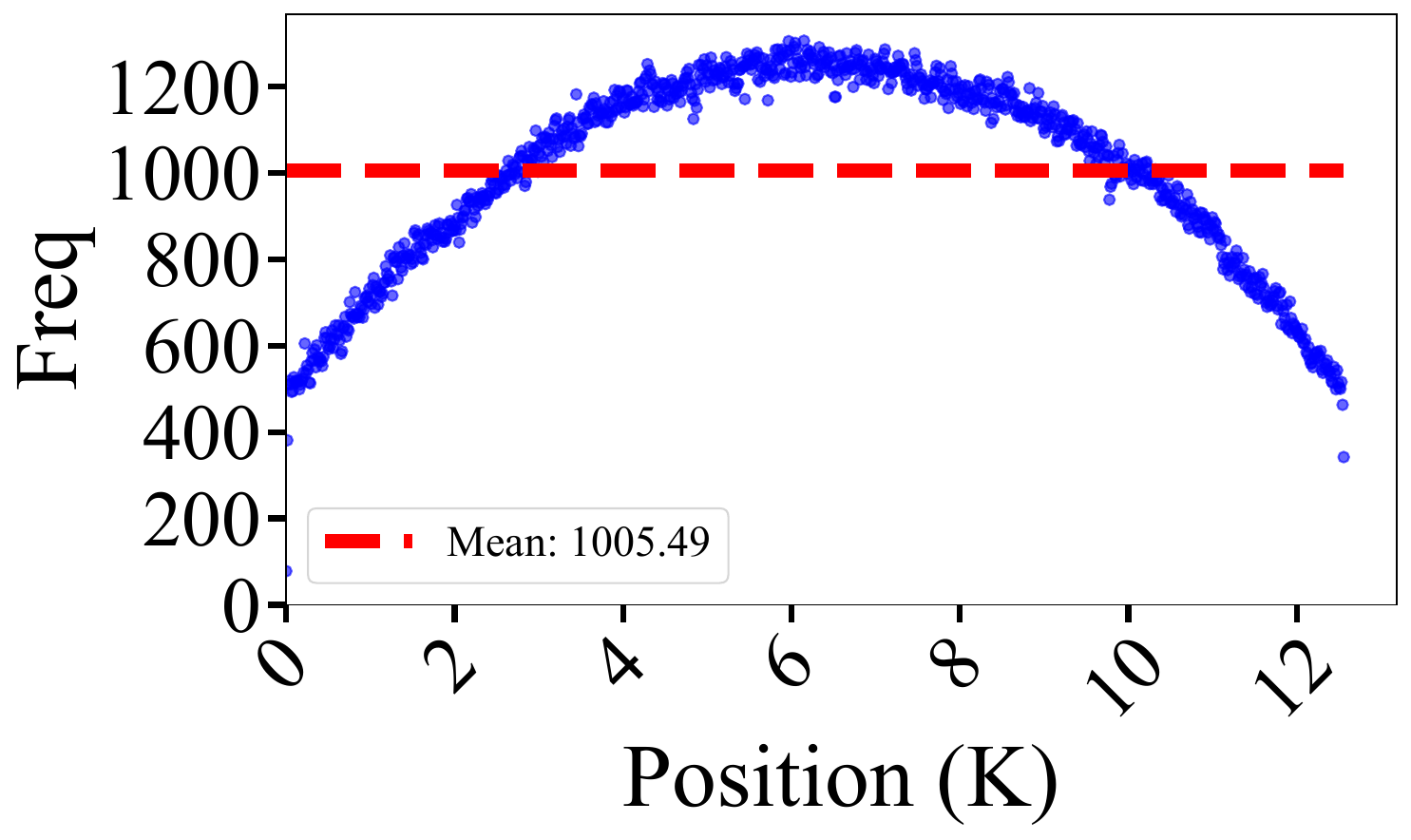}
        \caption{Pdftotext}
        \label{fig:pdftotext_1_2}
    \end{subfigure}
    \hfill
    \begin{subfigure}{0.18\textwidth}
        \centering
        \includegraphics[width=\linewidth]{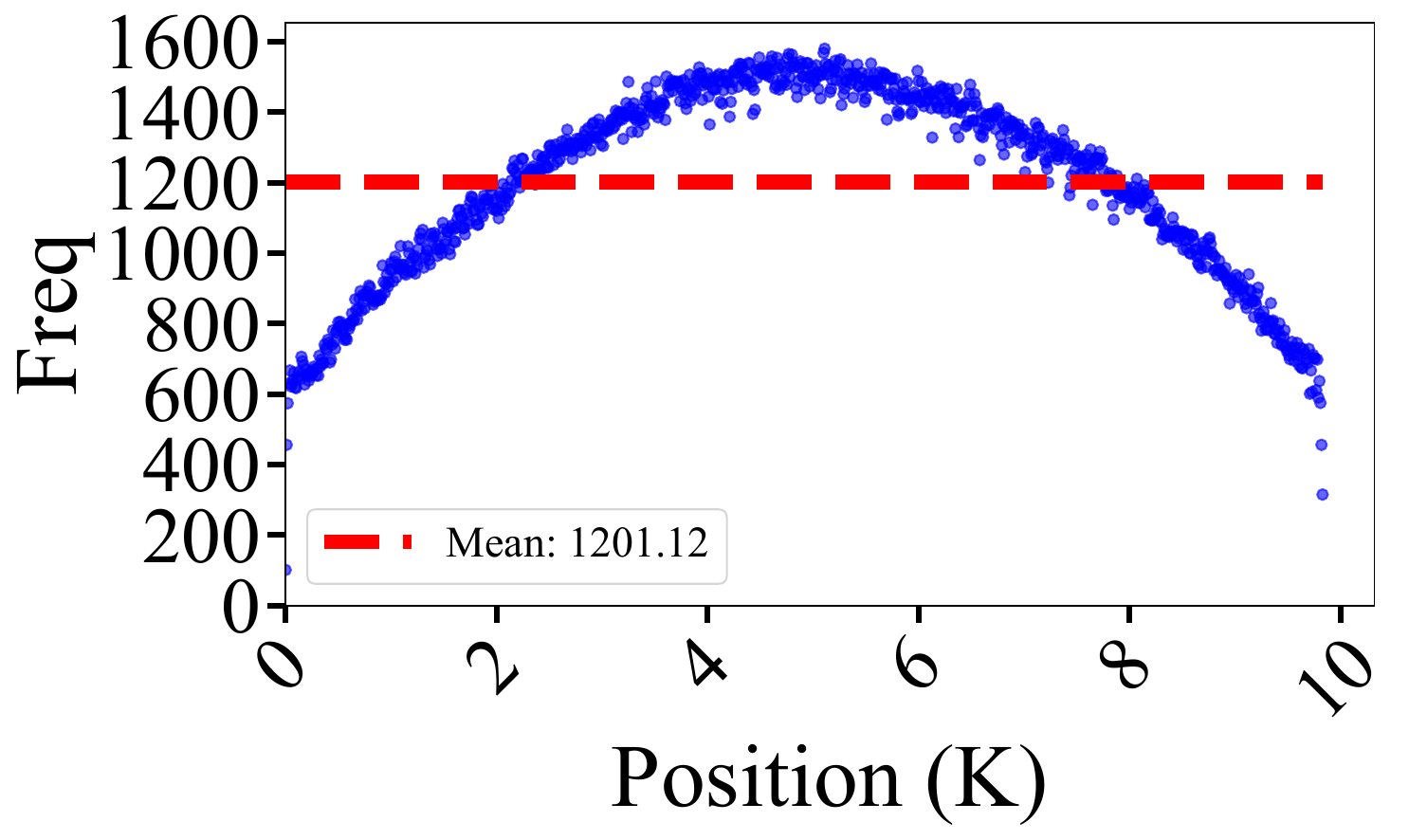}
        \caption{Tiffsplit}
        \label{fig:tiffsplit_1_2}
    \end{subfigure}
    \hfill
    \begin{subfigure}{0.18\textwidth}
        \centering
        \includegraphics[width=\linewidth]{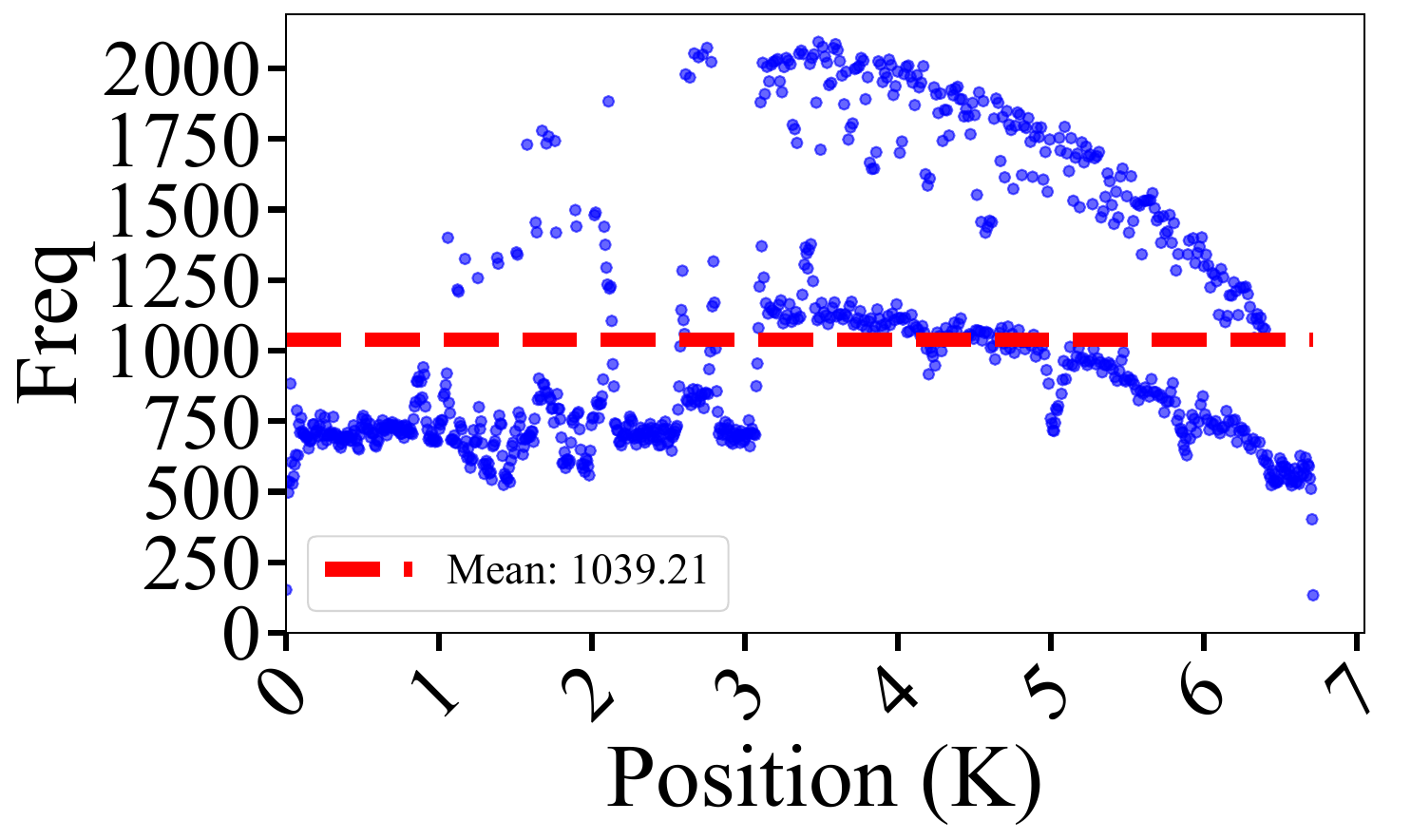}
        \caption{FFmpeg}
        \label{fig:ffmpeg_1_2}
    \end{subfigure}


    \begin{subfigure}{0.18\textwidth}
        \centering
        \includegraphics[width=\linewidth]{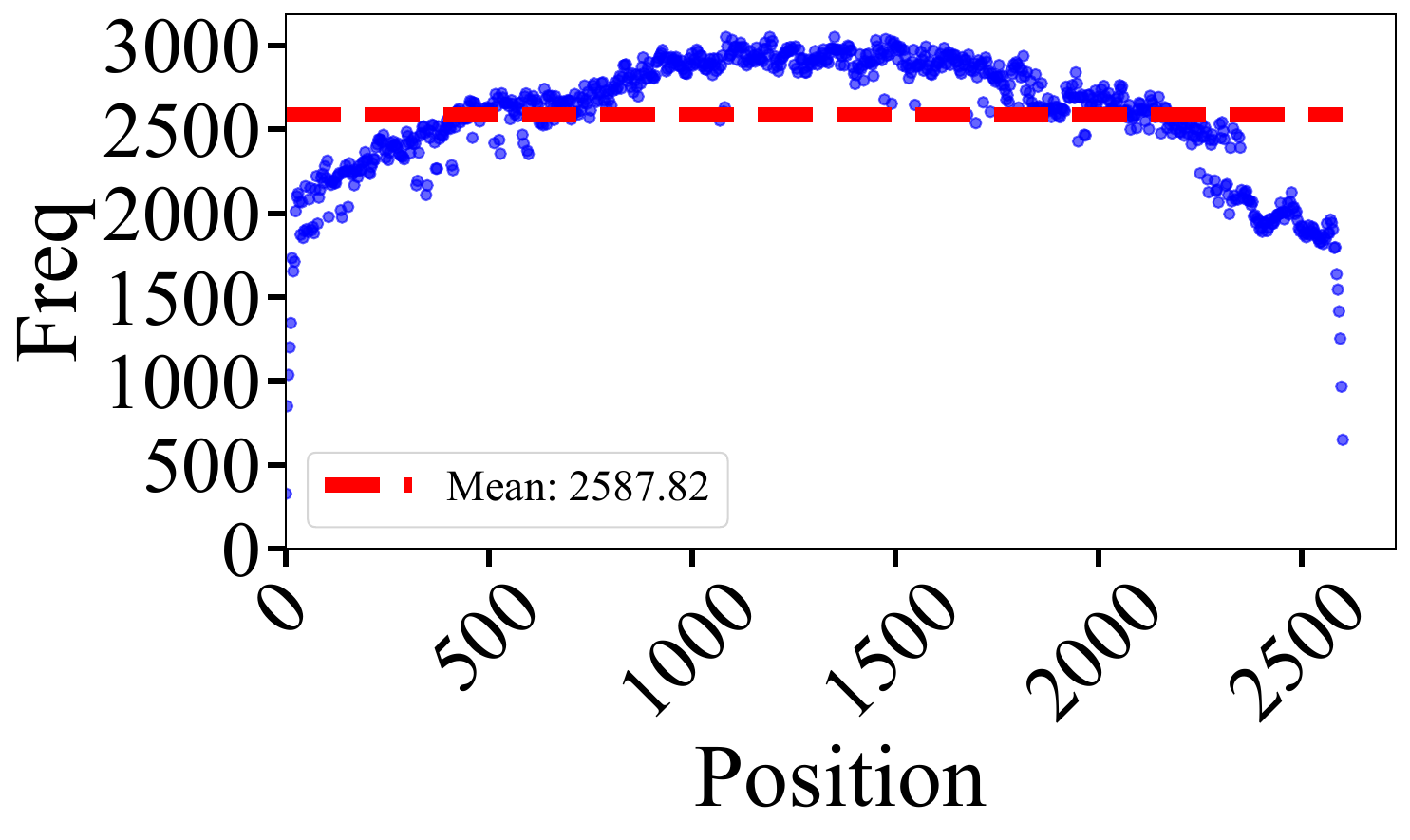}
        \caption{Vorbis}
        \label{fig:vorbis_1_2}
    \end{subfigure}
    \hfill
    \begin{subfigure}{0.18\textwidth}
        \centering
        \includegraphics[width=\linewidth]{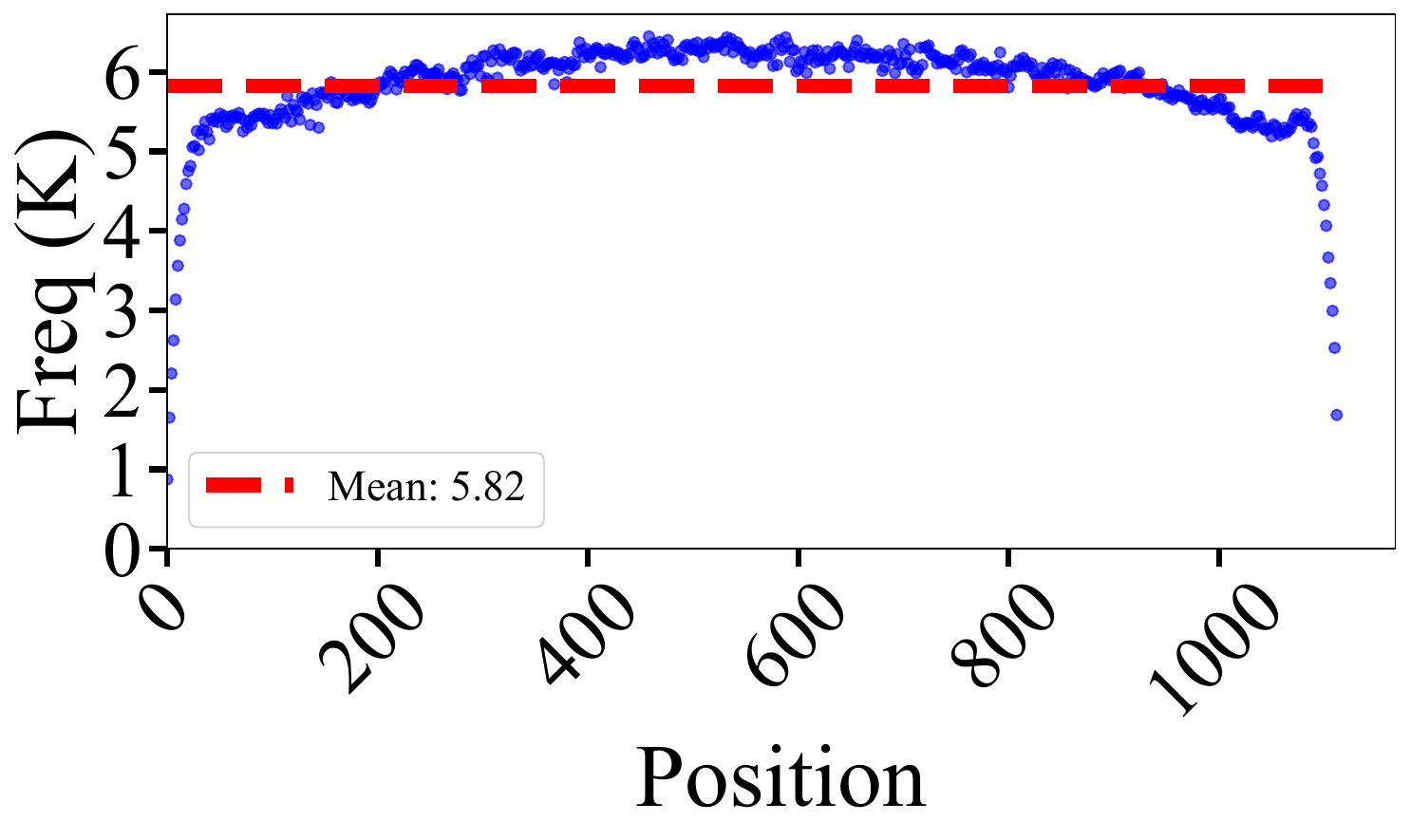}
        \caption{OpenSSL}
        \label{fig:openssl_1_2}
    \end{subfigure}
    \hfill
    \begin{subfigure}{0.18\textwidth}
        \centering
        \includegraphics[width=\linewidth]{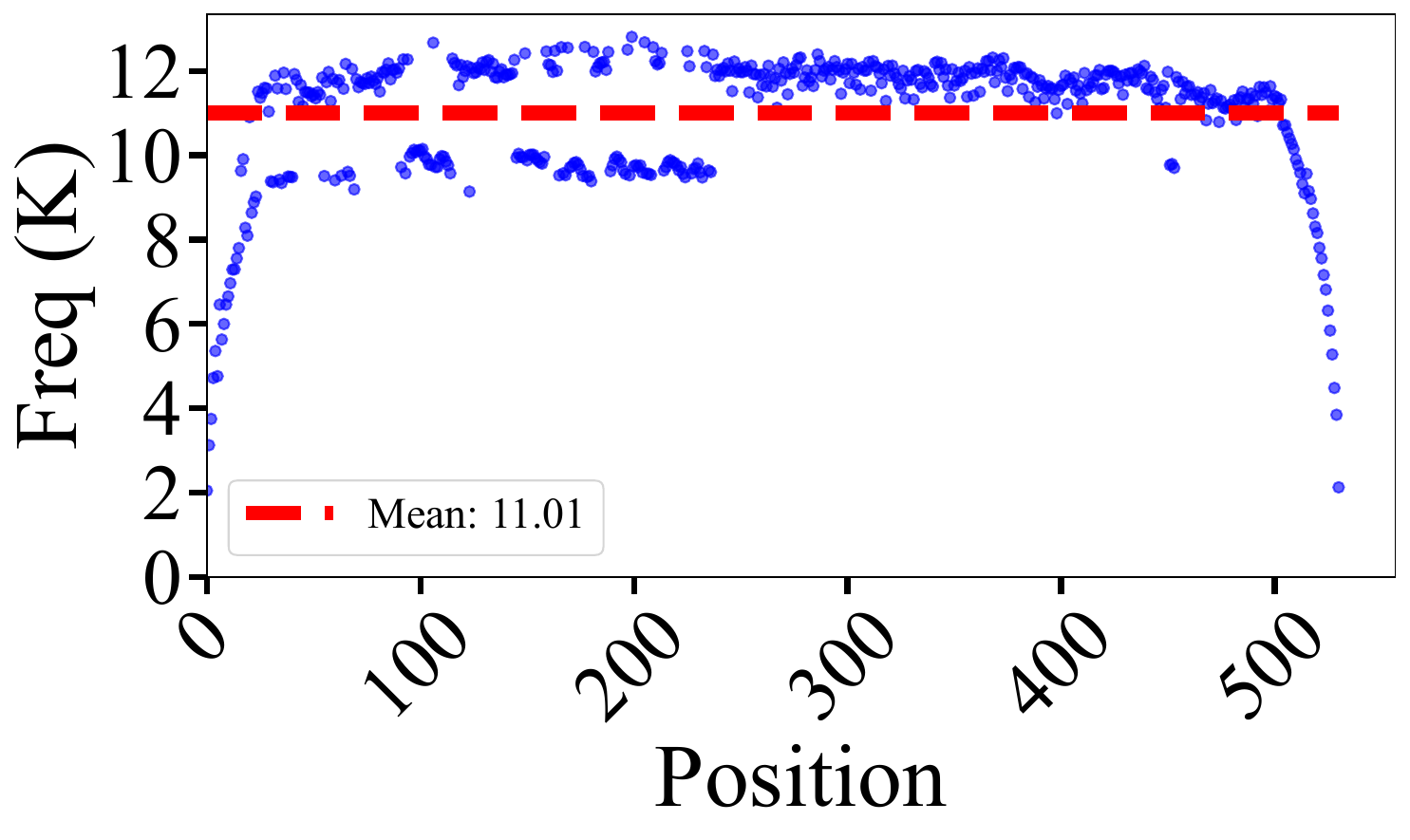}
        \caption{Tcpdump}
        \label{fig:tcpdump_1_2}
    \end{subfigure}
    \hfill
    \begin{subfigure}{0.18\textwidth}
        \centering
        \includegraphics[width=\linewidth]{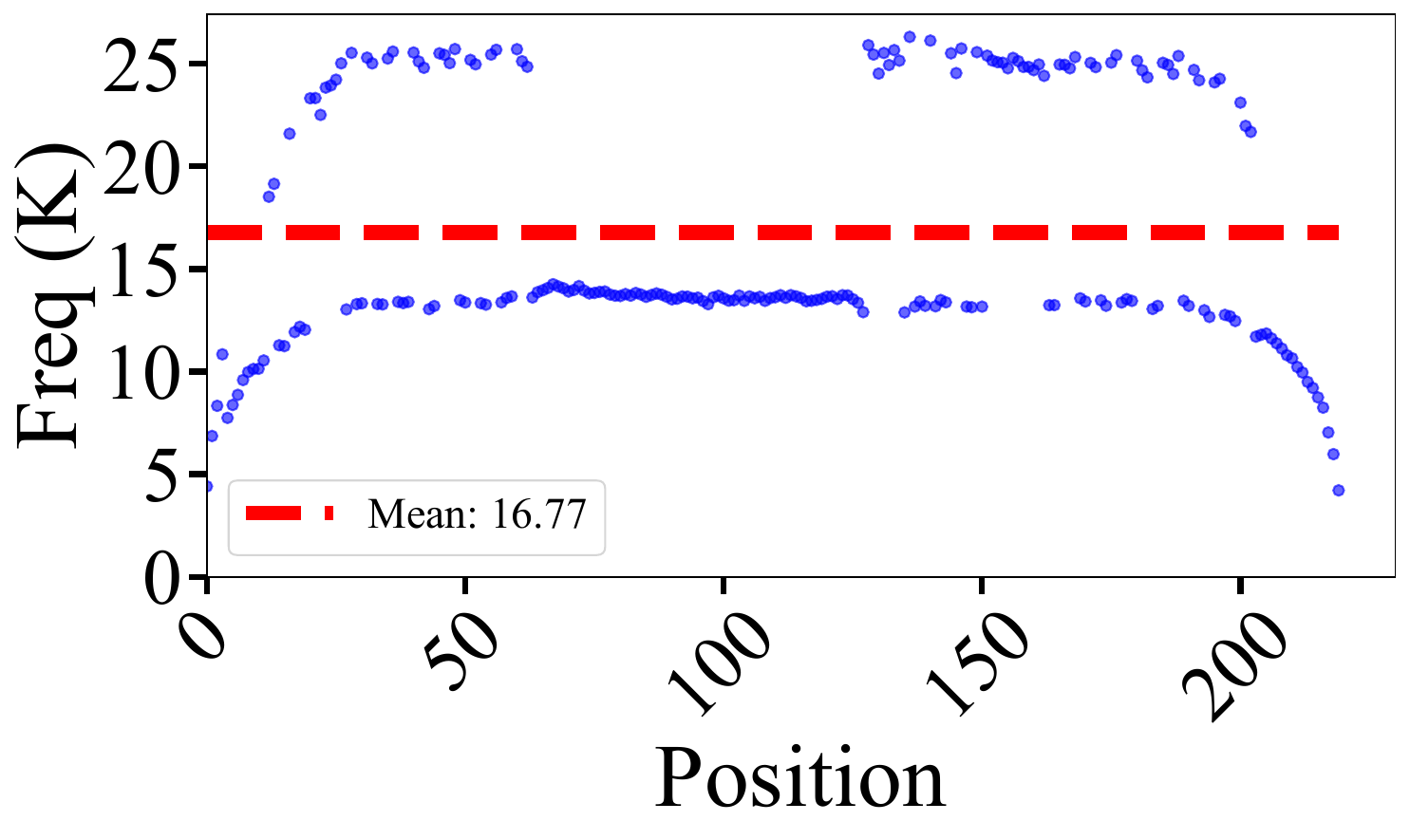}
        \caption{Libpcap}
        \label{fig:libpcap_1_2}
    \end{subfigure}
    \hfill
    \begin{subfigure}{0.18\textwidth}
        \centering
        \includegraphics[width=\linewidth]{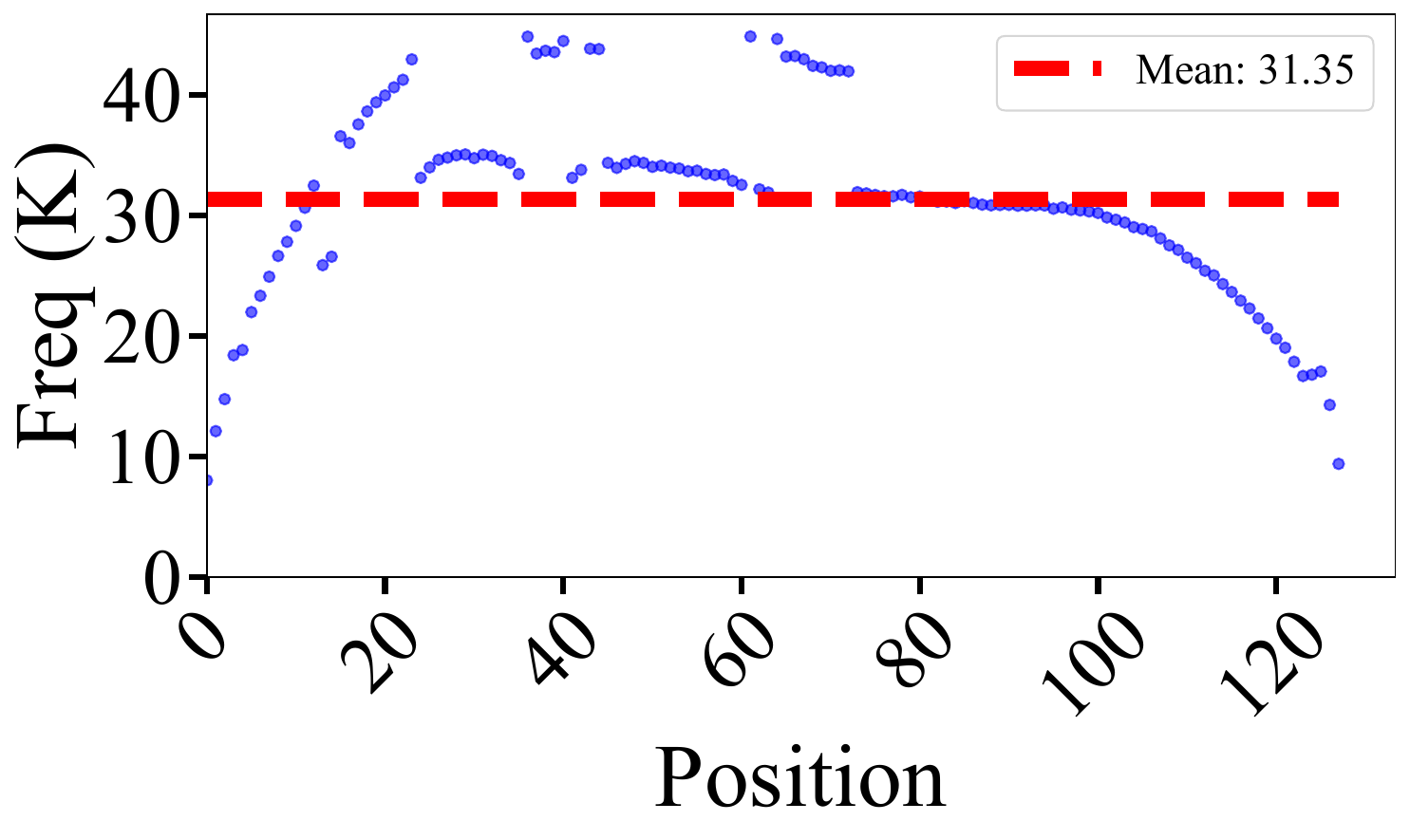}
        \caption{LCMS}
        \label{fig:lcms_1_2}
    \end{subfigure}

    \caption{\textbf{Distribution of influenced bytes (bytes that differ between the original seed and mutated input) across ten target programs. The x-axis represents the byte position in input space, while the y-axis shows the accumulated frequency of bytes being influenced over 10 independent runs, demonstrating the distribution pattern without threshold constraints. The red line indicates the average frequency across all positions.}}
    \label{fig:even2}
\end{figure}

Figure \ref{fig:even2} presents the results of the first part of our second experiment, which examined the distribution of influenced bytes during the AFL++ havoc stage without applying any threshold. For most targets, the distribution exhibits a pronounced peak, resembling a mountain-like shape rather than a uniform distribution. This pattern clearly demonstrates that in practice, influenced bytes are not uniformly distributed across the input space during havoc mutations. Some regions of the input are significantly more likely to be affected by mutations than others. The irregular shape of these distributions varies across different seeds, as it is highly dependent on the specific content of each seed input.

\begin{figure}[!htb]
    \centering
    \begin{subfigure}{0.18\textwidth}
        \centering
        \includegraphics[width=\linewidth]{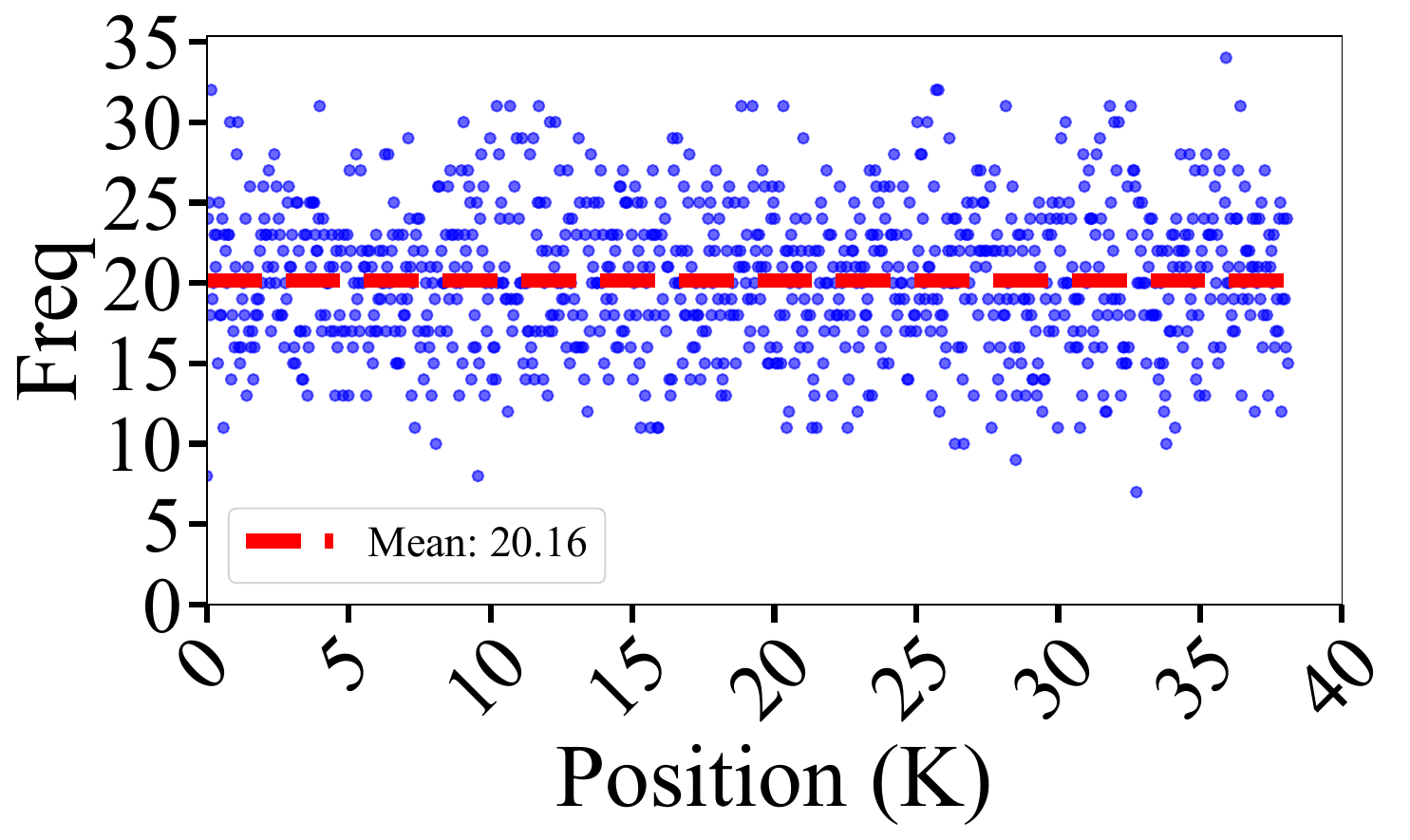}
        \caption{Cflow}
        \label{fig:cflow_1_3}
    \end{subfigure}
    \hfill
    \begin{subfigure}{0.18\textwidth}
        \centering
        \includegraphics[width=\linewidth]{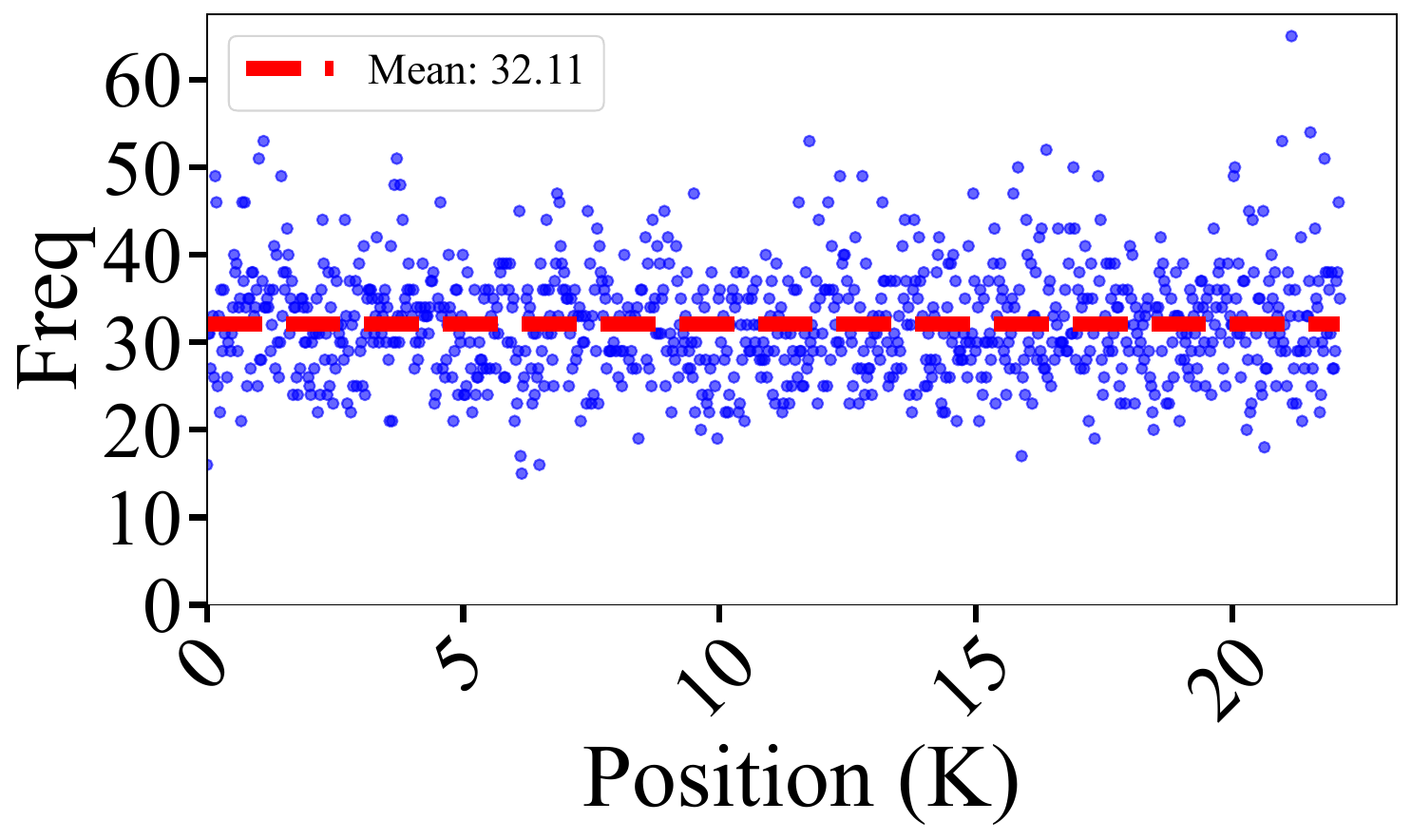}
        \caption{Bloaty}
        \label{fig:bloaty_1_3}
    \end{subfigure}
    \hfill
    \begin{subfigure}{0.18\textwidth}
        \centering
        \includegraphics[width=\linewidth]{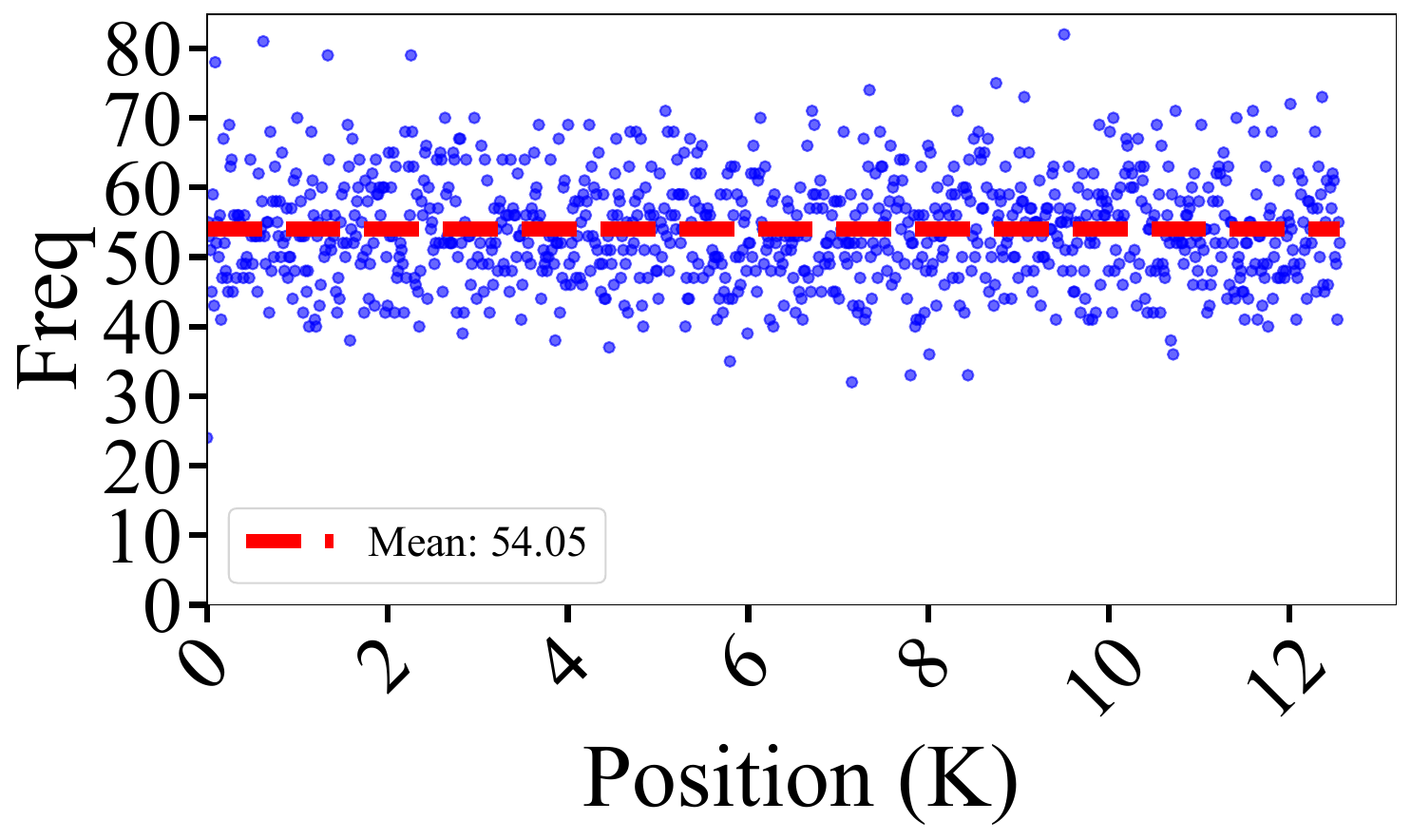}
        \caption{Pdftotext}
        \label{fig:pdftotext_1_3}
    \end{subfigure}
    \hfill
    \begin{subfigure}{0.18\textwidth}
        \centering
        \includegraphics[width=\linewidth]{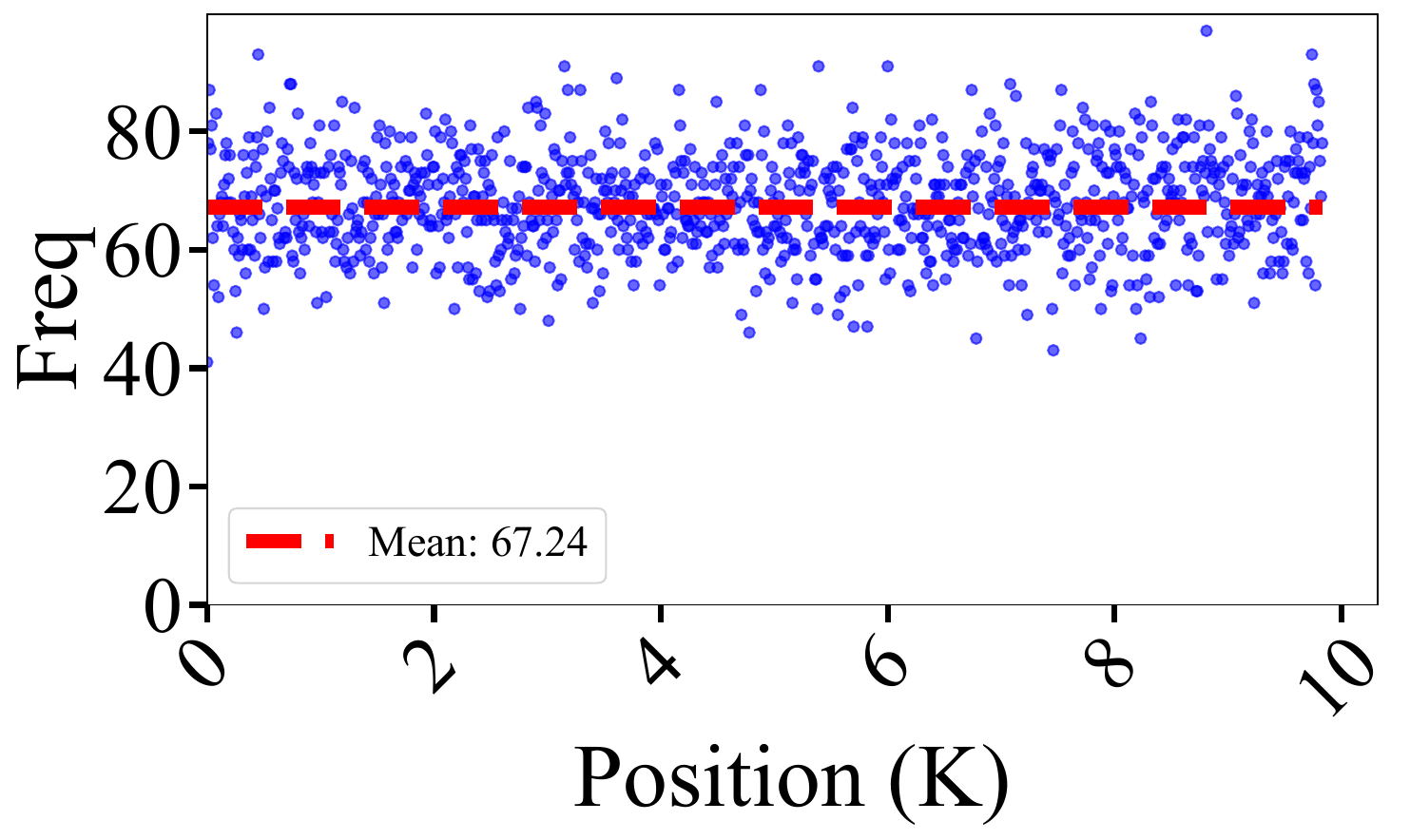}
        \caption{Tiffsplit}
        \label{fig:tiffsplit_1_3}
    \end{subfigure}
    \hfill
    \begin{subfigure}{0.18\textwidth}
        \centering
        \includegraphics[width=\linewidth]{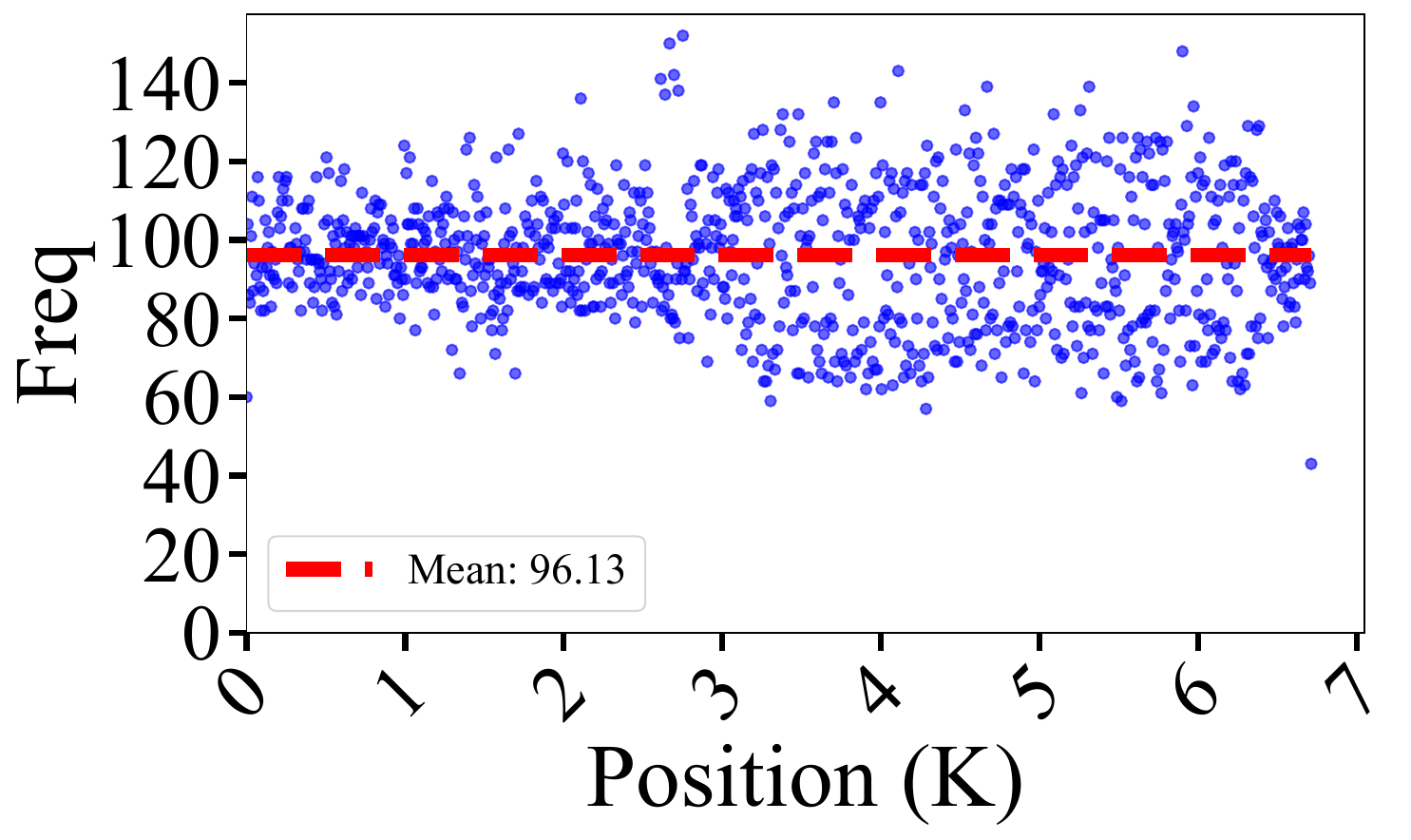}
        \caption{FFmpeg}
        \label{fig:ffmpeg_1_3}
    \end{subfigure}


    \begin{subfigure}{0.18\textwidth}
        \centering
        \includegraphics[width=\linewidth]{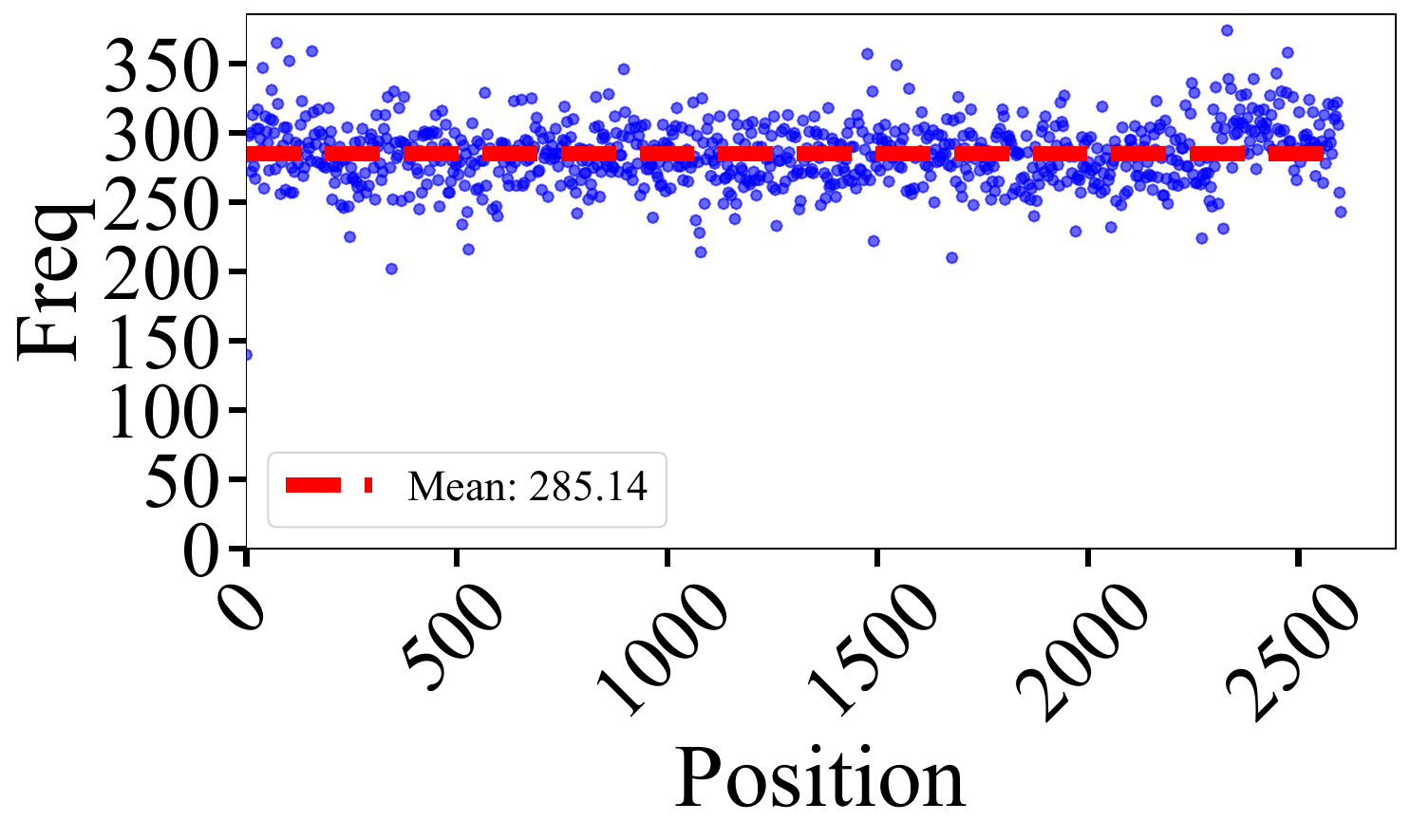}
        \caption{Vorbis}
        \label{fig:vorbis_1_3}
    \end{subfigure}
    \hfill
    \begin{subfigure}{0.18\textwidth}
        \centering
        \includegraphics[width=\linewidth]{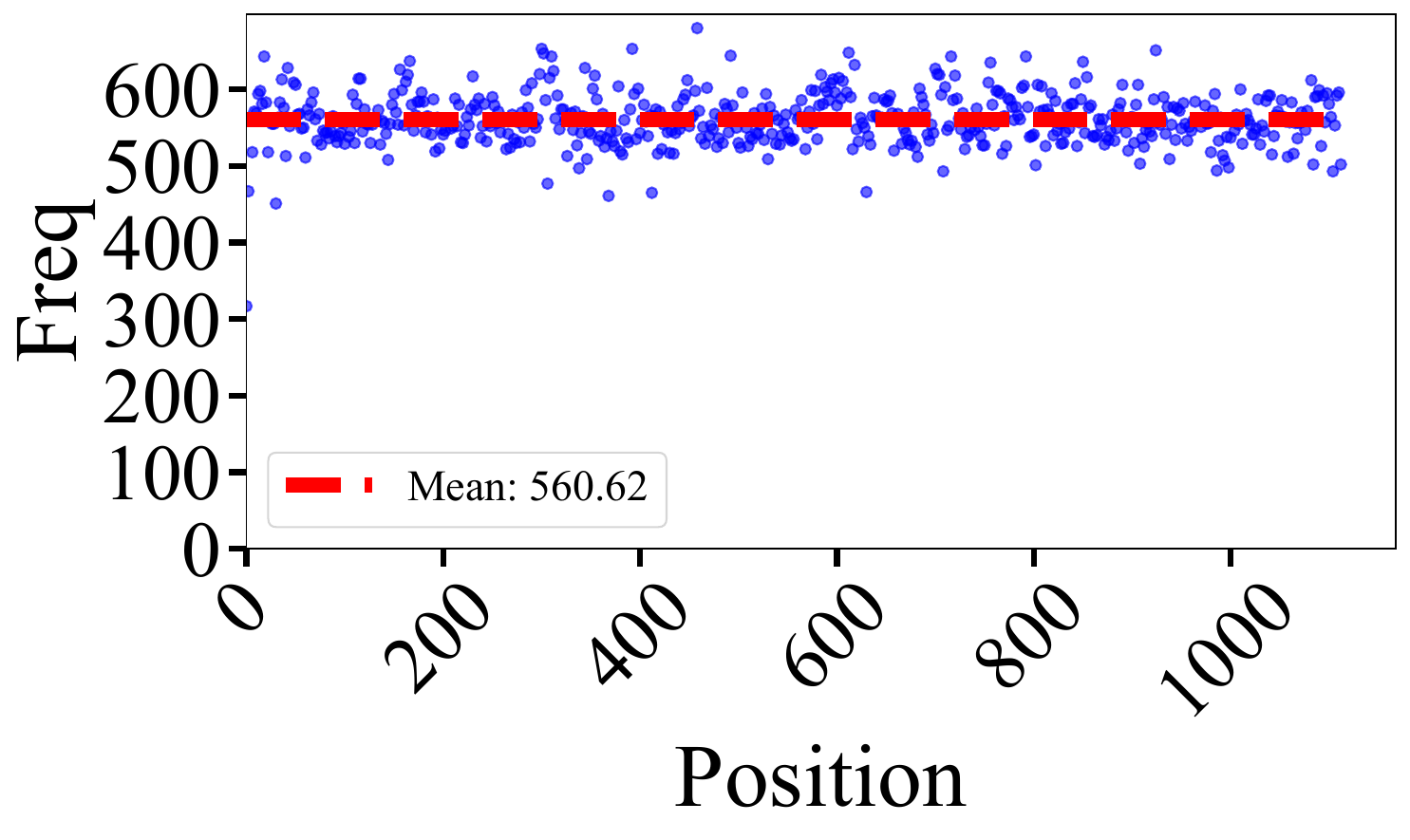}
        \caption{OpenSSL}
        \label{fig:openssl_1_3}
    \end{subfigure}
    \hfill
    \begin{subfigure}{0.18\textwidth}
        \centering
        \includegraphics[width=\linewidth]{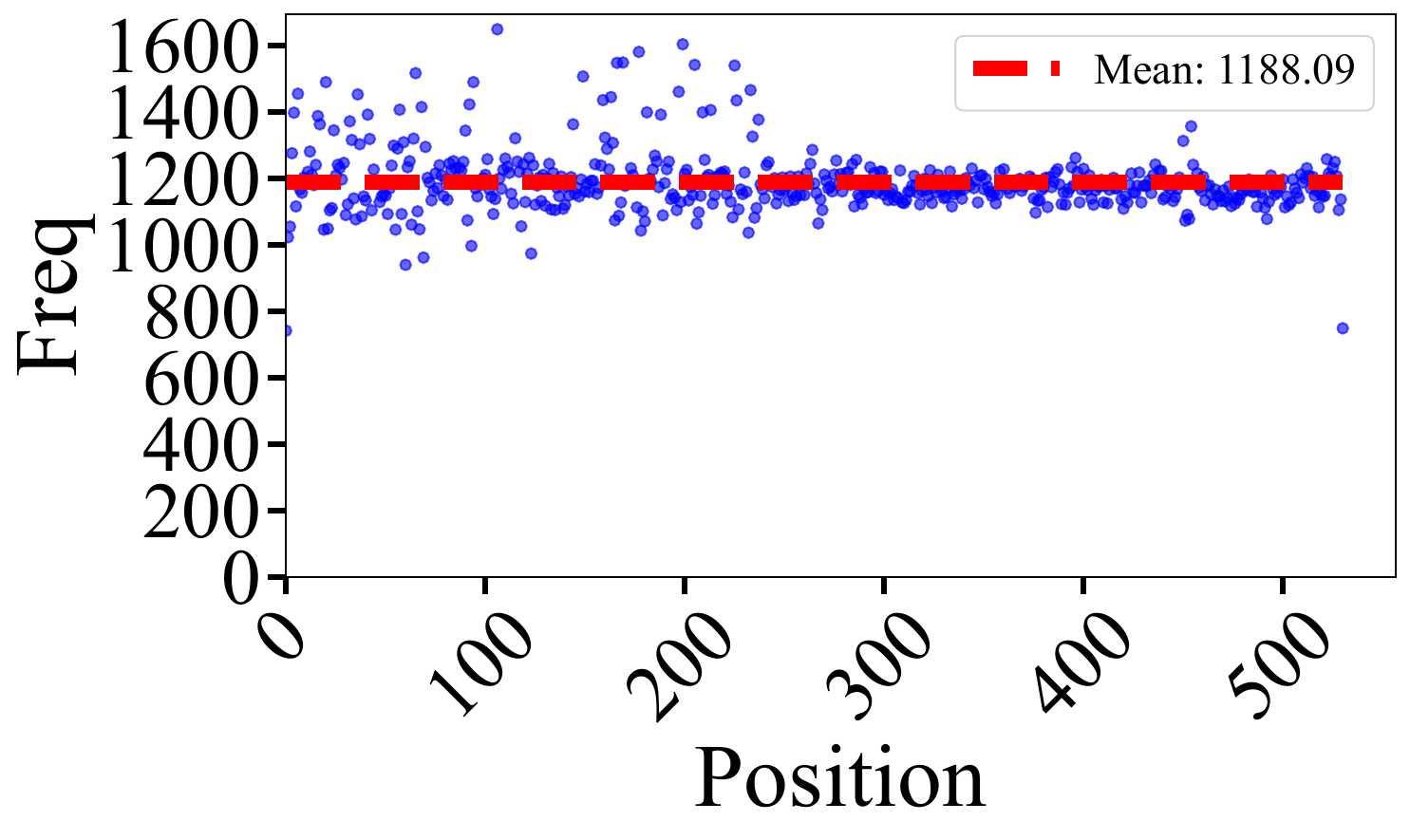}
        \caption{Tcpdump}
        \label{fig:tcpdump_1_3}
    \end{subfigure}
    \hfill
    \begin{subfigure}{0.18\textwidth}
        \centering
        \includegraphics[width=\linewidth]{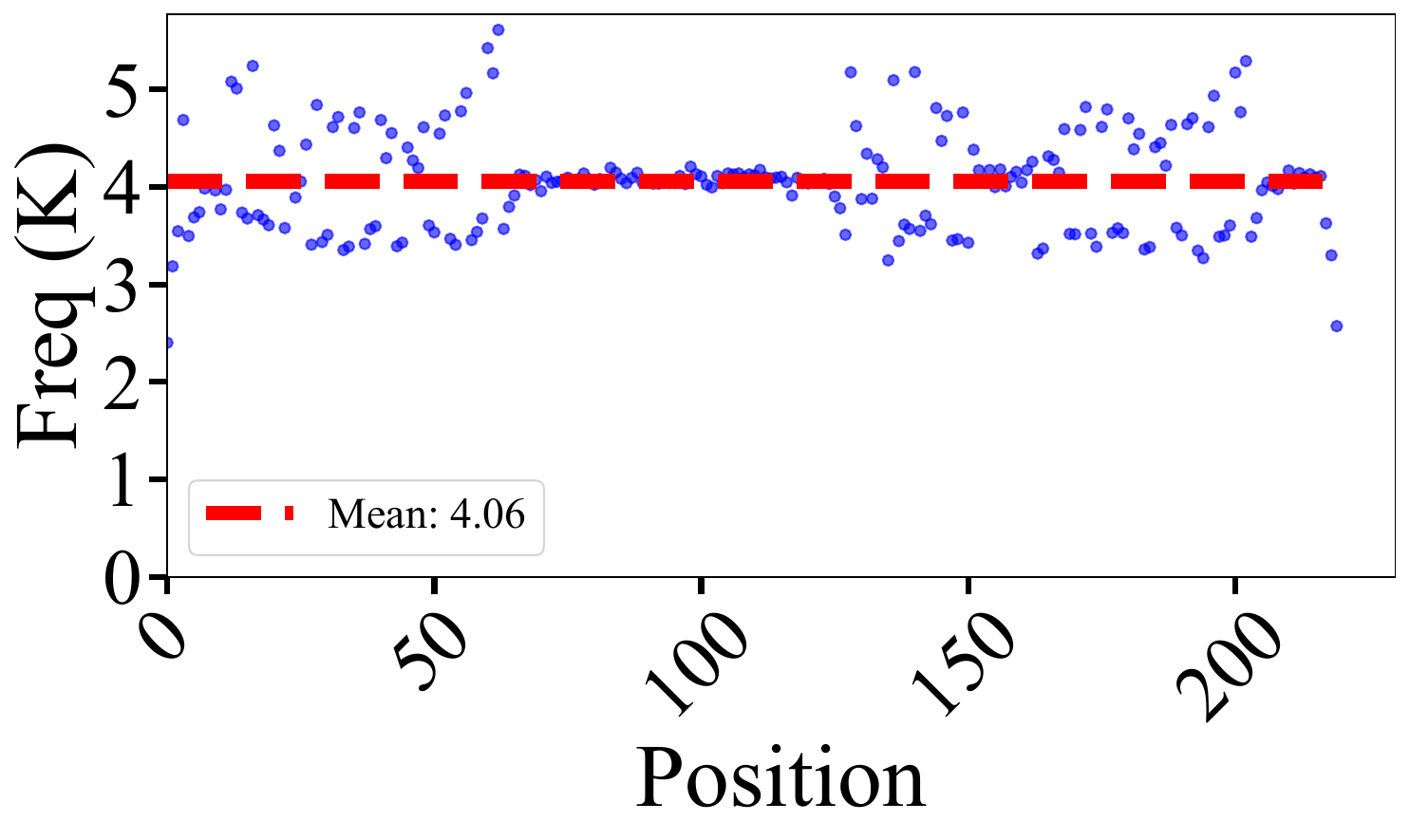}
        \caption{Libpcap}
        \label{fig:libpcap_1_3}
    \end{subfigure}
    \hfill
    \begin{subfigure}{0.18\textwidth}
        \centering
        \includegraphics[width=\linewidth]{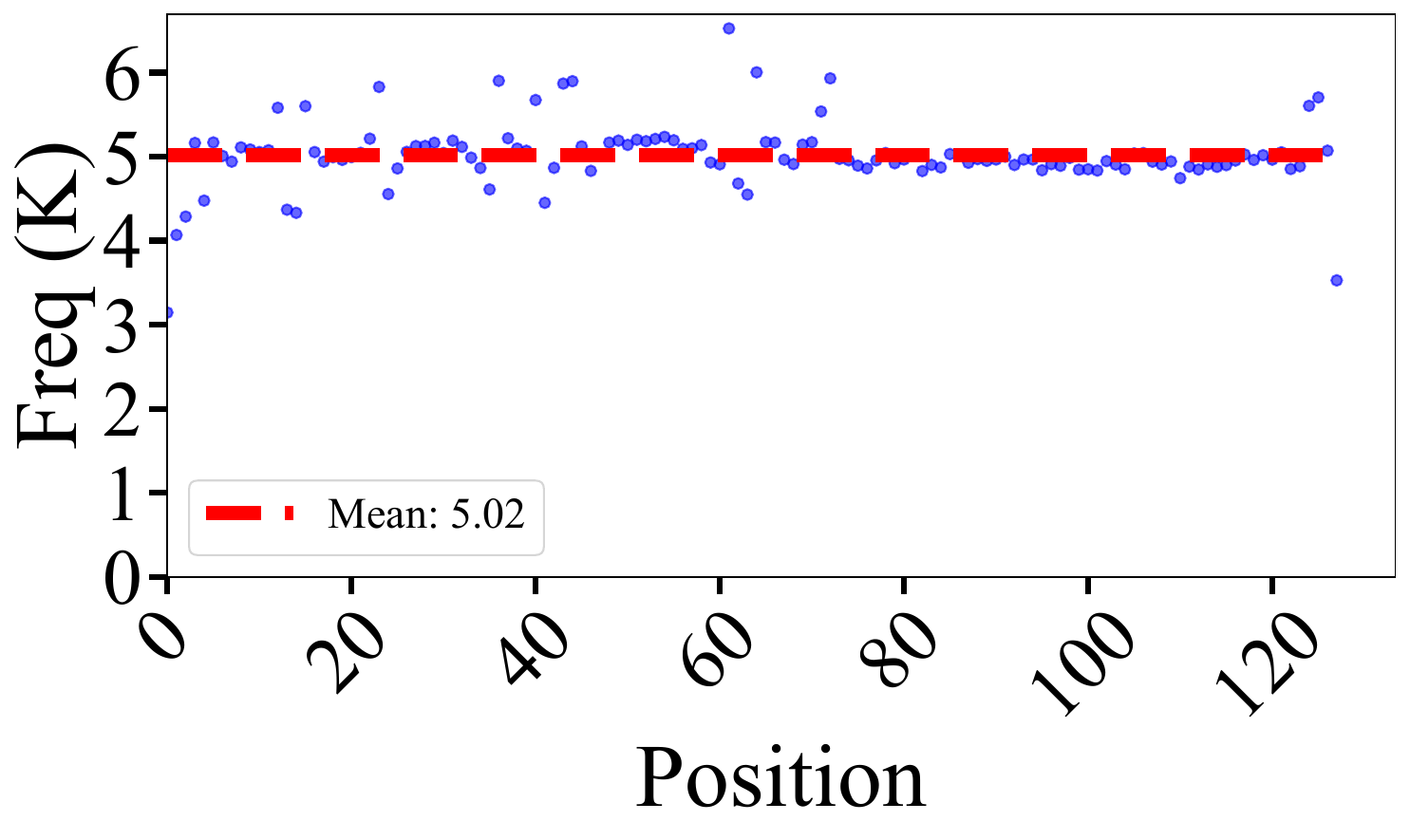}
        \caption{LCMS}
        \label{fig:lcms_1_3}
    \end{subfigure}

    \caption{\textbf{Distribution of influenced bytes across ten target programs with a perturbation threshold of k=5. The x-axis represents the byte position in input space, while the y-axis shows the accumulated frequency of bytes being influenced over 10 independent runs, demonstrating the distribution pattern under controlled mutation impact. The red line indicates the average frequency across all positions. }}
    \label{fig:even3}
\end{figure}

The second part of our experiment investigated the distribution of influenced bytes after implementing a threshold mechanism. Figure \ref{fig:even3} illustrates these results. In stark contrast to the previous findings, the distribution with the threshold applied exhibits a markedly more uniform pattern across most of the input space. The pronounced peaks observed in the non-threshold scenario have been largely eliminated, with only narrow regions at the beginning and end of the input showing slight deviations from uniformity. This nearly uniform distribution for most of the input demonstrates the effectiveness of our threshold approach in equalizing the influence of mutations across the input space. The results clearly indicate that our threshold mechanism successfully mitigates the bias towards specific input regions observed in standard havoc mutations.

\vspace{0cm}

\begin{longfbox}
\textbf{Result 1:} 
Both the havoc's starting positions and the influenced bytes (with threshold) exhibit approximately uniform distributions across the input space.
\end{longfbox}

\subsection{RQ2: Effect of Uniform Sampling}
\label{text:rq1.1}
\subsubsection{Experimental Setup}
For this experiment, because we use coverage as the metric, we choose all targets in FuzzBench and UniBench datasets to obtain a comprehensive evaluation. For our evaluation, we selected moderate-length seeds and well-formed structures from the seed corpora of AFL++, FuzzBench, and UniBench. We compared two configurations: vanilla AFL++ and AFL++ with normally distributed starting positions centered at randomly chosen positions in the seed (with the center fixed at fuzzer initialization). Each target was evaluated for 1 hour with 10 independent runs to account for the inherent randomness in fuzzing. To eliminate the impact of any overhead introduced by our modifications, we compare the coverage with the same number of executions by using the smaller execution count between the two configurations in the 1-hour runs. 

\subsubsection{Observations}
Based on our experimental results across all tested targets from FuzzBench and UniBench datasets, we observe that vanilla AFL++ (using uniform distribution for selecting mutation positions) consistently achieves higher coverage compared to using normally distributed starting positions, with an average improvement of 4.69\%. This suggests that the uniform selection of mutation positions provides better exploration of the input space and more adaptiveness across different program structures, rather than concentrating mutations around fixed points.

\begin{longfbox}
\textbf{Result 2:} 
Uniformly choosing mutation starting position contributes to the effectiveness and adaptiveness for the havoc mode.
\end{longfbox}

\begin{table}[]
\small
    \centering
    \caption{\textbf{Coverage comparison between vanilla AFL++ and AFL++ with normally distributed starting positions (Normal) across FuzzBench and UniBench targets. Each number represents the average edge coverage from 10 one-hour independent runs, with bold numbers indicating higher coverage.}}
    
    \setlength{\tabcolsep}{0.8pt}
    \renewcommand{\arraystretch}{0.9}
    \begin{tabular}{lrr|lrr|lrr}
    \toprule
    \textbf{Targets} & \textbf{Vanilla} & \textbf{Normal} & 
    \textbf{Targets} & \textbf{Vanilla} & \textbf{Normal} & 
    \textbf{Targets} & \textbf{Vanilla} & \textbf{Normal} \\
    \midrule
    \multicolumn{3}{c|}{\textbf{FuzzBench Targets}} & 
    sql & \textbf{9887} & 9817 &
    tcpdump       & \textbf{924} & 893 \\ 
    \cline{1-3}
    bloaty        & \textbf{6261} & 6108 &
    systemd       & \textbf{2122} & 2106 &
    jq            & \textbf{1689} & 1666 \\
    zlib          & \textbf{699} & 690 &
    jsoncpp       & \textbf{906} & 870 &
    sqlite3       & \textbf{27416} & 27253 \\	
    lcms          & \textbf{461} & 195 &
    libjpeg-t     & \textbf{2749} & 2722 &
    cflow         & \textbf{2018} & 2011 \\
    libpcap       & \textbf{42} & 36 &
    libpng        & \textbf{1549} & 1440 &
    exiv2         & \textbf{1947} & 1894 \\
    libxml2       & \textbf{4595} & 4283 &
    libxslt       & \textbf{2761} & 2686 &
    ffmpeg        & \textbf{20621} & 20509 \\
    openh264      & \textbf{6209} & 5959 &
    openssl       & \textbf{11034} & 11031 &
    infotocap     & \textbf{1105} & 1093 \\
    \cline{4-6}
    re2           & \textbf{4044} & 3896 &
    \multicolumn{3}{c|}{\multirow{2}{*}{\textbf{UniBench Targets}}} &
    nm-new        & \textbf{1393} & 1351 \\
    stbi          & \textbf{1239} & 1016 &
    \multicolumn{3}{c|}{} & 
    objdump       & \textbf{4411} & 4331 \\
    \cline{4-6}
    vorbis        & \textbf{1517} & 1478 &
    mp42aac       & \textbf{1362} & 1320 &
    pdftotext     & \textbf{5417} & 5399 \\
    woff2         & \textbf{1794} & 1752 &
    lame          & \textbf{5379} & 5184 &
    mp3gain       & \textbf{1260} & 1250 \\
    curl          & \textbf{8186} & 7790 &
    mujs          & \textbf{4145} & 4018 &
    tiffsplit     & \textbf{1207} & 1175 \\
    harfbuzz      & \textbf{24300} & 23998 &
    flvmeta       & \textbf{375} & 364 &
    jhead         & \textbf{334} & 327 \\
    \bottomrule
    \end{tabular}
    \label{tab:programs_distribution}
\end{table}
\subsection{RQ3: Effect of Mutation Stack}

\subsubsection{Experimental Setup}
In order to exclude other influences, we use the same targets, initial seed, and number of executions used in Section \ref{text:rq1}. We use five sets of experimental setups, that is, the havoc stack is 1, 2, 4, 8, and 16 for five cases. For each target program, we conducted 10 independent measurements to ensure statistical reliability. We measure the average perturbation distance using four different metrics: L0-norm, L1-norm, L2-norm, and edit distance, calculated between the initial seed and the mutant. All experiments are conducted using AFL++ as our base fuzzer.

\subsubsection{Observation}
\begin{figure}[h]
    \centering
    \begin{subfigure}{0.45\textwidth}
        \centering
        \includegraphics[width=\linewidth]{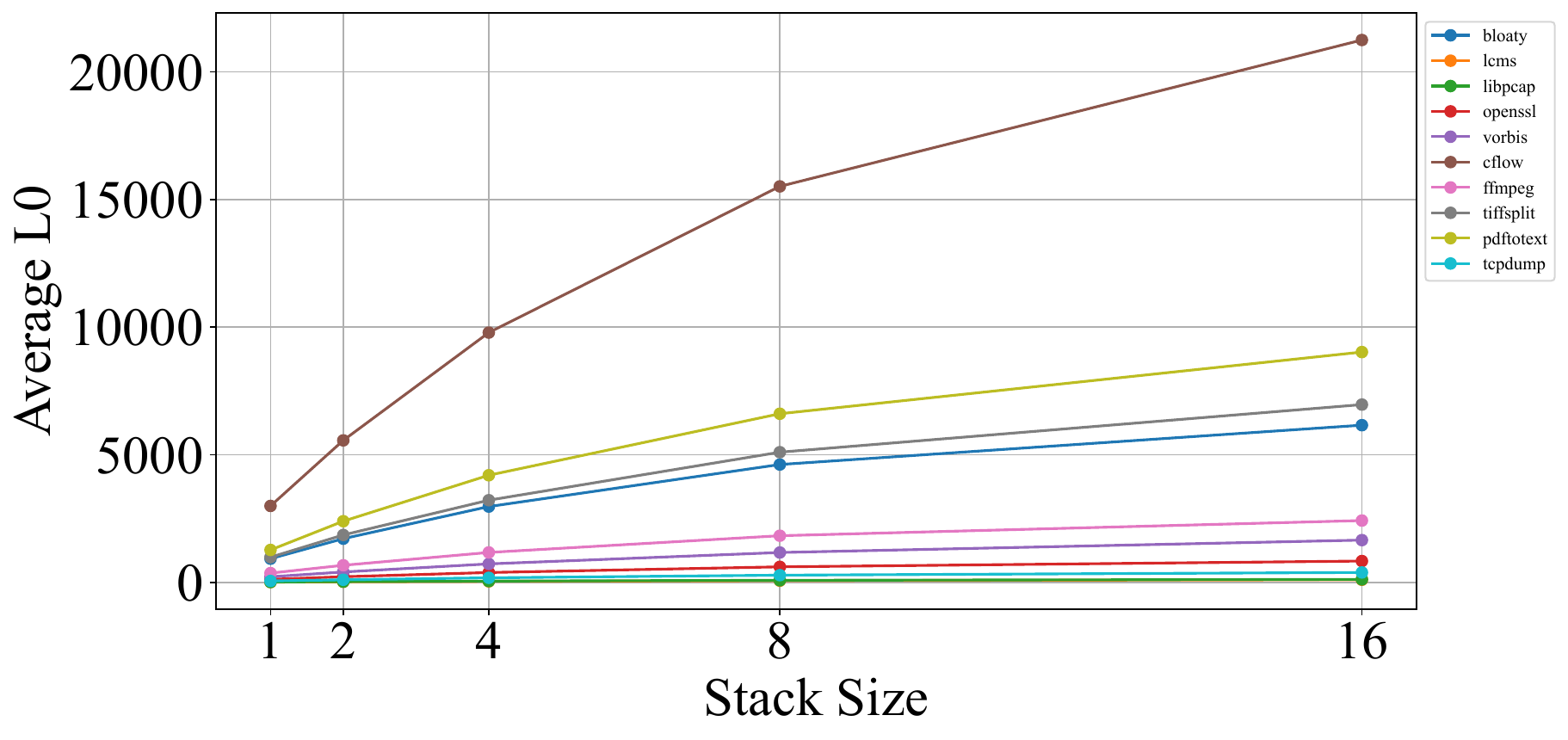}
        \caption{L0 Norm}
        \label{fig:l0}
    \end{subfigure}
    \hfill
    \begin{subfigure}{0.45\textwidth}
        \centering
        \includegraphics[width=\linewidth]{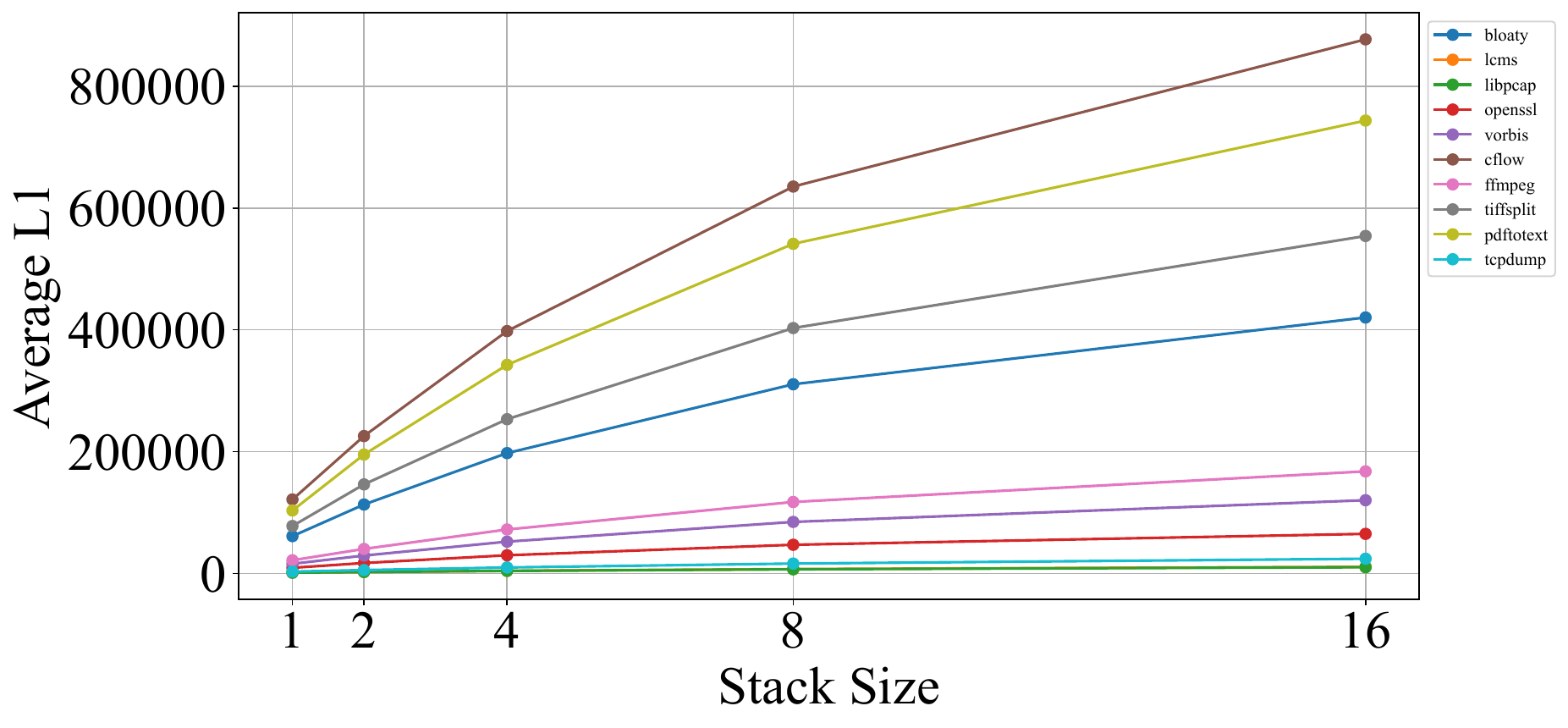}
        \caption{L1 Norm}
        \label{fig:l1}
    \end{subfigure}
    \hfill
    \begin{subfigure}{0.45\textwidth}
        \centering
        \includegraphics[width=\linewidth]{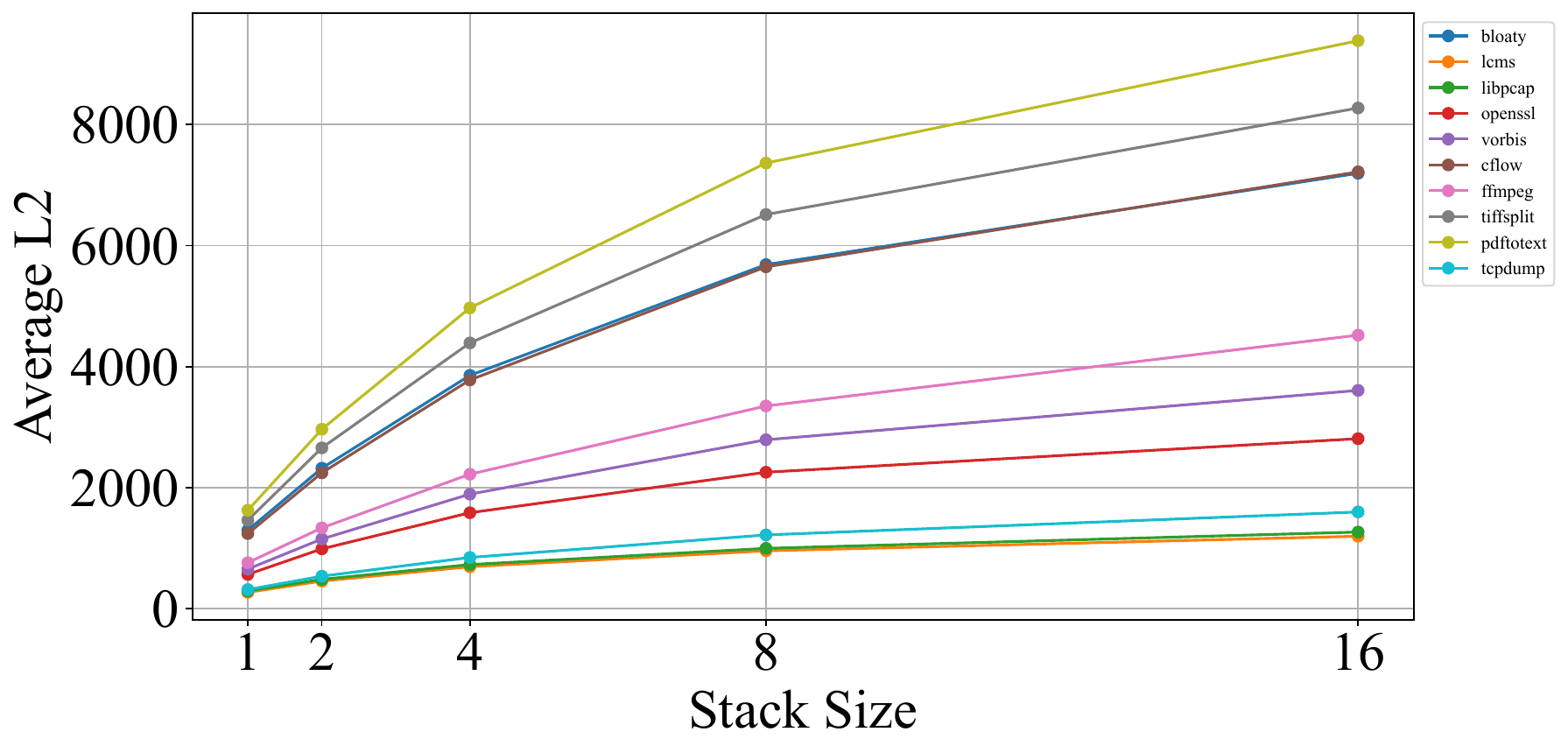}
        \caption{L2 Norm}
        \label{fig:l2}
    \end{subfigure}
    \hfill
    \begin{subfigure}{0.45\textwidth}
        \centering
        \includegraphics[width=\linewidth]{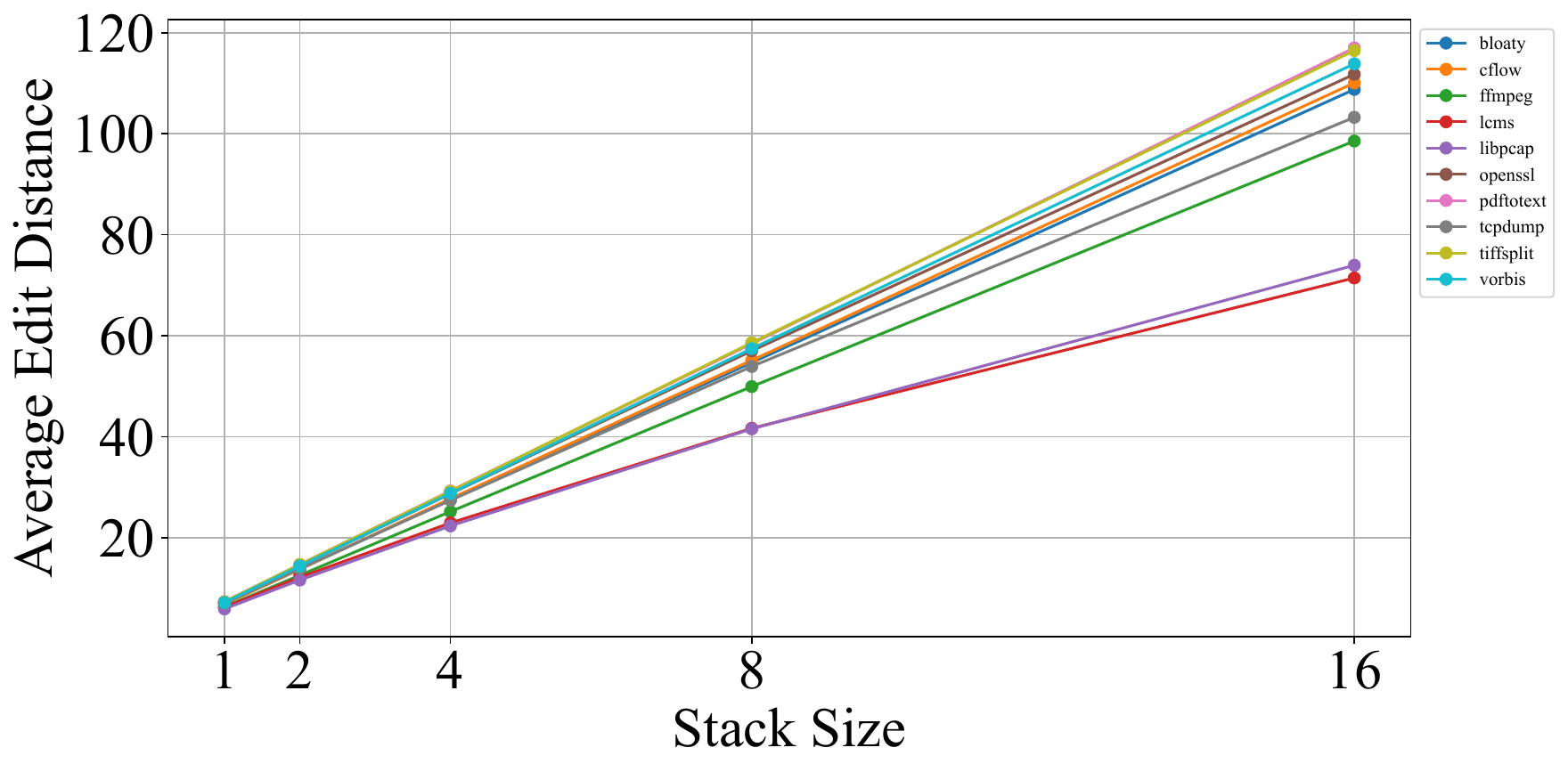}
        \caption{Edit Distance}
        \label{fig:edit}
    \end{subfigure}
    \caption{\textbf{Changes in perturbation distances across different havoc stack sizes (1, 2, 4, 8, 16). Each line represents a target program's perturbation distance measured by L0, L1, L2 norms and edit distance. Results from all ten targets are plotted individually.}}
    \label{fig:stack}
\end{figure}

The results are presented in Figure \ref{fig:stack}, which includes separate graphs for each metric. In these ten targets, the perturbation distance exhibits an increase with the increase in the size of the havoc stack. The results indicate a clear correlation, demonstrating that the havoc stack effectively amplifies the perturbation distance.

\vspace{0cm}

\begin{longfbox}
\textbf{Result 3:} 
The perturbation distance increases with larger havoc stack size.
\end{longfbox}

\subsection{RQ4: Effect of Splicing}
\subsubsection{Experimental Setup}
We designed our experiments to thoroughly investigate the effects of the splicing operator and the splicing stage on perturbation distance. We chose AFL++ as our measurement fuzzer due to its implementation of both the splicing operator and the splicing stage. To maintain consistency and minimize external variances, we used the same targets and initial seeds as in our previous experiments. For each target program, we conducted 10 independent measurements to ensure statistical reliability.

For testing splicing stage, we set the havoc stack to vanilla settings. The experiment consisted of 20,000 executions: the first 10,000 in normal stage to generate a diverse seed pool, followed by 10,000 in splicing stage. This design ensures the availability of two seeds for the splicing operation. We then measured the average perturbation distance of the resulting mutants. To evaluate the splicing operator, we set the havoc stack to 1 and compared the average perturbation distance produced by the splicing operator against other operators. 
In both experiments, we maintained the same measurement dimensions as in our previous studies, utilizing four metrics to quantify perturbation distances: L0-norm, L1-norm, L2-norm, and edit distance.

\subsubsection{Observation}
\textbf{Splicing Stage} The results are presented in Table \ref{tab:splicing_mode}. Using all four metrics, our analysis reveals that across all 10 target programs, the perturbation distances in the splicing stage consistently exceed those in normal stage. This observation confirms that the splicing stage introduces larger mutations compared to the normal stage.

\textbf{Splicing Operator} The results are presented in Table \ref{tab:splicing_operator}. The data show that the splicing operator generates consistently larger perturbation distances compared to other operators across all targets. 

\begin{longfbox}
\textbf{Result 4:} 
Both splicing stage and splicing operator generate larger perturbation distances compared to normal stage and other operators, respectively.
\end{longfbox}

\begin{table}[h]
\small
    \centering
    \caption{\textbf{L0-norm, L1-norm, L2-norm and edit distance in normal stage (-N suffix) and splicing stage (-S suffix). Bold numbers indicate the higher value between normal and splicing stages.}}
    \label{tab:splicing_mode}
    \setlength{\tabcolsep}{0.8pt}
    \renewcommand{\arraystretch}{0.9}
    \begin{tabular}{ccccccccc}
    \toprule
    \textbf{Targets} & \textbf{L0-N} & \textbf{L0-S} & \textbf{L1-N} & \textbf{L1-S} & \textbf{L2-N} & \textbf{L2-S} & \textbf{Edit-N} & \textbf{Edit-S} \\ \midrule
    \texttt{cflow} & 10,238 & \textbf{15,041} & 418,382 & \textbf{673,027} & 3,802 & \textbf{6,168} & 36 & \textbf{8,898} \\ 
    \texttt{bloaty} & 2,864 & \textbf{10,627} & 191,824 & \textbf{1,020,463} & 3,626 & \textbf{11,615} & 33 & \textbf{7,581} \\ 
    \texttt{pdftotext} & 4,088 & \textbf{5,981} & 334,316 & \textbf{500,484} & 4,697 & \textbf{7,556} & 34 & \textbf{4,287} \\ 
    \texttt{tiffsplit} & 3,745 & \textbf{5,606} & 295,852 & \textbf{476,705} & 4,852 & \textbf{7,468} & 44 & \textbf{3,013} \\ 
    \texttt{ffmpeg} & 897 & \textbf{3,124} & 55,101 & \textbf{461,475} & 1,719 & \textbf{9,800} & 19 & \textbf{2,159} \\ 
    \texttt{vorbis} & 736 & \textbf{1,587} & 53,111 & \textbf{139,519} & 1,841 & \textbf{4,081} & 34 & \textbf{1,194} \\ 
    \texttt{openssl} & 402 & \textbf{741} & 30,973 & \textbf{60,058} & 1,562 & \textbf{2,714} & 35 & \textbf{489} \\ 
    \texttt{tcpdump} & 149 & \textbf{304} & 8,345 & \textbf{23,111} & 699 & \textbf{1,594} & 26 & \textbf{232} \\ 
    \texttt{libpcap} & 52 & \textbf{81} & 4,238 & \textbf{7,291} & 688 & \textbf{1,067} & 25 & \textbf{66} \\ 
    \texttt{lcms} & 40 & \textbf{75} & 4,221 & \textbf{7,291} & 633 & \textbf{949} & 24 & \textbf{70} \\ \bottomrule
    \end{tabular}
\end{table}\vspace{0cm}

\begin{table}[h]
\small
    \centering
    \caption{\textbf{L0-norm, L1-norm, L2-norm and edit distance in other mutation operators (-O suffix) and splicing operator (-S suffix). Bold numbers indicate the higher value between other operators and splicing operator.}}
    \label{tab:splicing_operator}
    \setlength{\tabcolsep}{0.8pt}
    \renewcommand{\arraystretch}{0.9}
    \begin{tabular}{ccccccccc}
    \toprule
    \textbf{Targets} & \textbf{L0-O} & \textbf{L0-S} & \textbf{L1-O} & \textbf{L1-S} & \textbf{L2-O} & \textbf{L2-S} & \textbf{Edit-O} & \textbf{Edit-S} \\ \midrule
    \texttt{cflow} & 2,308 & \textbf{7,565} & 93,400 & \textbf{306,402} & 971 & \textbf{3,047} & 6 & \textbf{15} \\ 
    \texttt{bloaty} & 721 & \textbf{2,307} & 47,398 & \textbf{152,006} & 1,025 & \textbf{3,113} & 6 & \textbf{16} \\ 
    \texttt{pdftotext} & 1,014 & \textbf{3,239} & 82,464 & \textbf{263,318} & 1,306 & \textbf{4,038} & 6 & \textbf{16} \\ 
    \texttt{tiffsplit} & 780 & \textbf{2,472} & 61,489 & \textbf{195,378} & 1,169 & \textbf{3,586} & 6 & \textbf{17} \\ 
    \texttt{ffmpeg} & 286 & \textbf{898} & 16,942 & \textbf{53,505} & 619 & \textbf{1,721} & 5 & \textbf{14} \\ 
    \texttt{vorbis} & 172 & \textbf{555} & 12,380 & \textbf{39,841} & 524 & \textbf{1,555} & 6 & \textbf{16} \\ 
    \texttt{openssl} & 93 & \textbf{301} & 7,124 & \textbf{23,232} & 453 & \textbf{1,335} & 6 & \textbf{16} \\ 
    \texttt{tcpdump} & 44 & \textbf{140} & 2,318 & \textbf{7,300} & 264 & \textbf{685} & 6 & \textbf{16} \\ 
    \texttt{libpcap} & 13 & \textbf{38} & 1,031 & \textbf{2,867} & 259 & \textbf{551} & 5 & \textbf{13} \\ 
    \texttt{lcms} & 9 & \textbf{30} & 992 & \textbf{3,143} & 227 & \textbf{575} & 5 & \textbf{14} \\ \bottomrule
    \end{tabular}
\end{table}\vspace{0cm}
\subsection{RQ5: Zero-Execution \added{Fuzzing-Driven} Taint Inference}
\label{rq:ztaint}
\subsubsection{Experimental Setup}
To achieve a comprehensive evaluation, we leverage all targets from the FuzzBench and UniBench datasets. Since GreyOne\cite{gan2020greyone} and ProFuzz\cite{you2019profuzzer} are not open-source, we compare our approach with state-of-the-art fuzzers\cite{aflpp_fuzzbench_report} to demonstrate the effectiveness of our method. We assess the coverage of AFL++ and \sys under two different configurations: 1) default settings with no additional features, and 2) enhanced settings with both dictionary and AFL++ cmplog mode\cite{Aschermann2019} enabled. For each target, following previous research and typical fuzzing evaluation methodologies\cite{klees2018evaluating, liu2023vd, zhang2023shapfuzz, zhang2025low}, we conduct 10 campaigns of 24 hours for each pair of fuzzer programs. To maintain fairness in our comparison, each fuzzer operates on a single core per fuzzing campaign. We use the same seeds as Section \ref{text:rq1.1}. Our analysis includes standard statistical summaries, such as the mean over our evaluation metrics. Furthermore, we employ the Mann-Whitney U test to verify that the observed performance differences are statistically significant. This nonparametric test does not rely on assumptions about distribution, making it a popular choice in software testing studies to evaluate randomized algorithms\cite{tests2011arcuri} and fuzzers\cite{Aschermann2019}. 
\subsubsection{Observation}

\textbf{FuzzBench Targets} 
\replaced{Across FuzzBench targets (see results in \autoref{eval:fuzzbench}), \sys demonstrates superior performance compared to \aflpp in 12 out of 19 programs, achieving mean coverage improvements. For instance, \texttt{lcms} shows a remarkable 20.38\% coverage increase over \aflpp. When comparing \syscmplogdict to \aflppcmplogdict, the performance gap widens further, with \syscmplogdict outperforming \aflppcmplogdict on 13 out of 19 programs. Notable improvements include a 13.91\% increase for \texttt{libpcap}.}{Across FuzzBench targets (see results in \autoref{eval:fuzzbench}), \sys demonstrates superior performance compared to \aflpp in 13 out of 19 programs, achieving mean coverage improvements. For instance, \texttt{lcms} shows a remarkable 33.71\% coverage increase over \aflpp. When comparing \syscmplogdict to \aflppcmplogdict, the performance gap remains consistent, with \syscmplogdict outperforming \aflppcmplogdict on 13 out of 19 programs. Notable improvements include a 8.63\% increase for \texttt{libxslt}.}
The mean gain of \sys over AFL++ is \replaced{1.32\%}{2.97\%}, while \syscmplogdict achieves a \replaced{1.67}{1.65}\% mean gain over \aflppcmplogdict. As noted in previous research\cite{she2024fox}, the performance gap on FuzzBench is typically smaller than on standalone programs due to FuzzBench's well-optimized baseline implementations.
However, in some cases such as \replaced{\texttt{openssl}, \texttt{curl}, and \texttt{sqlite3}}{\texttt{stbi}}, \aflpp slightly outperforms \sys. These exceptions may be attributed to challenges with nested conditions, instrumentation overhead that affects throughput, and the lack of specialized constraint solvers for taint-guided mutation.

\textbf{Standalone Targets}
Comparison of mean coverage achieved by \sys against \aflpp, \syscmplogdict against \aflppcmplogdict is presented in \autoref{eval:standalone}. Across standalone programs, \sys exhibits improvements over \aflpp on \replaced{13}{11} programs, achieving up to \replaced{4.91\%}{6.12\%} more code coverage on average across the standalone target set. One of the targets where we see notable improvement is \texttt{nm-new}, where \sys uncovers \replaced{31.65\%}{51.12\%} more edges than \aflpp. Similarly, \syscmplogdict outperforms \aflppcmplogdict on 10 programs, including a remarkable \replaced{35.69\%}{41.53\%} improvement on \texttt{mp42aac}.
But in some cases such as \replaced{\texttt{ sqlite3}}{\texttt{infotocap}}, \aflpp slightly outperforms \sys. These exceptions may be attributed to the same reason as discussed before.

\begin{table*}[!ht]
\footnotesize

\caption{\small\textbf{Mean edge coverage of \sys and \aflpp, as well as \syscmplogdict (\syscmplogdictshort) and \aflppcmplogdict (\aflppcmplogdictshort) on \numfuzzbench FuzzBench programs (left) and \numstandalone standalone programs (right) for 24 hours over 10 runs. We mark the highest number in bold for each comparison.* indicates statistically significant differences.}}
\begin{minipage}{0.48\textwidth}
    \centering
    \setlength{\tabcolsep}{0.8pt}
    \label{eval:fuzzbench}
    \begin{tabular}{lrr|rr}
    \toprule
    \textbf{Targets} & \textbf{\sys} & \textbf{\aflpp} & \textbf{\syscmplogdictshort} & \textbf{\aflppcmplogdictshort} \\ 
    \midrule
    bloaty & \textbf{3,165} & 3,105 & \textbf{2,763} & 2,736 \\
    curl & \textbf{14,089} & 14,049 & 16,158 & \textbf{16,408} \\
    libxslt & \textbf{4,755} & 4,682 & \textbf{5,613*} & 5,167 \\
    harfbuzz & \textbf{23,696} & 23,432 & 25,039 & \textbf{25,627} \\
    jsoncpp & 1,338 & 1,338 & 1,343 & 1,343 \\
    lcms & \textbf{1,652} & 1,235 & \textbf{2,713} & 2,547 \\
    libjpeg & \textbf{3,244} & 3,177 & \textbf{3,240} & 3,235 \\
    libpcap & 45 & \textbf{50} & \textbf{4,438} & 4,307 \\
    libpng & \textbf{2,306} & 2,281 & 2,656 & \textbf{2,670} \\
    libxml2 & \textbf{10,896} & 9,128 & \textbf{13,332} & 12,764 \\
    vorbis & 2,044 & \textbf{2,045} & \textbf{2,039} & 2,033 \\
    openh264 & 13,543 & \textbf{13,567} & \textbf{11,560} & 10,716 \\
    openssl & \textbf{4,651} & 4,625 & 4,628 & \textbf{4,634} \\
    woff2 & \textbf{2,395} & 2,356 & \textbf{2,550*} & 2,489 \\
    zlib & 881 & \textbf{884} & \textbf{905} & 889 \\
    re2 & \textbf{6,260} & 6,253 & \textbf{6,261*} & 6,199 \\
    sqlite3 & \textbf{14,087} & 13,880 & \textbf{16,670} & 16,617 \\
    stbi & 2,112 & \textbf{2,132} & 2,842 & \textbf{2,911} \\
    systemd & \textbf{2,533} & 2,525 & \textbf{2,490} & 2,474 \\
    \midrule
    \multicolumn{2}{c|}{Mean gain}   & 2.97\% & \multicolumn{2}{c}{1.65\%} \\
    \bottomrule
    \end{tabular}
\end{minipage}
\hfill
\begin{minipage}{0.48\textwidth}
    \centering
    \setlength{\tabcolsep}{0.8pt}
    \label{eval:standalone}
    \begin{tabular}{lrr|rr}
    \toprule
        \textbf{Targets} & \textbf{\sys} & \textbf{\aflpp} & \textbf{\syscmplogdictshort} & \textbf{\aflppcmplogdictshort} \\
                \midrule
exiv2 & \textbf{5,688} & 4,776 & \textbf{7,130} & 6,623 \\
tiffsplit & \textbf{2,503*} & 2,173 & \textbf{2,338*} & 2,156 \\
mp3gain & \textbf{1,554} & 1,552 & \textbf{1,578} & 1,574 \\
mujs & 6,616 & \textbf{6,696} & \textbf{6,921} & 6,838 \\
pdftotext & \textbf{6,838} & 6,771 & 6,936 & \textbf{7,063} \\
infotocap & 2,139 & \textbf{2,317} & 2,263 & \textbf{2,448*} \\
mp42aac & \textbf{1,843*} & 1,739 & \textbf{2,903*} & 2,051 \\
flvmeta & 477 & 477 & 476 & 476 \\
objdump & \textbf{5,578*} & 5,429 & 4,973 & 4,973 \\
tcpdump & \textbf{8,078*} & 7,146 & \textbf{6,362} & 5,851 \\
ffmpeg & \textbf{22,221} & 21,734 & \textbf{18,986} & 18,502 \\
jq & 3,529 & \textbf{3,548} & 3,368 & \textbf{3,383} \\
cflow & 2,118 & \textbf{2,122} & 2,078 & \textbf{2,080} \\
nm-new & \textbf{3,386*} & 2,240 & 1,431 & \textbf{1,441} \\
sqlite3 & \textbf{15,682} & 15,220 & \textbf{14,637} & 14,273 \\
lame & \textbf{5,732} & 5,731 & \textbf{5,559} & 5,467 \\
jhead & 361 & 361 & \textbf{378} & 372 \\
        \midrule
        \multicolumn{2}{c|}{Mean gain} & 6.12\% & \multicolumn{2}{c}{3.86\%} \\
        \bottomrule
    \end{tabular}
\end{minipage}
\end{table*}

\begin{longfbox}
\textbf{Result 5:} 
\replaced{\sys outperforms existing state-of-the-art fuzzers with average improvements of 4.91\% on UniBench and 1.32\% on FuzzBench targets.}{\sys outperforms existing state-of-the-art fuzzers with average improvements of 6.12\% on UniBench and 2.97\% on FuzzBench targets.}
\end{longfbox}

\subsection{RQ6: Threshold parameter $k$}
\subsubsection{Experimental Setup}
\replaced{To evaluate the impact of parameter $k$ on \sys's performance, we select five programs from our benchmark suite. These programs are chosen from Section~\autoref{rq:ztaint} where \sys demonstrates statistically significant improvements over the baseline, allowing us to better observe the effects of different parameter settings. We configure three groups with different threshold values: $k=1$, $k=5$, and $k=20$.}{To achieve a comprehensive evaluation, we leverage all targets from the FuzzBench and UniBench datasets. To investigate the impact of different k values on \sys's performance, we configure six different settings: five fixed k values (k=1, k=10, k=100, k=1000, and k=10000) and one adaptive k setting.}
 For each target program, we conduct concurrent runs across all settings using the same seed inputs as in previous sections. Each configuration is tested with 10 independent instances running for 8 hours.

\subsubsection{Observation}
\replaced{As shown in Table~\ref{tab:threshold}, the optimal threshold value $k$ varies across different programs. For instance, while mp42aac and nm-new achieve their best coverage with $k=20$ (1795 and 2277 edges respectively), tcpdump performs optimally with $k=5$ (6447 edges), and libxml2\_xml and lcms reach their peak performance with $k=1$ (9392 and 1577 edges respectively). This variance aligns with our theoretical analysis: a small $k$ value may lead to undertaint issues where potential byte dependencies are missed, while a large $k$ value can result in overtaint problems and reduced efficiency due to excessive mutation attempts on irrelevant bytes. Empirically, $k=5$ achieves the best overall performance across all programs (with total edge coverage of 20970), thus we choose $k=5$ as the default setting in our experiments.}{As shown in Table~\ref{tab:threshold}, the adaptive k setting achieves the highest global wins (glo\_win) across all targets, where glo\_win represents the number of targets for which a particular setting achieves the highest average coverage. For constant k values, there is no clear trend in glo\_win as k increases or decreases: k=1, k=10, k=100, k=1000, and k=10000 achieve glo\_win values of 5, 5, 2, 9, and 7 respectively. This observation suggests that no single fixed k value consistently outperforms others across different targets. The results indicate that adaptive k effectively adjusts to the specific characteristics of each target program, eliminating the need for manual parameter tuning.}

\added{We attribute these observations to two main factors. First, as k increases, the overhead for tracking byte-level differences becomes more significant, leading to reduced overall throughput in the fuzzing process. Second, the choice of k value presents a fundamental trade-off: a small k may result in undertaint issues where potential byte dependencies are missed, while a large k can lead to overtaint problems where excessive bytes are incorrectly identified as related, causing inefficient mutation attempts on irrelevant bytes.}

\begin{longfbox}
\textbf{Result 6:} 
\replaced{The performance of \sys varies with different threshold values $k$, with $k=5$ empirically showing the best overall performance.}{The performance of \sys varies with different k values, with the adaptive k setting demonstrating better applicability across diverse targets compared to fixed k values.}
\end{longfbox}

\begin{table}[]

    \centering
    \caption{\textbf{\added{Coverage comparison between different k values (k=1, 10, 100, 1000, 10000, and adaptive) across FuzzBench and UniBench targets. Each number represents the average edge coverage from 10 eight-hour independent runs. Glo\_Win shows the number of targets where each setting achieves the highest average coverage. The highest number for each target is marked in bold.}}}
    \setlength{\tabcolsep}{1.5pt}
    \footnotesize
\begin{tabular}{lrrrrr|r|lrrrrr|r}
\toprule
\textbf{Targets} & \textbf{1} & \textbf{10} & \textbf{100} & \textbf{1K} & \textbf{10K} & \textbf{Ada} &
\textbf{Targets} & \textbf{1} & \textbf{10} & \textbf{100} & \textbf{1K} & \textbf{10K} & \textbf{Ada} \\
\midrule
curl & 13,832 & 13,840 & 13,900 & \textbf{14,059} & 13,414 & 13,935 &
woff2 & \textbf{2,409} & 2,383 & 2,396 & 2,405 & 2,334 & 2,404 \\
bloaty & 3,088 & \textbf{3,120} & 3,050 & 3,086 & 3,103 & 3,093 &
jhead & 357 & 357 & 358 & \textbf{360} & 357 & \textbf{360} \\
harfbuzz & 22,675 & \textbf{22,820} & 22,024 & 22,244 & 22,404 & 22,671 &
jq & 3,378 & 3,348 & 3,351 & 3,352 & 3,335 & \textbf{3,451} \\
zlib & \textbf{877} & \textbf{877} & \textbf{877} & \textbf{877} & \textbf{877} & \textbf{877} &
sqlite3 & 12,835 & 12,995 & 11,849 & \textbf{13,096} & 12,642 & 12,574 \\
jsoncpp & \textbf{1,337} & 1,331 & 1,328 & 1,332 & 1,333 & 1,333 &
cflow & 2,073 & 2,073 & 2,071 & 2,072 & 2,072 & \textbf{2,097} \\
lcms & 1,226 & 1,588 & 1,643 & 1,485 & 1,259 & \textbf{1,683} &
exiv2 & 2,103 & 2,233 & 2,116 & 2,106 & \textbf{3,236} & 2,662 \\
libjpeg & 3,208 & 3,213 & 3,154 & 3,071 & \textbf{3,214} & 3,198 &
ffmpeg & 7,327 & 9,129 & 7,606 & 6,855 & 6,813 & \textbf{13,933} \\
libpcap & 47 & 47 & 42 & 44 & 47 & \textbf{49} &
infotocap & 1,672 & 1,695 & 1,672 & 1,668 & 1,656 & \textbf{1,718} \\
libpng & \textbf{1,405} & 1,366 & 1,360 & 1,361 & 1,361 & 1,360 &
nm-new & 1,505 & 1,567 & 1,514 & 1,538 & 1,543 & \textbf{1,965} \\
libxml2 & 8,808 & 8,658 & 9,044 & 8,781 & 8,848 & \textbf{9,546} &
objdump & 4,944 & 4,990 & 5,099 & 4,997 & 4,990 & \textbf{5,215} \\
libxslt & 4,681 & 4,732 & \textbf{4,742} & 4,645 & 4,639 & 4,659 &
pdftotext & 6,239 & 6,252 & 6,256 & \textbf{6,279} & 6,229 & 6,267 \\
openh264 & 13,640 & 13,608 & 13,148 & \textbf{13,647} & 13,628 & 13,591 &
mp3gain & 1,429 & 1,415 & 1,412 & 1,406 & \textbf{1,453} & 1,424 \\
openssl & \textbf{4,576} & 4,554 & 4,550 & 4,569 & 4,550 & 4,559 &
tcpdump & 2,916 & 3,029 & 3,283 & 2,998 & 3,082 & \textbf{4,294} \\
re2 & 6,255 & 6,240 & 6,223 & \textbf{6,269} & 6,259 & 6,259 &
mujs & 5,117 & 4,939 & 5,039 & 4,900 & 4,914 & \textbf{5,456} \\
stbi & 2,224 & 2,221 & 2,164 & \textbf{2,332} & 1,931 & 2,086 &
tiffsplit & 1,096 & 1,355 & 1,293 & 1,201 & 1,206 & \textbf{1,412} \\
sql & 11,774 & 11,727 & 11,899 & 11,616 & 11,745 & \textbf{11,932} &
mp42aac & 1,615 & 1,676 & 1,667 & 1,662 & \textbf{1,682} & 1,587 \\
systemd & 2,511 & \textbf{2,522} & 2,521 & 2,516 & 2,518 & \textbf{2,522} &
flvmeta & 442 & 445 & 442 & \textbf{446} & \textbf{446} & 436 \\ \cline{8-14}
vorbis & 2,042 & \textbf{2,055} & 2,049 & 2,049 & 2,043 & 2,037 &  \multicolumn{1}{c}{\multirow{2}{*}{\textbf{Glo\_Win}}} &  \multicolumn{1}{c}{\multirow{2}{*}{\textbf{5}}} &  \multicolumn{1}{c}{\multirow{2}{*}{\textbf{5}}} &  \multicolumn{1}{c}{\multirow{2}{*}{\textbf{2}}} &  \multicolumn{1}{c}{\multirow{2}{*}{\textbf{9}}} &  \multicolumn{1}{c|}{\multirow{2}{*}{\textbf{7}}} &  \multicolumn{1}{c}{\multirow{2}{*}{\textbf{16}}}  \\
lame & 5,639 & 5,654 & 5,653 & 5,611 & \textbf{5,682} & 5,679 & & & & & & & \\
\bottomrule
\end{tabular}
    \label{tab:threshold}
\end{table}\vspace{0cm}

\subsection{\added{RQ7: Evaluation of \sys's Runtime Overhead\label{eval:overhead}}}
\subsubsection{Experimental Setup}
\added{To evaluate the runtime overhead introduced by \sys's fuzzing-driven taint inference mechanism, we directly utilize the execution count data from our experiments in Section \ref{rq:ztaint}. Specifically, we compare the number of executions achieved by \sys and \aflpp over 10 independent 24-hour runs across both FuzzBench and UniBench targets. This metric serves as a direct indicator of the runtime overhead, as any additional computational cost from fuzzing-driven taint inference would manifest as reduced execution throughput. The experimental configuration, including the hardware setup, seed inputs, and single-core execution environment, remains identical to that described in Section \ref{rq:ztaint}.}

\subsubsection{Observation}
\added{
As shown in Table~\ref{tab:throuput}, the overhead varies between the two benchmark suites: UniBench shows a lower overhead of 3.84\%, while FuzzBench exhibits a higher overhead of 12.58\%. These results indicate that the combined overhead from instrumentation and fuzzer-side logic is relatively modest.
The higher overhead observed in FuzzBench can be attributed to its generally higher execution throughput. Since FuzzBench targets typically achieve more executions per second, the fuzzer-side tracking operations become more prominent, leading to increased relative overhead. Nevertheless, this level of overhead remains acceptable for practical fuzzing applications.
}

\begin{table}[]
\small
    \centering
    \caption{\added{\textbf{Mean number of executions (in thousands) of \sys (\sysshort) and \aflpp across FuzzBench and UniBench targets over 10 24-hour independent runs. Overhead shows the percentage reduction in executions of \sys compared to \aflpp for each dataset.}}}
    \footnotesize
    \setlength{\tabcolsep}{4pt}
\begin{tabular}{lrr|lrr|lrr}
\toprule
\textbf{Targets} & \textbf{\sysshort} & \textbf{\aflpp} & 
\textbf{Targets} & \textbf{\sysshort} & \textbf{\aflpp} & 
\textbf{Targets} & \textbf{\sysshort} & \textbf{\aflpp} \\
\midrule
\multicolumn{3}{c|}{\textbf{FuzzBench Targets}} & 
sql & 88,947 & 110,594 &
tcpdump & 8,193 & 8,129 \\ 
\cline{1-3}
bloaty & 31,982 & 29,848 &
systemd & 24,597 & 27,272 &
jq & 2,309 & 2,659 \\
zlib & 12,451 & 12,346 &
jsoncpp & 104,963 & 135,009 &
sqlite3 & 1,896 & 1,901 \\	
lcms & 215,976 & 306,414 &
libjpeg & 651,025 & 708,923 &
cflow & 3,666 & 3,686 \\
libpcap & 89,014 & 92,063 &
libpng & 1,059,813 & 800,547 &
exiv2 & 4,769 & 5,828 \\
libxml2 & 229,220 & 249,728 &
openssl & 737,275 & 925,637 &
ffmpeg & 877 & 1,124 \\ \cline{4-6}
openh264 & 3,998 & 3,997 &
\multirow{2}{*}{\normalsize\textbf{Overhead}} &  \multicolumn{2}{c|}{\multirow{2}{*}{{\normalsize\textbf{3.84\%}}}} &
infotocap & 4,614 & 4,408 \\
libxslt & 497,771 & 530,097 &
\multicolumn{3}{c|}{} &
jhead & 12,860 & 12,642 \\ \cline{4-6}
re2 & 123,943 & 221,797 &
\multicolumn{3}{c|}{{\textbf{UniBench Targets}}} & 
nm-new & 6,446 & 7,183 \\ \cline{4-6}
stbi & 35,938 & 44,769 &
objdump & 5,947 & 6,110 &
tiffsplit & 13,499 & 14,177 \\
vorbis & 273,842 & 304,640 &
mp42aac & 9,063 & 9,835 &
pdftotext & 4,328 & 4,417 \\
woff2 & 305,926 & 444,382 &
lame & 2,498 & 2,419 &
mp3gain & 7,933 & 7,473 \\ \cline{7-9}
curl & 514,058 & 711,715 &
mujs & 8,685 & 8,894 &
\multirow{2}{*}{\normalsize\textbf{Overhead}} &  \multicolumn{2}{c}{\multirow{2}{*}{{\normalsize\textbf{12.58\%}}}} \\
harfbuzz & 332,316 & 411,216 &
flvmeta & 16,235 & 15,883 &
\multicolumn{3}{c}{} \\
\bottomrule
\end{tabular}
    \label{tab:throuput}
\end{table}
\vspace{0cm}

\begin{longfbox}
\textbf{Result 7:} 
\added{\sys shows a runtime overhead of 3.84\% on UniBench and 12.58\% on FuzzBench, demonstrating that the additional computational cost remains within an acceptable range.}
\end{longfbox}

\subsection{\added{RQ8: Comparision with \cmplog \label{eval:cmplog}}}
\subsubsection{Experimental Setup}
\added{\sys is designed to enhance the foundational mutation algorithm - havoc mode without any branch-specific solver support. To facilitate a fair comparison with branch-specific approaches like \cmplog, we implement \syssolver, a variant of \sys equipped with a simple gradient descent-based branch solver, similar to previous work such as \cite{chen2018angora, gan2020greyone, Aschermann2019, patafuzz, you2019profuzzer, she2024fox}. \syssolver incorporates specialized branch handling similar to \cmplog's approach. Using all targets from FuzzBench and UniBench datasets, we conduct 10 independent 24-hour runs with the same seeds as Section \ref{text:rq1.1}. The remaining experimental settings follow those described in Section \ref{rq:ztaint}.}

\subsubsection{Observation}
\added{On FuzzBench targets, \syssolver outperforms \cmplog on 10 out of 19 programs, achieving a mean coverage gain of 3.06\%. A notable example is \texttt{bloaty}, where \syssolver demonstrates a significant 59.6\% improvement over \cmplog. For UniBench targets, \syssolver shows superior performance over \cmplog on 13 out of 17 programs, with an average improvement of 13.54\% across all targets. A remarkable case is \texttt{tcpdump}, where \syssolver achieves 154.61\% improvement in coverage compared to \cmplog. However, on certain targets like \texttt{lcms}, \cmplog maintains better performance. This can be attributed to our solver implementation being a simple prototype, lacking some specialized handling mechanisms present in \cmplog, such as floating-point number processing and string comparisons where both operands are variables. This does not indicate a design limitation of zero-execution FTI, and we plan to incorporate these features in future work to further enhance our solver's capabilities.}

\begin{table}[]
    \centering
    \caption{\added{\textbf{Mean edge coverage of \syssolver (\syssolvershort) and \cmplog across FuzzBench and UniBench targets over 10 24-hour independent runs. Gain shows the percentage improvement in coverage of \syssolver compared to \cmplog for each dataset. The highest number for each target is marked in bold. * indicates statistically significant differences.}}}
    \footnotesize
    \setlength{\tabcolsep}{4pt}
    \begin{tabular}{lrr|lrr|lrr}
    \toprule
    \textbf{Targets} & \textbf{\syssolvershort} & \textbf{\cmplog} & 
    \textbf{Targets} & \textbf{\syssolvershort} & \textbf{\cmplog} & 
    \textbf{Targets} & \textbf{\syssolvershort} & \textbf{\cmplog} \\
    \midrule
    \multicolumn{3}{c|}{\textbf{FuzzBench Targets}} & 
    sql & 11,797 & \textbf{17,111*} &
    tcpdump & \textbf{7,046*} & 2,767 \\ 
    \cline{1-3}
    bloaty & \textbf{4,465*} & 2,798 &
    systemd & \textbf{2,491} & 2,488 &
    jq & \textbf{3,538} & 2,879 \\
    zlib & 903 & \textbf{907} &
    jsoncpp & \textbf{1,341} & 1,334 &
    sqlite3 & 15,092 & \textbf{15,415} \\	
    lcms & 1,565 & \textbf{2,684*} &
    libjpeg & 3,231 & 3,231 &
    cflow & \textbf{2,122*} & 2,071 \\
    libpcap & \textbf{4,782*} & 3,573 &
    libpng & 2,401 & \textbf{2,691*} &
    exiv2 & \textbf{5,391} & 5,063 \\
    libxml2 & \textbf{13,496*} & 9,246 &
    libxslt & \textbf{5,338} & 5,263 &
    ffmpeg & \textbf{25,626*} & 18,305 \\ \cline{4-6}
    openh264 & \textbf{13,689} & 13,522 &
    \multirow{2}{*}{\normalsize\textbf{Gain}} &  \multicolumn{2}{c|}{\multirow{2}{*}{{\normalsize\textbf{3.06\%}}}} &
    infotocap & 1,890 & \textbf{2,284*} \\
    openssl & \textbf{4,604*} & 4,320 &
    \multicolumn{3}{c|}{} &
    jhead & \textbf{395*} & 361 \\ \cline{4-6}
    re2 & \textbf{6,268*} & 6,249 &
    \multicolumn{3}{c|}{{\textbf{UniBench Targets}}} & 
    nm-new & \textbf{1,706} & 1,411 \\    \cline{4-6}
    stbi & 2,798 & \textbf{3,024*} &
    tiffsplit & 1,348 & \textbf{1,787*} &
    objdump & \textbf{5,811*} & 4,954 \\
    vorbis & 2,039 & \textbf{2,041} &
    mp42aac & 1,607 & \textbf{1,783} &
    pdftotext & \textbf{6,835} & 6,806 \\
    woff2 & 2,425 & \textbf{2,551*} &
    lame & \textbf{5722} & \textbf{5544} &
    mp3gain & \textbf{1,494*} & 1,416 \\ \cline{7-9}
    curl & 14,765 & \textbf{14,824*} &
    mujs & \textbf{6,008} & 5,974 &
    \multirow{2}{*}{\normalsize\textbf{Gain}} &  \multicolumn{2}{c}{\multirow{2}{*}{{\normalsize\textbf{13.54\%}}}} \\ 
    harfbuzz & \textbf{25,483*} & 24,169 &
    flvmeta & \textbf{458} & 456 &
    \multicolumn{3}{c}{} \\ 
    \bottomrule
    \end{tabular}
    \label{tab:solver_coverage}
\end{table}\vspace{0cm}

\begin{longfbox}
\textbf{Result 8:} 
\added{\syssolver demonstrates competitive performance against \cmplog with average improvements of 13.54\% on UniBench and 3.06\% on FuzzBench targets.}
\end{longfbox}

\section{Related Work}
AFL\cite{ZalewskiAFL} ranks among the most prominent fuzzers globally, employing coverage-guided fuzzing to efficiently detect a multitude of vulnerabilities. Numerous researchers have based their subsequent studies on AFL's approach. 
AFLFast\cite{bohme2016coverage}, for instance, incorporates a strategic power schedule to modify both seed selection and the frequency of seed executions. Several other studies\cite{She2022EffectiveSS, wang2020not, wang2021reinforcement, zhang2022path} have also concentrated on optimizing seed schedules. \replaced{Additionally, various efforts have been made to tackle constraints that arise during fuzzing, such as utilizing symbolic execution\cite{stephens2016driller, peng2018t, yun2018qsym} and employing certain insightful algorithms\cite{Aschermann2019, chen2018angora, she2024fox}.}{Additionally, various efforts have been made to tackle constraints that arise during fuzzing, such as utilizing symbolic execution\cite{stephens2016driller, peng2018t, yun2018qsym, wang2018safl, huang2020pangolin} and its variants, including grey-box concolic testing\cite{choi2019grey}, which combines elements of symbolic execution and grey-box fuzzing. Furthermore, certain insightful algorithms\cite{Aschermann2019, chen2018angora, she2024fox, chen2019matryoshka, rawat2017vuzzer} have also been proposed to improve constraint solving and fuzzing efficiency.} 

Havoc is often integrated into many fuzzing processes. MOPT\cite{lyu2019mopt} applies a tailored Particle Swarm Optimization to manage the havoc operator. DARWIN\cite{jauernig2022darwin}, as proposed by Patrick et al., employs an Evolution Strategy to systematically refine and adjust the mutation operators' probability distribution. Additional experimentation by Wu et al.\cite{wu2022one} has demonstrated the efficacy of Havoc and its synergistic interactions. At SBFT 2025 fuzzing competition\cite{sbft2025}, our team built an ensemble fuzzer--HFuzz\cite{xie2025hfuzz} using an early version prototype of ZTaint. This implementation achieved second place in the competition.

Furthermore, Marcel et al. introduced directed fuzzing in AFLGo\cite{bohme2017directed}, leveraging the control flow graph (CFG) to compute distances. Recent advancements in directed fuzzing include approaches like Beacon\cite{huang2022beacon}, Titan\cite{huang2023titan}, $MC^2$\cite{shah2022mc2} and DAFL\cite{kim2023dafl}. 

Fuzzing has demonstrated its effectiveness across diverse domains. In software testing\cite{wen23usenixsecurity, wen23ccs}, it has been instrumental in identifying vulnerabilities across various applications, with applications extending to code similarity analysis\cite{wang2017memory}. The blockchain domain has seen extensive application of fuzzing techniques for smart contract testing\cite{wu2024we, weimin24sp,wong2024confuzz,su2022effectively, choi2021smartian}. Most recently, fuzzing has emerged as a promising approach in testing large language models, helping identify potential weaknesses and limitations in these AI systems\cite{xia2023universal,zhang2024your, deng2024large, deng2023large, yang2023fuzzing}.

\section{Discussion}
Our method, although effective, encounters several limitations. Firstly, it requires feedback, which entails code instrumentation and can introduce overhead, potentially reducing execution speed. This performance impact must be considered against the advantages our method offers in practical scenarios. Additionally, the success of our approach depends on the selected threshold value. Choosing a threshold that is too low may result in undertaint, possibly missing crucial correlations between input bytes and program behavior. Moreover, our use of filters to minimize noise in taint analysis leads to a smaller sample size for fuzzing-driven taint inference. While this filtering enhances the quality of our taint data, it might also decrease the efficiency of the inference process.

\section{Conclusion}
\added{In conclusion, this paper presents a computational model of havoc mode and demonstrates how it can be leveraged for zero-execution FTI in coverage-guided fuzzing. Our extensive evaluation on FuzzBench and UniBench shows that \sys achieves 2.97\% and 6.12\% higher edge coverage respectively compared to AFL++, demonstrating the effectiveness of utilizing existing havoc mutations for FTI without additional execution.}
\section{Data Availability Statement}
Our research artifacts are publicly available at: \url{https://github.com/Yu3H0/ZTaint-Havoc}

\begin{acks}
We sincerely appreciate the anonymous reviewers for their valuable feedback and guidance. We also thank Kunpeng Zhang, Shuangjie Yao and Qiao Zhang for their helpful comments and discussions that improved the paper.
\end{acks}
\newpage
\bibliographystyle{ACM-Reference-Format}
\bibliography{paper}


\end{document}